\documentclass[twocolumn,prx,superscriptaddress,longbibliography]{revtex4-1}
\usepackage{amsfonts}
\usepackage{amsmath}
\usepackage{amssymb}
\usepackage{amsbsy}
\usepackage{graphicx}
\usepackage{bm}
\usepackage{color}
\usepackage{mathrsfs}
\usepackage{latexsym,epsfig,graphicx}
\usepackage{dcolumn}
\usepackage{subfigure}
\usepackage{comment}
\usepackage{xspace}
\usepackage{epstopdf}
\usepackage{tabularx}
\usepackage{longtable}
\usepackage[colorlinks=true, letterpaper=true, pdfstartview=FitV, linkcolor=blue, citecolor=blue, urlcolor=blue]{hyperref}
\usepackage[normalem]{ulem}
\renewcommand{\Re}{\operatorname{Re}}
\renewcommand{\Im}{\operatorname{Im}}

\newcommand{\RNum}[1]{\uppercase\expandafter{\romannumeral #1\relax}}

\begin{document}
\title{Universal non-Hermitian skin effect in two and higher dimensions}
\author{Kai Zhang}
\affiliation{Beijing National Laboratory for Condensed Matter Physics and Institute of Physics, Chinese Academy of Sciences, Beijing 100190, China}
\affiliation{University of Chinese Academy of Sciences, Beijing 100049, China}
\author{Zhesen Yang}
\email{yangzs@ucas.ac.cn}
\affiliation{Kavli Institute for Theoretical Sciences, Chinese Academy of Sciences, Beijing 100190, China}
\author{Chen Fang}
\email{cfang@iphy.ac.cn}
\affiliation{Beijing National Laboratory for Condensed Matter Physics and Institute of Physics, Chinese Academy of Sciences, Beijing 100190, China}
\affiliation{Kavli Institute for Theoretical Sciences, Chinese Academy of Sciences, Beijing 100190, China}
\affiliation{Songshan Lake Materials Laboratory, Dongguan, Guangdong 523808, China}

\begin{abstract}
Skin effect, experimentally discovered in one dimension, describes the physical phenomenon that on an open chain, an extensive number of eigenstates of a non-Hermitian hamiltonian are localized at the end(s) of the chain. 
Here in two and higher dimensions, we establish a theorem that the skin effect exists, if and only if periodic-boundary spectrum of the hamiltonian covers a finite area on the complex plane. 
This theorem establishes the universality of the effect, because the above condition is satisfied in almost every generic non-Hermitian hamiltonian, and, unlike in one dimension, is compatible with all spatial symmetries. 
We propose two new types of skin effect in two and higher dimensions: the corner-skin effect where all eigenstates are localized at one corner of the system, and the geometry-dependent-skin effect where skin modes disappear for systems of a particular shape, but appear on generic polygons.
An immediate corollary of our theorem is that any non-Hermitian system having exceptional points (lines) in two (three) dimensions exhibits skin effect, making this phenomenon accessible to experiments in photonic crystals, Weyl semimetals, and Kondo insulators.
\end{abstract}
\maketitle

\section*{Introduction}

The study of non-Hermitian hamiltonians, which can be regarded as the effective description of dissipative processes, can be traced back to the investigation of alpha decay, where real and imaginary parts of the complex energy are related to the experimentally observed energy level and decay rate~\cite{Gamow1928}. 
When a lattice system is coupled with environments and has dissipations, e.g. photonic crystals having radiational loss~\cite{FengNP2017,Ganainy2018,Ozawa2019} and electronic systems having finite quasiparticle lifetime~\cite{FuLiang2017_arXiv,FuPRL2018,FuPRB2019,FuLiang2020_PRL}, the non-Hermitian band theory becomes a conceptually simple and efficient approach~\cite{Nori2017,XuYongPRL,FuLiang2018_PRL,Kunst2019,ZongPingGongPRX,AshvinPRL2019,Ueda2020_arXiv,Sato2019_PRX}. 

Skin effect~\cite{Yao2018,WangZhong2018,Kunst2018_PRL,Torres2018,ChingHua2019,Murakami2019_PRL,Kai2020,Okuma2020_PRL,Zhesen2020_aGBZ,ZhesenYangOnSite,LinhuLiPRL2020,BorgniaPRL2020,LonghiPRR2019,LonghiPRL2020,LiNC2020}, a phenomenon unique to the non-Hermitian band theory, refers to the localization of eigenstates at the boundary, the number of which scales with the volume of the system. 
For example, in one dimension, all eigenstates of a non-Hermitian hamiltonian can be localized at the ends of a chain~\cite{Yao2018}. 
This suggests the failure of Bloch's theorem~\cite{Xiong2018,Yang2020_arXiv}, which states that eigenstates in the bulk are modulated plane waves.
As Bloch's theorem plays a fundamental role in the development of condensed-matter physics~\cite{Ashcroft76}, the emergence of skin effect indicates a new and possibly revolutionary direction. 
Especially, the skin effect has been experimentally observed in one-dimensional classical systems~\cite{XuePeng2020,LeiXiao_PT,Thomale2020,FunnelingScience2020,BrandenbourgerNC2019,GhatakPNAS2020,WangScience2021}, inspiring further studies on their higher dimensional generalizations~\cite{WangZhong2018,ChingHua2019_Hybrid,EzawaInterface2019,EzawaPRB2019,Titus2020,Nori2019,Yoshida2020,Kawabata2020,Vincenzo2020,SonicSOTI_2019,ChuanweiPRL2019,KunstPRB2019,HughesQSHE2020,YokomizoPRB2020,Lucas2020,SyntheticPhcPRA_2020,YongxuPRB2021,GennadyPRB2021,SongFei2021}. 
However, a general theory for the higher-dimensional skin effect has not been established. 

Apart form the skin effect, another focus topic in non-Hermitian band systems is the exceptional point (or line)~\cite{Kato2013,Miri2019,YangLan2019,Lin2011_Invisibility,LeePRL2016,MolinaPRL2018,BergholtzPRA2018,Cerjan2018,Sato2019_EP,ThomasPRB2019,Zhesen2019_EP,Chong2020,ZhenBo2015,Zhou2018,ZhangXufeng2019,Cerjan2019,HopflinkPRB2019,Tang2020,ETI_2020,JonesPRL2020} that refers to stable point-type (or line-type) non-Hermitian band degeneracy in the Brillouin zone. 
At the exceptional point, not only eigenvalues but also eigenstates of the Bloch hamiltonian coalesce~\cite{Miri2019}. 
Many a novel phenomenon related to exceptional points has been predicted and observed~\cite{Dembowski2001,RichterPRL2003,Regensburger2012,FengNM2013,WiersigPRL2014,Gao2015,XuNature2016,Doppler2016,FrancoPRL2016,Hodaei2017,ChenNature2017,YoonNature2018,Tang2020}, such as the emergence of bulk-Fermi arc terminated at the exceptional points~\cite{FuLiang2017_arXiv,Zhou2018}. 
Since the bulk-boundary correspondence plays a central role in the development of topological phases~\cite{Kane2010}, it is natural to ask if there exists a non-Hermitian bulk-boundary correspondence in bands having exceptional points, analogous to the surface Fermi arc in the Weyl semimetals in the Hermitian counterpart~\cite{Ashvin2018}.

\begin{figure*}
	\begin{centering}
		\includegraphics[width=1\linewidth]{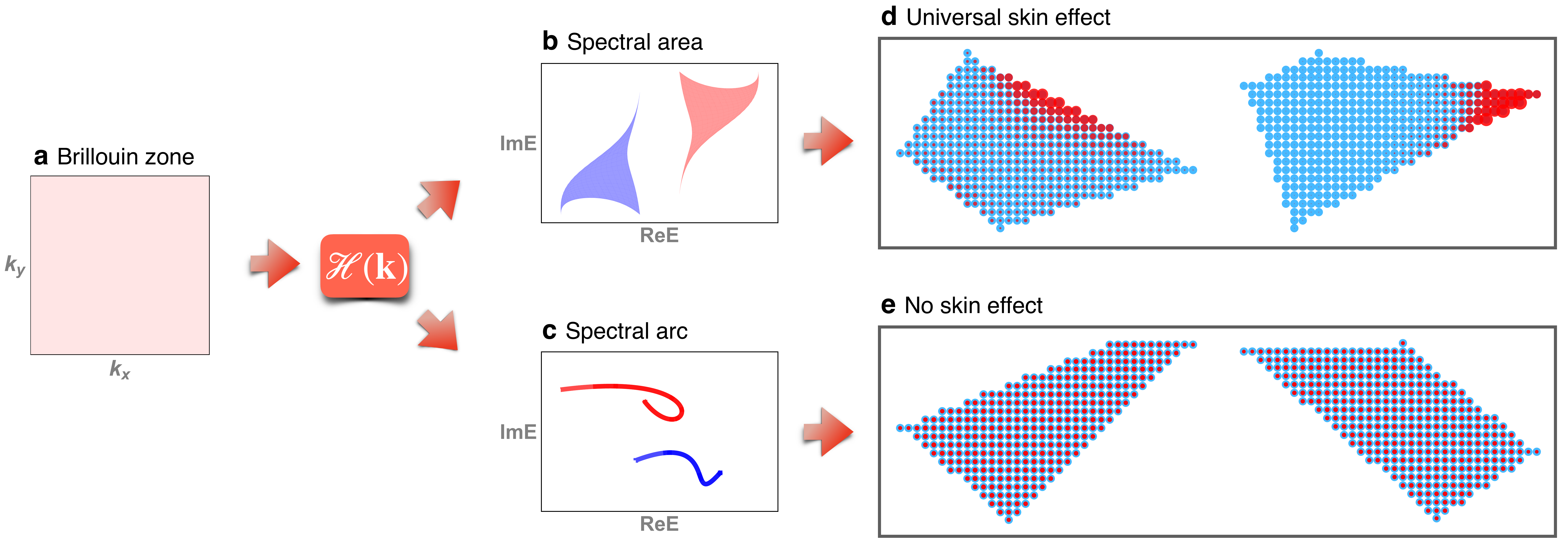}
		\par\end{centering}
	\protect\caption{\label{fig:1} The theorem of universal skin effect. (a) represents the Brillouin zone. (b)(d) shows that when the spectral area of $\mathcal{H}(\mathbf{k})$ is nonzero, the system on generic geometries must have universal skin effect. (c)(e) shows that when the spectral area of $\mathcal{H}(\mathbf{k})$ is zero, or forming one or several arcs on the complex plane, there is no skin effect under any geometry.} 
\end{figure*}

In this paper, we establish a theorem that reveals a universal bulk-boundary correspondence in two and higher dimensional non-Hermitian bands, as shown in Fig.~\ref{fig:1}.
The ``bulk'' refers to the area of the spectrum of the hamiltonian on the complex plane with periodic boundary condition, and ``boundary'' the presence (absence) of the skin effect for open-boundary system of an arbitrary shape.
The theorem states that the skin effect appears if and only if the spectral area is nonzero.
This skin effect is ``universal'' for three reasons: (i) a randomly generated local non-Hermitian hamiltonian has the skin effect with probability one; (ii) the skin effect is, unlike in one dimension, compatible with all spatial symmetries and time-reversal symmetry; and (iii) it does not require any special geometry of the open-boundary system.
We also propose two manifestations, restricted to two and higher dimensions, of the universal skin effect, i.e., the corner-skin effect and the geometry-dependent skin effect. 

A surprising corollary of our theorem is that all stable exceptional points~\cite{FuLiang2018_PRL,Sato2019_EP,Zhesen2019_EP} imply the presence of skin effect.
Because exceptional points have been either observed or proposed in meta-materials as well as in condensed matter, this corollary makes skin effect observable in known systems.
We predict the geometry-dependent skin effect in the two-dimensional photonic crystal studied in Ref.~\cite{Zhou2018}, and propose to observe this effect in the anomalous dynamics of wave packets.

\section*{Theorem: an equivalence between spectral area and skin effect}

For generic one-dimensional non-Hermitian systems, the correspondence between the spectral shape and the skin effect has been derived~\cite{Kai2020,Okuma2020_PRL}, i.e., when the Bloch spectrum is a loop-type (an arc-type), the skin effect appears (disappears). 

Generalizing the correspondence to two dimensions, we note two main differences. 
One difference is in the periodic-boundary spectrum, $E_i(\mathbf{k})$, where $i$ is the band index and $\mathbf{k}$ the crystal momentum in the first Brillouin zone (BZ).
Generally speaking, $E_i(\mathbf{k})$ is a mapping from the $d$-dimensional torus to the complex plane, $\mathbb{C}$.
When $d=1$, the image of $E_i(k)$ forms a loop; but when $d>1$, the image is generically a continuum on $\mathbb{C}$, denoted by $E_i(\rm{BZ})$.
The area covered by $E_i(\rm{BZ})$ on the complex plane is called the {\em spectral area}, denoted by $A_i$.
Another difference is in the variety of open-boundary condition.
There is only one geometry for an open system in one dimension, i.e., an open chain; but there are an infinite number of geometries in two dimensions such as triangle, rectangle and pentagon.

Now we are ready to state the theorem of universal skin effect: in the thermodynamic limit, the skin effect is present in a hamiltonian having open boundary of arbitrary geometry, if the spectral area is nonzero ($A_i\neq 0$); vice versa, the skin effect is absent for all possible geometries, if the spectral area is zero ($A_i=0$).
As the periodic-boundary hamiltonian describes the dynamics in the bulk, the theorem relates a bulk property (spectral area) to a boundary one (existence of skin modes).
Fig.~\ref{fig:1} shows some schematic examples. 
In the Supplemental Material Sec.~\ref{secI}, a complete proof of the theorem has been provided. 

The above theorem has implied the universality of skin effect in two and higher dimensions.
As $E_i(\rm{BZ})$ is the image of the $d\geq 2$-dimensional torus on the complex plane, it takes fine tuning of parameters to make $A_i=0$ for every $i$.
In fact, for single-band hamiltonian, we can prove that $A=0$ if and only if $\mathcal{H}(\mathbf{k})=P[h(\mathbf{k})]$, where $h(\mathbf{k})$ is a Hermitian hamiltonian and $P$ is a polynomial.
In other words, a randomly generated non-Hermitian hamiltonian $\mathcal{H}(\mathbf{k})$ has skin effect: the first meaning of universality.
In previous studies, other types of skin effect, such as the line-skin and the high-order-skin effect, in two and higher dimensions have been proposed~\cite{ChingHua2019_Hybrid,Kawabata2020}.
These types all require the open-boundary system take a special geometry (usually a rectangle) and are hence considered special and non-generic.
The skin effect when $A_i\neq0$ assumes a completely generic geometry of boundary: the second meaning of universality.
The third meaning of universality lies in the fact that, unlike in one dimension, higher-dimensional skin effect is compatible with all spatial symmetries.
A standing wave explanation for the above theorem is provided in the Supplemental Material Sec.~\ref{secII}. 

\begin{figure*}[t]
	\begin{centering}
	\includegraphics[width=.98\linewidth]{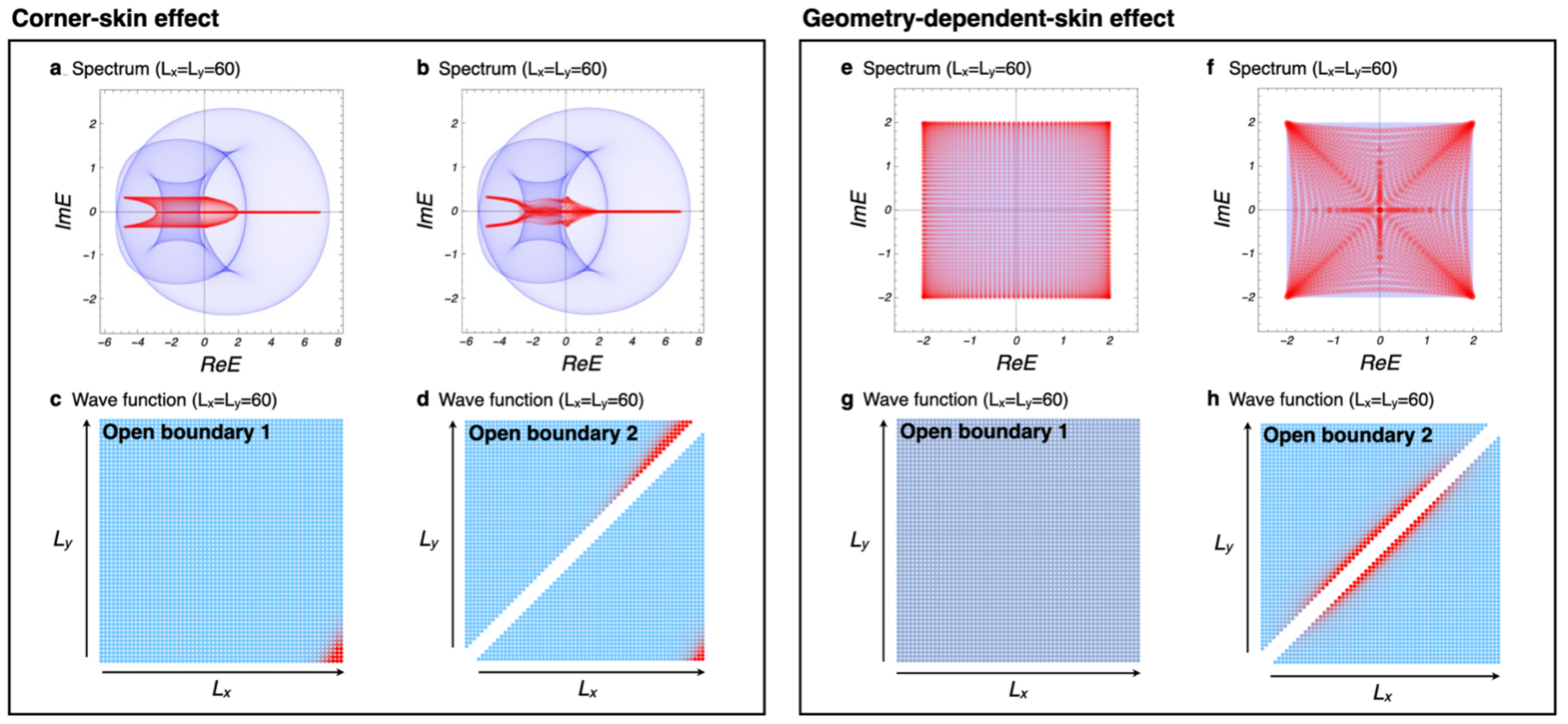}
	\par\end{centering}
	\protect\caption{\label{fig:2} Two manifestations of skin effect. One is the CSE (a)-(d), the other is the GDSE (e)-(h). In in (a)(b)(e)(f), the light blue regions represent the spectrum under periodic boundary, and the red points represent the eigenvalues under different open-boundary geometries. The spatial distributions of eigenstates $W(\bm{x})$ are plotted in (c)(d)(g)(h). The GDSE disappears under square geometry (open boundary 1) in (g), and reappears under triangle geometry (open boundary 2) in (h).}
\end{figure*}

\section*{The corner-skin and the geometry-dependent-skin effect}

While the theorem shows that the skin effect is universal, it does not specify what skin modes look like in higher dimensions. 
Here, we report two types of the universal skin effect, the corner-skin effect (CSE) and the geometry-dependent skin effect (GDSE). Note that the CSE we mentioned in this paper has the nature of non-reciprocity, similar to the one-dimensional skin effect.

The hamiltonian of the example for CSE is
\begin{equation}
	\begin{split}
	\mathcal{H}(\mathbf{k})=&[5(\cos{k_x}+\cos{2k_x})-i(\sin{k_x}+3\sin{2k_x}) \\
	&+5\cos{k_y}+i\sin{k_y}]/2,
	\end{split}
\end{equation}
of which the spectral area under square geometry and triangle geometry is shown in Fig.~\ref{fig:2} (a)(b) with light blue color. 
Because of the nonzero spectral area, the theorem tells us that the hamiltonian must have the universal skin effect.
This is verified in Fig.~\ref{fig:2} (c)(d), where the spatial distributions of all the eigenstates 
\begin{equation}
	W(\bm{x})=\frac{1}{N}\sum_n|\psi_n(\bm{x})|^2
\end{equation}
under different open boundaries are plotted. 
Here $\psi_n(x)$ is a normalized eigenstate and $N$ is the number of eigenstates. 
It is found that the wave functions are always localized at the corner of the boundary in Fig.~\ref{fig:2}(c), even if the open-boundary geometry is changed in Fig.~\ref{fig:2}(d). 
We also plot the corresponding eigenvalue spectra under different open boundaries, as shown in Fig.~\ref{fig:2} (a)(b) with red color. One can notice that the spectral areas under periodic and open boundaries do not equal. 
This kind of skin effect is called CSE, which can be explained by a nonzero current functional
\begin{equation}
J_\alpha[n]=\sum_i\oint_{\rm{BZ}}{dk^d}n(E_i,E^\ast_i)\partial_{k_\alpha}{E}_i(\mathbf{k})
\end{equation}
under the periodic-boundary condition, where $n(E,E^\ast)$ is any smooth function~\cite{Kai2020}.
The current functional is shown to vanish in two and three dimensions under point groups $C_i$, $D_{2,3,4,6}$, $C_{2h,3h,4h,6h}$, $D_{2d,3d,2h,3h,4h,6h}$, $T$, $T_{d,h}$, $O$ and $O_h$.
Therefore, the CSE is only compatible with point groups $C_m$ and $C_{2,3,4,6,2v,3v,4v,6v}$ (see details in the Supplemental Material Sec.~\ref{secIII}).

\begin{figure*}[t]
	\begin{centering}
	\includegraphics[width=1\linewidth]{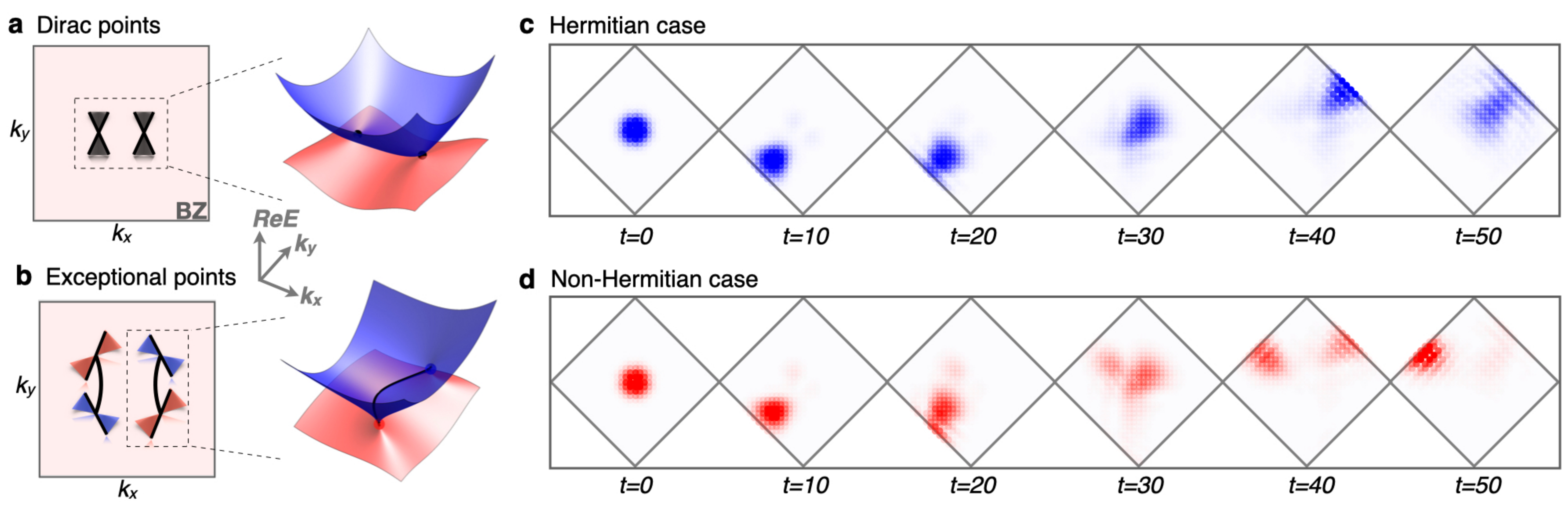}
	\par\end{centering}
	\protect\caption{\label{fig:3} Two Dirac points (a) of a two-dimensional photonic crystal model are split into four exceptional points (b) upon adding non-Hermitian term, such as radiational loss. Correspondingly, the evolution of Gaussian wave packet with initial velocity at the center of a diamond geometry for each ten time intervals is shown in (c) with $\gamma=0$ (Hermitian) and (d) with $\gamma=1/4$ (GDSE). }
\end{figure*}

The hamiltonian of the example for GDSE reads 
\begin{equation}
	\mathcal{H}(\mathbf{k})=2\cos{k_x}+2i\cos{k_y}.
\end{equation}
Since the spectral area is nonzero, our theorem tells us that the system must have skin effect for generic geometry, such as a random polygon. 
However, an interesting phenomenon in this example is that the skin effect disappears under the square geometry due to the existence of two mirror symmetries shown in Fig.~\ref{fig:2}~(g). 
Once we choose other types of boundaries where mirror symmetries are broken, the skin effect reappears as shown in Fig.~\ref{fig:2}~(h).
Since the emergence of the skin effect and the position of localization depend on the geometry, it is called the GDSE. 
Besides the distribution of the eigenstates, another feature of the GDSE is that area of the open-boundary spectrum seems to be the same as $A_i$.
However, the corresponding density of states is dependent by the choice of geometry as shown in Fig.~\ref{fig:2} (e)(f). 
We conjecture this is a universal phenomenon for the GDSE. 
In the Supplemental Material Sec.~\ref{secIII}, we have provided some additional examples to illustrate this new type of skin effect, and shown that the increase in the number of skin modes is proportional to the increase in the system volume. 
For GDSE, there is at least one spatial geometry such that skin modes vanish, and as such is mutually exclusive with CSE.
Additionally, GDSE is compatible with all point groups, in contrast to CSE. 

\section*{Corollary: skin effect from exceptional points}

An immediate corollary of our theorem is that all lattice hamiltonians having stable exceptional points have universal skin effect, connecting two unique phenomena in the non-Hermitian band theory. 
Consider a stable exceptional point $\mathbf{k}_0$ in two dimensions. Due to the branch point structure of exceptional point, the dispersion around $\bm{k_0}$ can be expressed as $E_{\pm}(\mathbf{k}) = \pm c_0 \sqrt{q_x+c_1 q_y} + O(|\mathbf{k}-\mathbf{k}_0|)$, where $q_{i=x,y}$ denotes a small derivation from exceptional point in $x$ or $y$ direction, that is, $q_{i}=k_{i}-k_{0i}$. Here $c_0, c_1$ are nonzero complex numbers and $c_1\notin \mathbb{R}$. 
Suppose the range of the expansion is $r_0$, then it is clear that $A_\pm\ge|c_0|\pi{r_0}^2/2\neq0$.
By the theorem, the system must have skin effect (see more details in the Supplementary Material Sec.~\ref{secIV}). 

Now we use the photonic crystal model that has been experimentally realized in Ref.~\cite{Zhou2018} to demonstrate our corollary. The tight-binding model hamiltonian with periodic boundary can be written as
\begin{equation}
	H(k) = \bm{d}(k)\cdot \bm{\sigma} - i\gamma/2 (\sigma_0-\sigma_z),
\end{equation}
where $\bm{\sigma}=(\sigma_0,\sigma_x,\sigma_y,\sigma_z)$ is a vector of the Pauli matrices and $\bm{d}(k)$ is a vector with four components, that is, 
\begin{equation}
	\begin{split}
		\bm{d}(k) = &\{\mu_0-(t_2+t_3)(\cos{k_x}+\cos{k_y}), \\
		&t_1 [1-\cos{k_x}-\cos{k_y}+\cos{(k_x-k_y)}], \\
		&t_1 [\sin{k_x}-\sin{k_y}-\sin{(k_x-k_y)}], \\
		&\mu_z+(t_2-t_3)(\cos{k_x}-\cos{k_y})\}.
	\end{split}
\end{equation}
The parameters are chosen as follows, $(t_1,t_2,t_3,\mu_0,\mu_z)=(0.4,-0.1,0.5,1.35,-0.02)$. As shown in Fig.~\ref{fig:3} (a), in the Hermitian limit, i.e. $\gamma=0$, the system has two Dirac points along the $x$-axis. When external dissipation or radiational loss is added, i.e., $\gamma\neq 0$, each Dirac point splits into two exceptional points shown in Fig.~\ref{fig:2} (b), connected by the bulk Fermi arc. According to our theorem, the system must have the universal skin effect, more precisely, the GDSE shown in the Supplementary Material Sec.~\ref{secIV}. The skin effect disappears under square geometry but reappears under diamond geometry shown in Fig.~\ref{fig:3} (c)(d). In this case, the majority of the eigenstates are concentrated on the four edges. 

As mentioned in the previous discussion, the appearance of the skin effect can be reflected in the dynamical properties. In order to show this, we simulate the time evolution of the wave packets with initial velocity at the center of the diamond geometry. Here the initial state is chosen to be Gaussian form $|\psi_0\rangle = \mathcal{N}$exp[${-(x-x_0)^2/10-(y-y_0)^2/10-i2x-i2y}]$$(1,1)^T$, where $\mathcal{N}$ is the normalization factor and $x_0=y_0=21$ is the center coordinate of the diamond geometry. 
We plot the corresponding normalized final states $|\psi(t_f)\rangle=\mathcal{N}(t_f)e^{-i\mathcal{H}_{\rm{OBC}}t_f}|\psi_0\rangle$ for every ten time intervals, where $\mathcal{H}_{\rm{OBC}}$ represents the open-boundary hamiltonian on the diamond geometry. 
As shown in Fig.~\ref{fig:3} (c), in the Hermitian case, the center of the wave packets obeys the reflection rule. Since the initial velocity of the wave packet is perpendicular to the two edges, the center of the wave packet just bounces between the two edges regardless of the dispersion of the wave packet. However, in the non-Hermitian case ($\gamma=1/4$) with GDSE, besides the normal reflection channel, the wave packet will evolve into the upper left boundary as shown in Fig.~\ref{fig:3} (d). This anomalous dynamical behavior is a experimental signature of GDSE. 

We also propose the realization for CSE in a three-dimensional system with exceptional lines in the Supplemental Material Sec.~\ref{secIV}. Experimentally, the non-reciprocity of the CSE can be detected by two-point Green function approach. 

\section*{Discussion}

Our work has built a bridge between two distinct phenomena that only exist in non-Hermitian systems, i.e., the exceptional points (lines) and the non-Hermitian skin effect, by establishing the correspondence between bulk (spectral area) and boundary (universal skin effect). We prove that the skin effect be universal and compatible with all spatial symmetries and reciprocity in two and higher dimensions. Due to the universality, it is expected that the skin effect is observable in a wide range of platforms, such as photonic crystals with natural radiational loss, acoustic metamaterials and circuit networks with lossy components such as resistors. Beyond these classical systems, the skin effect can also be realized in condensed matter, e.g., the heavy-fermion material with finite quasiparticle lifetime and the Weyl-exceptional-ring semimetal. The latter is realizable in Weyl semimetals made from inverting bands that have disparate effective masses, such as d- and f-bands.

One should be reminded, however, that the results in this paper assume the coherent dynamics of the constituent degrees of freedom, which is unlikely the case in macroscopic condensed-matter systems where the coherence length is shorter than the system size. On the contrary, for the systems where the system size and the coherent length are comparable, as in mesoscopic systems, we believe that the universal skin effect has a significant contribution to the transport properties, a subject for future exploration.

\bibliography{UnivSkin}

\begin{thebibliography}{93}%
\makeatletter
\providecommand \@ifxundefined [1]{%
 \@ifx{#1\undefined}
}%
\providecommand \@ifnum [1]{%
 \ifnum #1\expandafter \@firstoftwo
 \else \expandafter \@secondoftwo
 \fi
}%
\providecommand \@ifx [1]{%
 \ifx #1\expandafter \@firstoftwo
 \else \expandafter \@secondoftwo
 \fi
}%
\providecommand \natexlab [1]{#1}%
\providecommand \enquote  [1]{``#1''}%
\providecommand \bibnamefont  [1]{#1}%
\providecommand \bibfnamefont [1]{#1}%
\providecommand \citenamefont [1]{#1}%
\providecommand \href@noop [0]{\@secondoftwo}%
\providecommand \href [0]{\begingroup \@sanitize@url \@href}%
\providecommand \@href[1]{\@@startlink{#1}\@@href}%
\providecommand \@@href[1]{\endgroup#1\@@endlink}%
\providecommand \@sanitize@url [0]{\catcode `\\12\catcode `\$12\catcode
  `\&12\catcode `\#12\catcode `\^12\catcode `\_12\catcode `\%12\relax}%
\providecommand \@@startlink[1]{}%
\providecommand \@@endlink[0]{}%
\providecommand \url  [0]{\begingroup\@sanitize@url \@url }%
\providecommand \@url [1]{\endgroup\@href {#1}{\urlprefix }}%
\providecommand \urlprefix  [0]{URL }%
\providecommand \Eprint [0]{\href }%
\providecommand \doibase [0]{http://dx.doi.org/}%
\providecommand \selectlanguage [0]{\@gobble}%
\providecommand \bibinfo  [0]{\@secondoftwo}%
\providecommand \bibfield  [0]{\@secondoftwo}%
\providecommand \translation [1]{[#1]}%
\providecommand \BibitemOpen [0]{}%
\providecommand \bibitemStop [0]{}%
\providecommand \bibitemNoStop [0]{.\EOS\space}%
\providecommand \EOS [0]{\spacefactor3000\relax}%
\providecommand \BibitemShut  [1]{\csname bibitem#1\endcsname}%
\let\auto@bib@innerbib\@empty
\bibitem [{\citenamefont {Gamow}(1928)}]{Gamow1928}%
  \BibitemOpen
  \bibfield  {author} {\bibinfo {author} {\bibfnamefont {G.}~\bibnamefont
  {Gamow}},\ }\bibfield  {title} {\enquote {\bibinfo {title} {{Zur
  Quantentheorie des Atomkernes}},}\ }\href {\doibase 10.1007/BF01343196}
  {\bibfield  {journal} {\bibinfo  {journal} {Zeitschrift f{\"u}r Physik}\
  }\textbf {\bibinfo {volume} {51}},\ \bibinfo {pages} {204--212} (\bibinfo
  {year} {1928})}\BibitemShut {NoStop}%
\bibitem [{\citenamefont {Feng}\ \emph {et~al.}(2017)\citenamefont {Feng},
  \citenamefont {El-Ganainy},\ and\ \citenamefont {Ge}}]{FengNP2017}%
  \BibitemOpen
  \bibfield  {author} {\bibinfo {author} {\bibfnamefont {Liang}\ \bibnamefont
  {Feng}}, \bibinfo {author} {\bibfnamefont {Ramy}\ \bibnamefont {El-Ganainy}},
  \ and\ \bibinfo {author} {\bibfnamefont {Li}~\bibnamefont {Ge}},\ }\bibfield
  {title} {\enquote {\bibinfo {title} {{Non-Hermitian photonics based on
  parity–time symmetry}},}\ }\href {\doibase 10.1038/s41566-017-0031-1}
  {\bibfield  {journal} {\bibinfo  {journal} {Nature Photonics}\ }\textbf
  {\bibinfo {volume} {11}},\ \bibinfo {pages} {752--762} (\bibinfo {year}
  {2017})}\BibitemShut {NoStop}%
\bibitem [{\citenamefont {El-Ganainy}\ \emph {et~al.}(2018)\citenamefont
  {El-Ganainy}, \citenamefont {Makris}, \citenamefont {Khajavikhan},
  \citenamefont {Musslimani}, \citenamefont {Rotter},\ and\ \citenamefont
  {Christodoulides}}]{Ganainy2018}%
  \BibitemOpen
  \bibfield  {author} {\bibinfo {author} {\bibfnamefont {Ramy}\ \bibnamefont
  {El-Ganainy}}, \bibinfo {author} {\bibfnamefont {Konstantinos~G.}\
  \bibnamefont {Makris}}, \bibinfo {author} {\bibfnamefont {Mercedeh}\
  \bibnamefont {Khajavikhan}}, \bibinfo {author} {\bibfnamefont {Ziad~H.}\
  \bibnamefont {Musslimani}}, \bibinfo {author} {\bibfnamefont {Stefan}\
  \bibnamefont {Rotter}}, \ and\ \bibinfo {author} {\bibfnamefont
  {Demetrios~N.}\ \bibnamefont {Christodoulides}},\ }\bibfield  {title}
  {\enquote {\bibinfo {title} {{Non-Hermitian physics and PT symmetry}},}\
  }\href {\doibase 10.1038/nphys4323} {\bibfield  {journal} {\bibinfo
  {journal} {Nature Physics}\ }\textbf {\bibinfo {volume} {14}},\ \bibinfo
  {pages} {11--19} (\bibinfo {year} {2018})}\BibitemShut {NoStop}%
\bibitem [{\citenamefont {Ozawa}\ \emph {et~al.}(2019)\citenamefont {Ozawa},
  \citenamefont {Price}, \citenamefont {Amo}, \citenamefont {Goldman},
  \citenamefont {Hafezi}, \citenamefont {Lu}, \citenamefont {Rechtsman},
  \citenamefont {Schuster}, \citenamefont {Simon}, \citenamefont {Zilberberg},\
  and\ \citenamefont {Carusotto}}]{Ozawa2019}%
  \BibitemOpen
  \bibfield  {author} {\bibinfo {author} {\bibfnamefont {Tomoki}\ \bibnamefont
  {Ozawa}}, \bibinfo {author} {\bibfnamefont {Hannah~M.}\ \bibnamefont
  {Price}}, \bibinfo {author} {\bibfnamefont {Alberto}\ \bibnamefont {Amo}},
  \bibinfo {author} {\bibfnamefont {Nathan}\ \bibnamefont {Goldman}}, \bibinfo
  {author} {\bibfnamefont {Mohammad}\ \bibnamefont {Hafezi}}, \bibinfo {author}
  {\bibfnamefont {Ling}\ \bibnamefont {Lu}}, \bibinfo {author} {\bibfnamefont
  {Mikael~C.}\ \bibnamefont {Rechtsman}}, \bibinfo {author} {\bibfnamefont
  {David}\ \bibnamefont {Schuster}}, \bibinfo {author} {\bibfnamefont
  {Jonathan}\ \bibnamefont {Simon}}, \bibinfo {author} {\bibfnamefont {Oded}\
  \bibnamefont {Zilberberg}}, \ and\ \bibinfo {author} {\bibfnamefont {Iacopo}\
  \bibnamefont {Carusotto}},\ }\bibfield  {title} {\enquote {\bibinfo {title}
  {{Topological photonics}},}\ }\href {\doibase 10.1103/RevModPhys.91.015006}
  {\bibfield  {journal} {\bibinfo  {journal} {Rev. Mod. Phys.}\ }\textbf
  {\bibinfo {volume} {91}},\ \bibinfo {pages} {015006} (\bibinfo {year}
  {2019})}\BibitemShut {NoStop}%
\bibitem [{\citenamefont {Kozii}\ and\ \citenamefont
  {Fu}(2017)}]{FuLiang2017_arXiv}%
  \BibitemOpen
  \bibfield  {author} {\bibinfo {author} {\bibfnamefont {Vladyslav}\
  \bibnamefont {Kozii}}\ and\ \bibinfo {author} {\bibfnamefont {Liang}\
  \bibnamefont {Fu}},\ }\bibfield  {title} {\enquote {\bibinfo {title}
  {{Non-Hermitian Topological Theory of Finite-Lifetime Quasiparticles:
  Prediction of Bulk Fermi Arc Due to Exceptional Point}},}\ }\href@noop {}
  {\bibfield  {journal} {\bibinfo  {journal} {arXiv:1708.05841}\ } (\bibinfo
  {year} {2017})}\BibitemShut {NoStop}%
\bibitem [{\citenamefont {Shen}\ and\ \citenamefont {Fu}(2018)}]{FuPRL2018}%
  \BibitemOpen
  \bibfield  {author} {\bibinfo {author} {\bibfnamefont {Huitao}\ \bibnamefont
  {Shen}}\ and\ \bibinfo {author} {\bibfnamefont {Liang}\ \bibnamefont {Fu}},\
  }\bibfield  {title} {\enquote {\bibinfo {title} {{Quantum Oscillation from
  In-Gap States and a Non-Hermitian Landau Level Problem}},}\ }\href {\doibase
  10.1103/PhysRevLett.121.026403} {\bibfield  {journal} {\bibinfo  {journal}
  {Phys. Rev. Lett.}\ }\textbf {\bibinfo {volume} {121}},\ \bibinfo {pages}
  {026403} (\bibinfo {year} {2018})}\BibitemShut {NoStop}%
\bibitem [{\citenamefont {Papaj}\ \emph {et~al.}(2019)\citenamefont {Papaj},
  \citenamefont {Isobe},\ and\ \citenamefont {Fu}}]{FuPRB2019}%
  \BibitemOpen
  \bibfield  {author} {\bibinfo {author} {\bibfnamefont {Micha\l{}}\
  \bibnamefont {Papaj}}, \bibinfo {author} {\bibfnamefont {Hiroki}\
  \bibnamefont {Isobe}}, \ and\ \bibinfo {author} {\bibfnamefont {Liang}\
  \bibnamefont {Fu}},\ }\bibfield  {title} {\enquote {\bibinfo {title} {{Nodal
  arc of disordered Dirac fermions and non-Hermitian band theory}},}\ }\href
  {\doibase 10.1103/PhysRevB.99.201107} {\bibfield  {journal} {\bibinfo
  {journal} {Phys. Rev. B}\ }\textbf {\bibinfo {volume} {99}},\ \bibinfo
  {pages} {201107} (\bibinfo {year} {2019})}\BibitemShut {NoStop}%
\bibitem [{\citenamefont {Nagai}\ \emph {et~al.}(2020)\citenamefont {Nagai},
  \citenamefont {Qi}, \citenamefont {Isobe}, \citenamefont {Kozii},\ and\
  \citenamefont {Fu}}]{FuLiang2020_PRL}%
  \BibitemOpen
  \bibfield  {author} {\bibinfo {author} {\bibfnamefont {Yuki}\ \bibnamefont
  {Nagai}}, \bibinfo {author} {\bibfnamefont {Yang}\ \bibnamefont {Qi}},
  \bibinfo {author} {\bibfnamefont {Hiroki}\ \bibnamefont {Isobe}}, \bibinfo
  {author} {\bibfnamefont {Vladyslav}\ \bibnamefont {Kozii}}, \ and\ \bibinfo
  {author} {\bibfnamefont {Liang}\ \bibnamefont {Fu}},\ }\bibfield  {title}
  {\enquote {\bibinfo {title} {{DMFT Reveals the Non-Hermitian Topology and
  Fermi Arcs in Heavy-Fermion Systems}},}\ }\href {\doibase
  10.1103/PhysRevLett.125.227204} {\bibfield  {journal} {\bibinfo  {journal}
  {Phys. Rev. Lett.}\ }\textbf {\bibinfo {volume} {125}},\ \bibinfo {pages}
  {227204} (\bibinfo {year} {2020})}\BibitemShut {NoStop}%
\bibitem [{\citenamefont {Leykam}\ \emph {et~al.}(2017)\citenamefont {Leykam},
  \citenamefont {Bliokh}, \citenamefont {Huang}, \citenamefont {Chong},\ and\
  \citenamefont {Nori}}]{Nori2017}%
  \BibitemOpen
  \bibfield  {author} {\bibinfo {author} {\bibfnamefont {Daniel}\ \bibnamefont
  {Leykam}}, \bibinfo {author} {\bibfnamefont {Konstantin~Y.}\ \bibnamefont
  {Bliokh}}, \bibinfo {author} {\bibfnamefont {Chunli}\ \bibnamefont {Huang}},
  \bibinfo {author} {\bibfnamefont {Y.~D.}\ \bibnamefont {Chong}}, \ and\
  \bibinfo {author} {\bibfnamefont {Franco}\ \bibnamefont {Nori}},\ }\bibfield
  {title} {\enquote {\bibinfo {title} {{Edge Modes, Degeneracies, and
  Topological Numbers in Non-Hermitian Systems}},}\ }\href {\doibase
  10.1103/PhysRevLett.118.040401} {\bibfield  {journal} {\bibinfo  {journal}
  {Phys. Rev. Lett.}\ }\textbf {\bibinfo {volume} {118}},\ \bibinfo {pages}
  {040401} (\bibinfo {year} {2017})}\BibitemShut {NoStop}%
\bibitem [{\citenamefont {Xu}\ \emph {et~al.}(2017)\citenamefont {Xu},
  \citenamefont {Wang},\ and\ \citenamefont {Duan}}]{XuYongPRL}%
  \BibitemOpen
  \bibfield  {author} {\bibinfo {author} {\bibfnamefont {Yong}\ \bibnamefont
  {Xu}}, \bibinfo {author} {\bibfnamefont {Sheng-Tao}\ \bibnamefont {Wang}}, \
  and\ \bibinfo {author} {\bibfnamefont {L.-M.}\ \bibnamefont {Duan}},\
  }\bibfield  {title} {\enquote {\bibinfo {title} {{Weyl Exceptional Rings in a
  Three-Dimensional Dissipative Cold Atomic Gas}},}\ }\href {\doibase
  10.1103/PhysRevLett.118.045701} {\bibfield  {journal} {\bibinfo  {journal}
  {Phys. Rev. Lett.}\ }\textbf {\bibinfo {volume} {118}},\ \bibinfo {pages}
  {045701} (\bibinfo {year} {2017})}\BibitemShut {NoStop}%
\bibitem [{\citenamefont {Shen}\ \emph {et~al.}(2018)\citenamefont {Shen},
  \citenamefont {Zhen},\ and\ \citenamefont {Fu}}]{FuLiang2018_PRL}%
  \BibitemOpen
  \bibfield  {author} {\bibinfo {author} {\bibfnamefont {Huitao}\ \bibnamefont
  {Shen}}, \bibinfo {author} {\bibfnamefont {Bo}~\bibnamefont {Zhen}}, \ and\
  \bibinfo {author} {\bibfnamefont {Liang}\ \bibnamefont {Fu}},\ }\bibfield
  {title} {\enquote {\bibinfo {title} {{Topological Band Theory for
  Non-Hermitian Hamiltonians}},}\ }\href {\doibase
  10.1103/PhysRevLett.120.146402} {\bibfield  {journal} {\bibinfo  {journal}
  {Phys. Rev. Lett.}\ }\textbf {\bibinfo {volume} {120}},\ \bibinfo {pages}
  {146402} (\bibinfo {year} {2018})}\BibitemShut {NoStop}%
\bibitem [{\citenamefont {Bergholtz}\ \emph {et~al.}(2021)\citenamefont
  {Bergholtz}, \citenamefont {Budich},\ and\ \citenamefont
  {Kunst}}]{Kunst2019}%
  \BibitemOpen
  \bibfield  {author} {\bibinfo {author} {\bibfnamefont {Emil~J.}\ \bibnamefont
  {Bergholtz}}, \bibinfo {author} {\bibfnamefont {Jan~Carl}\ \bibnamefont
  {Budich}}, \ and\ \bibinfo {author} {\bibfnamefont {Flore~K.}\ \bibnamefont
  {Kunst}},\ }\bibfield  {title} {\enquote {\bibinfo {title} {{Exceptional
  topology of non-Hermitian systems}},}\ }\href {\doibase
  10.1103/RevModPhys.93.015005} {\bibfield  {journal} {\bibinfo  {journal}
  {Rev. Mod. Phys.}\ }\textbf {\bibinfo {volume} {93}},\ \bibinfo {pages}
  {015005} (\bibinfo {year} {2021})}\BibitemShut {NoStop}%
\bibitem [{\citenamefont {Gong}\ \emph {et~al.}(2018)\citenamefont {Gong},
  \citenamefont {Ashida}, \citenamefont {Kawabata}, \citenamefont {Takasan},
  \citenamefont {Higashikawa},\ and\ \citenamefont {Ueda}}]{ZongPingGongPRX}%
  \BibitemOpen
  \bibfield  {author} {\bibinfo {author} {\bibfnamefont {Zongping}\
  \bibnamefont {Gong}}, \bibinfo {author} {\bibfnamefont {Yuto}\ \bibnamefont
  {Ashida}}, \bibinfo {author} {\bibfnamefont {Kohei}\ \bibnamefont
  {Kawabata}}, \bibinfo {author} {\bibfnamefont {Kazuaki}\ \bibnamefont
  {Takasan}}, \bibinfo {author} {\bibfnamefont {Sho}\ \bibnamefont
  {Higashikawa}}, \ and\ \bibinfo {author} {\bibfnamefont {Masahito}\
  \bibnamefont {Ueda}},\ }\bibfield  {title} {\enquote {\bibinfo {title}
  {{Topological Phases of Non-Hermitian Systems}},}\ }\href {\doibase
  10.1103/PhysRevX.8.031079} {\bibfield  {journal} {\bibinfo  {journal} {Phys.
  Rev. X}\ }\textbf {\bibinfo {volume} {8}},\ \bibinfo {pages} {031079}
  (\bibinfo {year} {2018})}\BibitemShut {NoStop}%
\bibitem [{\citenamefont {Lee}\ \emph {et~al.}(2019{\natexlab{a}})\citenamefont
  {Lee}, \citenamefont {Ahn}, \citenamefont {Zhou},\ and\ \citenamefont
  {Vishwanath}}]{AshvinPRL2019}%
  \BibitemOpen
  \bibfield  {author} {\bibinfo {author} {\bibfnamefont {Jong~Yeon}\
  \bibnamefont {Lee}}, \bibinfo {author} {\bibfnamefont {Junyeong}\
  \bibnamefont {Ahn}}, \bibinfo {author} {\bibfnamefont {Hengyun}\ \bibnamefont
  {Zhou}}, \ and\ \bibinfo {author} {\bibfnamefont {Ashvin}\ \bibnamefont
  {Vishwanath}},\ }\bibfield  {title} {\enquote {\bibinfo {title} {{Topological
  Correspondence between Hermitian and Non-Hermitian Systems: Anomalous
  Dynamics}},}\ }\href {\doibase 10.1103/PhysRevLett.123.206404} {\bibfield
  {journal} {\bibinfo  {journal} {Phys. Rev. Lett.}\ }\textbf {\bibinfo
  {volume} {123}},\ \bibinfo {pages} {206404} (\bibinfo {year}
  {2019}{\natexlab{a}})}\BibitemShut {NoStop}%
\bibitem [{\citenamefont {Ashida}\ \emph {et~al.}(2020)\citenamefont {Ashida},
  \citenamefont {Gong},\ and\ \citenamefont {Ueda}}]{Ueda2020_arXiv}%
  \BibitemOpen
  \bibfield  {author} {\bibinfo {author} {\bibfnamefont {Yuto}\ \bibnamefont
  {Ashida}}, \bibinfo {author} {\bibfnamefont {Zongping}\ \bibnamefont {Gong}},
  \ and\ \bibinfo {author} {\bibfnamefont {Masahito}\ \bibnamefont {Ueda}},\
  }\bibfield  {title} {\enquote {\bibinfo {title} {{Non-Hermitian Physics}},}\
  }\href@noop {} {\bibfield  {journal} {\bibinfo  {journal} {arXiv:2006.01837}\
  } (\bibinfo {year} {2020})}\BibitemShut {NoStop}%
\bibitem [{\citenamefont {Kawabata}\ \emph
  {et~al.}(2019{\natexlab{a}})\citenamefont {Kawabata}, \citenamefont
  {Shiozaki}, \citenamefont {Ueda},\ and\ \citenamefont {Sato}}]{Sato2019_PRX}%
  \BibitemOpen
  \bibfield  {author} {\bibinfo {author} {\bibfnamefont {Kohei}\ \bibnamefont
  {Kawabata}}, \bibinfo {author} {\bibfnamefont {Ken}\ \bibnamefont
  {Shiozaki}}, \bibinfo {author} {\bibfnamefont {Masahito}\ \bibnamefont
  {Ueda}}, \ and\ \bibinfo {author} {\bibfnamefont {Masatoshi}\ \bibnamefont
  {Sato}},\ }\bibfield  {title} {\enquote {\bibinfo {title} {{Symmetry and
  Topology in Non-Hermitian Physics}},}\ }\href {\doibase
  10.1103/PhysRevX.9.041015} {\bibfield  {journal} {\bibinfo  {journal} {Phys.
  Rev. X}\ }\textbf {\bibinfo {volume} {9}},\ \bibinfo {pages} {041015}
  (\bibinfo {year} {2019}{\natexlab{a}})}\BibitemShut {NoStop}%
\bibitem [{\citenamefont {Yao}\ and\ \citenamefont {Wang}(2018)}]{Yao2018}%
  \BibitemOpen
  \bibfield  {author} {\bibinfo {author} {\bibfnamefont {Shunyu}\ \bibnamefont
  {Yao}}\ and\ \bibinfo {author} {\bibfnamefont {Zhong}\ \bibnamefont {Wang}},\
  }\bibfield  {title} {\enquote {\bibinfo {title} {{Edge States and Topological
  Invariants of Non-Hermitian Systems}},}\ }\href {\doibase
  10.1103/PhysRevLett.121.086803} {\bibfield  {journal} {\bibinfo  {journal}
  {Phys. Rev. Lett.}\ }\textbf {\bibinfo {volume} {121}},\ \bibinfo {pages}
  {086803} (\bibinfo {year} {2018})}\BibitemShut {NoStop}%
\bibitem [{\citenamefont {Yao}\ \emph {et~al.}(2018)\citenamefont {Yao},
  \citenamefont {Song},\ and\ \citenamefont {Wang}}]{WangZhong2018}%
  \BibitemOpen
  \bibfield  {author} {\bibinfo {author} {\bibfnamefont {Shunyu}\ \bibnamefont
  {Yao}}, \bibinfo {author} {\bibfnamefont {Fei}\ \bibnamefont {Song}}, \ and\
  \bibinfo {author} {\bibfnamefont {Zhong}\ \bibnamefont {Wang}},\ }\bibfield
  {title} {\enquote {\bibinfo {title} {{Non-Hermitian Chern Bands}},}\ }\href
  {\doibase 10.1103/PhysRevLett.121.136802} {\bibfield  {journal} {\bibinfo
  {journal} {Phys. Rev. Lett.}\ }\textbf {\bibinfo {volume} {121}},\ \bibinfo
  {pages} {136802} (\bibinfo {year} {2018})}\BibitemShut {NoStop}%
\bibitem [{\citenamefont {Kunst}\ \emph {et~al.}(2018)\citenamefont {Kunst},
  \citenamefont {Edvardsson}, \citenamefont {Budich},\ and\ \citenamefont
  {Bergholtz}}]{Kunst2018_PRL}%
  \BibitemOpen
  \bibfield  {author} {\bibinfo {author} {\bibfnamefont {Flore~K.}\
  \bibnamefont {Kunst}}, \bibinfo {author} {\bibfnamefont {Elisabet}\
  \bibnamefont {Edvardsson}}, \bibinfo {author} {\bibfnamefont {Jan~Carl}\
  \bibnamefont {Budich}}, \ and\ \bibinfo {author} {\bibfnamefont {Emil~J.}\
  \bibnamefont {Bergholtz}},\ }\bibfield  {title} {\enquote {\bibinfo {title}
  {{Biorthogonal Bulk-Boundary Correspondence in Non-Hermitian Systems}},}\
  }\href {\doibase 10.1103/PhysRevLett.121.026808} {\bibfield  {journal}
  {\bibinfo  {journal} {Phys. Rev. Lett.}\ }\textbf {\bibinfo {volume} {121}},\
  \bibinfo {pages} {026808} (\bibinfo {year} {2018})}\BibitemShut {NoStop}%
\bibitem [{\citenamefont {Martinez~Alvarez}\ \emph {et~al.}(2018)\citenamefont
  {Martinez~Alvarez}, \citenamefont {Barrios~Vargas},\ and\ \citenamefont
  {Foa~Torres}}]{Torres2018}%
  \BibitemOpen
  \bibfield  {author} {\bibinfo {author} {\bibfnamefont {V.~M.}\ \bibnamefont
  {Martinez~Alvarez}}, \bibinfo {author} {\bibfnamefont {J.~E.}\ \bibnamefont
  {Barrios~Vargas}}, \ and\ \bibinfo {author} {\bibfnamefont {L.~E.~F.}\
  \bibnamefont {Foa~Torres}},\ }\bibfield  {title} {\enquote {\bibinfo {title}
  {{Non-Hermitian robust edge states in one dimension: Anomalous localization
  and eigenspace condensation at exceptional points}},}\ }\href {\doibase
  10.1103/PhysRevB.97.121401} {\bibfield  {journal} {\bibinfo  {journal} {Phys.
  Rev. B}\ }\textbf {\bibinfo {volume} {97}},\ \bibinfo {pages} {121401(R)}
  (\bibinfo {year} {2018})}\BibitemShut {NoStop}%
\bibitem [{\citenamefont {Lee}\ and\ \citenamefont
  {Thomale}(2019)}]{ChingHua2019}%
  \BibitemOpen
  \bibfield  {author} {\bibinfo {author} {\bibfnamefont {Ching~Hua}\
  \bibnamefont {Lee}}\ and\ \bibinfo {author} {\bibfnamefont {Ronny}\
  \bibnamefont {Thomale}},\ }\bibfield  {title} {\enquote {\bibinfo {title}
  {{Anatomy of skin modes and topology in non-Hermitian systems}},}\ }\href
  {\doibase 10.1103/PhysRevB.99.201103} {\bibfield  {journal} {\bibinfo
  {journal} {Phys. Rev. B}\ }\textbf {\bibinfo {volume} {99}},\ \bibinfo
  {pages} {201103(R)} (\bibinfo {year} {2019})}\BibitemShut {NoStop}%
\bibitem [{\citenamefont {Yokomizo}\ and\ \citenamefont
  {Murakami}(2019)}]{Murakami2019_PRL}%
  \BibitemOpen
  \bibfield  {author} {\bibinfo {author} {\bibfnamefont {Kazuki}\ \bibnamefont
  {Yokomizo}}\ and\ \bibinfo {author} {\bibfnamefont {Shuichi}\ \bibnamefont
  {Murakami}},\ }\bibfield  {title} {\enquote {\bibinfo {title} {{Non-Bloch
  Band Theory of Non-Hermitian Systems}},}\ }\href {\doibase
  10.1103/PhysRevLett.123.066404} {\bibfield  {journal} {\bibinfo  {journal}
  {Phys. Rev. Lett.}\ }\textbf {\bibinfo {volume} {123}},\ \bibinfo {pages}
  {066404} (\bibinfo {year} {2019})}\BibitemShut {NoStop}%
\bibitem [{\citenamefont {Zhang}\ \emph {et~al.}(2020)\citenamefont {Zhang},
  \citenamefont {Yang},\ and\ \citenamefont {Fang}}]{Kai2020}%
  \BibitemOpen
  \bibfield  {author} {\bibinfo {author} {\bibfnamefont {Kai}\ \bibnamefont
  {Zhang}}, \bibinfo {author} {\bibfnamefont {Zhesen}\ \bibnamefont {Yang}}, \
  and\ \bibinfo {author} {\bibfnamefont {Chen}\ \bibnamefont {Fang}},\
  }\bibfield  {title} {\enquote {\bibinfo {title} {{Correspondence between
  Winding Numbers and Skin Modes in Non-Hermitian Systems}},}\ }\href {\doibase
  10.1103/PhysRevLett.125.126402} {\bibfield  {journal} {\bibinfo  {journal}
  {Phys. Rev. Lett.}\ }\textbf {\bibinfo {volume} {125}},\ \bibinfo {pages}
  {126402} (\bibinfo {year} {2020})}\BibitemShut {NoStop}%
\bibitem [{\citenamefont {Okuma}\ \emph {et~al.}(2020)\citenamefont {Okuma},
  \citenamefont {Kawabata}, \citenamefont {Shiozaki},\ and\ \citenamefont
  {Sato}}]{Okuma2020_PRL}%
  \BibitemOpen
  \bibfield  {author} {\bibinfo {author} {\bibfnamefont {Nobuyuki}\
  \bibnamefont {Okuma}}, \bibinfo {author} {\bibfnamefont {Kohei}\ \bibnamefont
  {Kawabata}}, \bibinfo {author} {\bibfnamefont {Ken}\ \bibnamefont
  {Shiozaki}}, \ and\ \bibinfo {author} {\bibfnamefont {Masatoshi}\
  \bibnamefont {Sato}},\ }\bibfield  {title} {\enquote {\bibinfo {title}
  {{Topological Origin of Non-Hermitian Skin Effects}},}\ }\href {\doibase
  10.1103/PhysRevLett.124.086801} {\bibfield  {journal} {\bibinfo  {journal}
  {Phys. Rev. Lett.}\ }\textbf {\bibinfo {volume} {124}},\ \bibinfo {pages}
  {086801} (\bibinfo {year} {2020})}\BibitemShut {NoStop}%
\bibitem [{\citenamefont {Yang}\ \emph
  {et~al.}(2020{\natexlab{a}})\citenamefont {Yang}, \citenamefont {Zhang},
  \citenamefont {Fang},\ and\ \citenamefont {Hu}}]{Zhesen2020_aGBZ}%
  \BibitemOpen
  \bibfield  {author} {\bibinfo {author} {\bibfnamefont {Zhesen}\ \bibnamefont
  {Yang}}, \bibinfo {author} {\bibfnamefont {Kai}\ \bibnamefont {Zhang}},
  \bibinfo {author} {\bibfnamefont {Chen}\ \bibnamefont {Fang}}, \ and\
  \bibinfo {author} {\bibfnamefont {Jiangping}\ \bibnamefont {Hu}},\ }\bibfield
   {title} {\enquote {\bibinfo {title} {{Non-Hermitian Bulk-Boundary
  Correspondence and Auxiliary Generalized Brillouin Zone Theory}},}\ }\href
  {\doibase 10.1103/PhysRevLett.125.226402} {\bibfield  {journal} {\bibinfo
  {journal} {Phys. Rev. Lett.}\ }\textbf {\bibinfo {volume} {125}},\ \bibinfo
  {pages} {226402} (\bibinfo {year} {2020}{\natexlab{a}})}\BibitemShut
  {NoStop}%
\bibitem [{\citenamefont {Yi}\ and\ \citenamefont
  {Yang}(2020)}]{ZhesenYangOnSite}%
  \BibitemOpen
  \bibfield  {author} {\bibinfo {author} {\bibfnamefont {Yifei}\ \bibnamefont
  {Yi}}\ and\ \bibinfo {author} {\bibfnamefont {Zhesen}\ \bibnamefont {Yang}},\
  }\bibfield  {title} {\enquote {\bibinfo {title} {{Non-Hermitian Skin Modes
  Induced by On-Site Dissipations and Chiral Tunneling Effect}},}\ }\href
  {\doibase 10.1103/PhysRevLett.125.186802} {\bibfield  {journal} {\bibinfo
  {journal} {Phys. Rev. Lett.}\ }\textbf {\bibinfo {volume} {125}},\ \bibinfo
  {pages} {186802} (\bibinfo {year} {2020})}\BibitemShut {NoStop}%
\bibitem [{\citenamefont {Li}\ \emph {et~al.}(2020{\natexlab{a}})\citenamefont
  {Li}, \citenamefont {Lee},\ and\ \citenamefont {Gong}}]{LinhuLiPRL2020}%
  \BibitemOpen
  \bibfield  {author} {\bibinfo {author} {\bibfnamefont {Linhu}\ \bibnamefont
  {Li}}, \bibinfo {author} {\bibfnamefont {Ching~Hua}\ \bibnamefont {Lee}}, \
  and\ \bibinfo {author} {\bibfnamefont {Jiangbin}\ \bibnamefont {Gong}},\
  }\bibfield  {title} {\enquote {\bibinfo {title} {{Topological Switch for
  Non-Hermitian Skin Effect in Cold-Atom Systems with Loss}},}\ }\href
  {\doibase 10.1103/PhysRevLett.124.250402} {\bibfield  {journal} {\bibinfo
  {journal} {Phys. Rev. Lett.}\ }\textbf {\bibinfo {volume} {124}},\ \bibinfo
  {pages} {250402} (\bibinfo {year} {2020}{\natexlab{a}})}\BibitemShut
  {NoStop}%
\bibitem [{\citenamefont {Borgnia}\ \emph {et~al.}(2020)\citenamefont
  {Borgnia}, \citenamefont {Kruchkov},\ and\ \citenamefont
  {Slager}}]{BorgniaPRL2020}%
  \BibitemOpen
  \bibfield  {author} {\bibinfo {author} {\bibfnamefont {Dan~S.}\ \bibnamefont
  {Borgnia}}, \bibinfo {author} {\bibfnamefont {Alex~Jura}\ \bibnamefont
  {Kruchkov}}, \ and\ \bibinfo {author} {\bibfnamefont {Robert-Jan}\
  \bibnamefont {Slager}},\ }\bibfield  {title} {\enquote {\bibinfo {title}
  {{Non-Hermitian Boundary Modes and Topology}},}\ }\href {\doibase
  10.1103/PhysRevLett.124.056802} {\bibfield  {journal} {\bibinfo  {journal}
  {Phys. Rev. Lett.}\ }\textbf {\bibinfo {volume} {124}},\ \bibinfo {pages}
  {056802} (\bibinfo {year} {2020})}\BibitemShut {NoStop}%
\bibitem [{\citenamefont {Longhi}(2019)}]{LonghiPRR2019}%
  \BibitemOpen
  \bibfield  {author} {\bibinfo {author} {\bibfnamefont {Stefano}\ \bibnamefont
  {Longhi}},\ }\bibfield  {title} {\enquote {\bibinfo {title} {{Probing
  non-Hermitian skin effect and non-Bloch phase transitions}},}\ }\href
  {\doibase 10.1103/PhysRevResearch.1.023013} {\bibfield  {journal} {\bibinfo
  {journal} {Phys. Rev. Research}\ }\textbf {\bibinfo {volume} {1}},\ \bibinfo
  {pages} {023013} (\bibinfo {year} {2019})}\BibitemShut {NoStop}%
\bibitem [{\citenamefont {Longhi}(2020)}]{LonghiPRL2020}%
  \BibitemOpen
  \bibfield  {author} {\bibinfo {author} {\bibfnamefont {S.}~\bibnamefont
  {Longhi}},\ }\bibfield  {title} {\enquote {\bibinfo {title} {{Non-Bloch-Band
  Collapse and Chiral Zener Tunneling}},}\ }\href {\doibase
  10.1103/PhysRevLett.124.066602} {\bibfield  {journal} {\bibinfo  {journal}
  {Phys. Rev. Lett.}\ }\textbf {\bibinfo {volume} {124}},\ \bibinfo {pages}
  {066602} (\bibinfo {year} {2020})}\BibitemShut {NoStop}%
\bibitem [{\citenamefont {Li}\ \emph {et~al.}(2020{\natexlab{b}})\citenamefont
  {Li}, \citenamefont {Lee}, \citenamefont {Mu},\ and\ \citenamefont
  {Gong}}]{LiNC2020}%
  \BibitemOpen
  \bibfield  {author} {\bibinfo {author} {\bibfnamefont {Linhu}\ \bibnamefont
  {Li}}, \bibinfo {author} {\bibfnamefont {Ching~Hua}\ \bibnamefont {Lee}},
  \bibinfo {author} {\bibfnamefont {Sen}\ \bibnamefont {Mu}}, \ and\ \bibinfo
  {author} {\bibfnamefont {Jiangbin}\ \bibnamefont {Gong}},\ }\bibfield
  {title} {\enquote {\bibinfo {title} {{Critical non-Hermitian skin effect}},}\
  }\href {\doibase 10.1038/s41467-020-18917-4} {\bibfield  {journal} {\bibinfo
  {journal} {Nature Communications}\ }\textbf {\bibinfo {volume} {11}},\
  \bibinfo {pages} {5491} (\bibinfo {year} {2020}{\natexlab{b}})}\BibitemShut
  {NoStop}%
\bibitem [{\citenamefont {Xiong}(2018)}]{Xiong2018}%
  \BibitemOpen
  \bibfield  {author} {\bibinfo {author} {\bibfnamefont {Ye}~\bibnamefont
  {Xiong}},\ }\bibfield  {title} {\enquote {\bibinfo {title} {{Why does bulk
  boundary correspondence fail in some non-hermitian topological models}},}\
  }\href {\doibase 10.1088/2399-6528/aab64a} {\bibfield  {journal} {\bibinfo
  {journal} {Journal of Physics Communications}\ }\textbf {\bibinfo {volume}
  {2}},\ \bibinfo {pages} {035043} (\bibinfo {year} {2018})}\BibitemShut
  {NoStop}%
\bibitem [{\citenamefont {Yang}(2020)}]{Yang2020_arXiv}%
  \BibitemOpen
  \bibfield  {author} {\bibinfo {author} {\bibfnamefont {Zhesen}\ \bibnamefont
  {Yang}},\ }\bibfield  {title} {\enquote {\bibinfo {title} {{Non-perturbative
  Breakdown of Bloch's Theorem and Hermitian Skin Effects}},}\ }\href@noop {}
  {\bibfield  {journal} {\bibinfo  {journal} {arXiv:2012.03333}\ } (\bibinfo
  {year} {2020})}\BibitemShut {NoStop}%
\bibitem [{\citenamefont {Ashcroft}\ and\ \citenamefont
  {Mermin}(1976)}]{Ashcroft76}%
  \BibitemOpen
  \bibfield  {author} {\bibinfo {author} {\bibfnamefont {N.~W.}\ \bibnamefont
  {Ashcroft}}\ and\ \bibinfo {author} {\bibfnamefont {N.~D.}\ \bibnamefont
  {Mermin}},\ }\href@noop {} {\emph {\bibinfo {title} {{S}olid {S}tate
  {P}hysics}}}\ (\bibinfo  {publisher} {Holt-Saunders},\ \bibinfo {year}
  {1976})\BibitemShut {NoStop}%
\bibitem [{\citenamefont {Xiao}\ \emph
  {et~al.}(2020{\natexlab{a}})\citenamefont {Xiao}, \citenamefont {Deng},
  \citenamefont {Wang}, \citenamefont {Zhu}, \citenamefont {Wang},
  \citenamefont {Yi},\ and\ \citenamefont {Xue}}]{XuePeng2020}%
  \BibitemOpen
  \bibfield  {author} {\bibinfo {author} {\bibfnamefont {Lei}\ \bibnamefont
  {Xiao}}, \bibinfo {author} {\bibfnamefont {Tianshu}\ \bibnamefont {Deng}},
  \bibinfo {author} {\bibfnamefont {Kunkun}\ \bibnamefont {Wang}}, \bibinfo
  {author} {\bibfnamefont {Gaoyan}\ \bibnamefont {Zhu}}, \bibinfo {author}
  {\bibfnamefont {Zhong}\ \bibnamefont {Wang}}, \bibinfo {author}
  {\bibfnamefont {Wei}\ \bibnamefont {Yi}}, \ and\ \bibinfo {author}
  {\bibfnamefont {Peng}\ \bibnamefont {Xue}},\ }\bibfield  {title} {\enquote
  {\bibinfo {title} {{Non-Hermitian bulk--boundary correspondence in quantum
  dynamics}},}\ }\href {\doibase 10.1038/s41567-020-0836-6} {\bibfield
  {journal} {\bibinfo  {journal} {Nature Physics}\ }\textbf {\bibinfo {volume}
  {16}},\ \bibinfo {pages} {761--766} (\bibinfo {year}
  {2020}{\natexlab{a}})}\BibitemShut {NoStop}%
\bibitem [{\citenamefont {Xiao}\ \emph
  {et~al.}(2020{\natexlab{b}})\citenamefont {Xiao}, \citenamefont {Deng},
  \citenamefont {Wang}, \citenamefont {Wang}, \citenamefont {Yi},\ and\
  \citenamefont {Xue}}]{LeiXiao_PT}%
  \BibitemOpen
  \bibfield  {author} {\bibinfo {author} {\bibfnamefont {Lei}\ \bibnamefont
  {Xiao}}, \bibinfo {author} {\bibfnamefont {Tianshu}\ \bibnamefont {Deng}},
  \bibinfo {author} {\bibfnamefont {Kunkun}\ \bibnamefont {Wang}}, \bibinfo
  {author} {\bibfnamefont {Zhong}\ \bibnamefont {Wang}}, \bibinfo {author}
  {\bibfnamefont {Wei}\ \bibnamefont {Yi}}, \ and\ \bibinfo {author}
  {\bibfnamefont {Peng}\ \bibnamefont {Xue}},\ }\bibfield  {title} {\enquote
  {\bibinfo {title} {{Observation of non-Bloch parity-time symmetry and
  exceptional points}},}\ }\href@noop {} {\  (\bibinfo {year}
  {2020}{\natexlab{b}})},\ \Eprint {http://arxiv.org/abs/arXiv:2009.07288}
  {arXiv:2009.07288} \BibitemShut {NoStop}%
\bibitem [{\citenamefont {Helbig}\ \emph {et~al.}(2020)\citenamefont {Helbig},
  \citenamefont {Hofmann}, \citenamefont {Imhof}, \citenamefont {Abdelghany},
  \citenamefont {Kiessling}, \citenamefont {Molenkamp}, \citenamefont {Lee},
  \citenamefont {Szameit}, \citenamefont {Greiter},\ and\ \citenamefont
  {Thomale}}]{Thomale2020}%
  \BibitemOpen
  \bibfield  {author} {\bibinfo {author} {\bibfnamefont {T.}~\bibnamefont
  {Helbig}}, \bibinfo {author} {\bibfnamefont {T.}~\bibnamefont {Hofmann}},
  \bibinfo {author} {\bibfnamefont {S.}~\bibnamefont {Imhof}}, \bibinfo
  {author} {\bibfnamefont {M.}~\bibnamefont {Abdelghany}}, \bibinfo {author}
  {\bibfnamefont {T.}~\bibnamefont {Kiessling}}, \bibinfo {author}
  {\bibfnamefont {L.~W.}\ \bibnamefont {Molenkamp}}, \bibinfo {author}
  {\bibfnamefont {C.~H.}\ \bibnamefont {Lee}}, \bibinfo {author} {\bibfnamefont
  {A.}~\bibnamefont {Szameit}}, \bibinfo {author} {\bibfnamefont
  {M.}~\bibnamefont {Greiter}}, \ and\ \bibinfo {author} {\bibfnamefont
  {R.}~\bibnamefont {Thomale}},\ }\bibfield  {title} {\enquote {\bibinfo
  {title} {{Generalized bulk--boundary correspondence in non-Hermitian
  topolectrical circuits}},}\ }\href {\doibase 10.1038/s41567-020-0922-9}
  {\bibfield  {journal} {\bibinfo  {journal} {Nature Physics}\ }\textbf
  {\bibinfo {volume} {16}},\ \bibinfo {pages} {747--750} (\bibinfo {year}
  {2020})}\BibitemShut {NoStop}%
\bibitem [{\citenamefont {Weidemann}\ \emph {et~al.}(2020)\citenamefont
  {Weidemann}, \citenamefont {Kremer}, \citenamefont {Helbig}, \citenamefont
  {Hofmann}, \citenamefont {Stegmaier}, \citenamefont {Greiter}, \citenamefont
  {Thomale},\ and\ \citenamefont {Szameit}}]{FunnelingScience2020}%
  \BibitemOpen
  \bibfield  {author} {\bibinfo {author} {\bibfnamefont {Sebastian}\
  \bibnamefont {Weidemann}}, \bibinfo {author} {\bibfnamefont {Mark}\
  \bibnamefont {Kremer}}, \bibinfo {author} {\bibfnamefont {Tobias}\
  \bibnamefont {Helbig}}, \bibinfo {author} {\bibfnamefont {Tobias}\
  \bibnamefont {Hofmann}}, \bibinfo {author} {\bibfnamefont {Alexander}\
  \bibnamefont {Stegmaier}}, \bibinfo {author} {\bibfnamefont {Martin}\
  \bibnamefont {Greiter}}, \bibinfo {author} {\bibfnamefont {Ronny}\
  \bibnamefont {Thomale}}, \ and\ \bibinfo {author} {\bibfnamefont {Alexander}\
  \bibnamefont {Szameit}},\ }\bibfield  {title} {\enquote {\bibinfo {title}
  {{Topological funneling of light}},}\ }\href {\doibase
  10.1126/science.aaz8727} {\bibfield  {journal} {\bibinfo  {journal}
  {Science}\ }\textbf {\bibinfo {volume} {368}},\ \bibinfo {pages} {311--314}
  (\bibinfo {year} {2020})}\BibitemShut {NoStop}%
\bibitem [{\citenamefont {Brandenbourger}\ \emph {et~al.}(2019)\citenamefont
  {Brandenbourger}, \citenamefont {Locsin}, \citenamefont {Lerner},\ and\
  \citenamefont {Coulais}}]{BrandenbourgerNC2019}%
  \BibitemOpen
  \bibfield  {author} {\bibinfo {author} {\bibfnamefont {Martin}\ \bibnamefont
  {Brandenbourger}}, \bibinfo {author} {\bibfnamefont {Xander}\ \bibnamefont
  {Locsin}}, \bibinfo {author} {\bibfnamefont {Edan}\ \bibnamefont {Lerner}}, \
  and\ \bibinfo {author} {\bibfnamefont {Corentin}\ \bibnamefont {Coulais}},\
  }\bibfield  {title} {\enquote {\bibinfo {title} {{Non-reciprocal robotic
  metamaterials}},}\ }\href {\doibase 10.1038/s41467-019-12599-3} {\bibfield
  {journal} {\bibinfo  {journal} {Nature Communications}\ }\textbf {\bibinfo
  {volume} {10}},\ \bibinfo {pages} {4608} (\bibinfo {year}
  {2019})}\BibitemShut {NoStop}%
\bibitem [{\citenamefont {Ghatak}\ \emph {et~al.}(2020)\citenamefont {Ghatak},
  \citenamefont {Brandenbourger}, \citenamefont {van Wezel},\ and\
  \citenamefont {Coulais}}]{GhatakPNAS2020}%
  \BibitemOpen
  \bibfield  {author} {\bibinfo {author} {\bibfnamefont {Ananya}\ \bibnamefont
  {Ghatak}}, \bibinfo {author} {\bibfnamefont {Martin}\ \bibnamefont
  {Brandenbourger}}, \bibinfo {author} {\bibfnamefont {Jasper}\ \bibnamefont
  {van Wezel}}, \ and\ \bibinfo {author} {\bibfnamefont {Corentin}\
  \bibnamefont {Coulais}},\ }\bibfield  {title} {\enquote {\bibinfo {title}
  {Observation of non-hermitian topology and its bulk{\textendash}edge
  correspondence in an active mechanical metamaterial},}\ }\href {\doibase
  10.1073/pnas.2010580117} {\bibfield  {journal} {\bibinfo  {journal}
  {Proceedings of the National Academy of Sciences}\ }\textbf {\bibinfo
  {volume} {117}},\ \bibinfo {pages} {29561--29568} (\bibinfo {year}
  {2020})}\BibitemShut {NoStop}%
\bibitem [{\citenamefont {Wang}\ \emph {et~al.}(2021)\citenamefont {Wang},
  \citenamefont {Dutt}, \citenamefont {Yang}, \citenamefont {Wojcik},
  \citenamefont {Vu{\v c}kovi{\'c}},\ and\ \citenamefont
  {Fan}}]{WangScience2021}%
  \BibitemOpen
  \bibfield  {author} {\bibinfo {author} {\bibfnamefont {Kai}\ \bibnamefont
  {Wang}}, \bibinfo {author} {\bibfnamefont {Avik}\ \bibnamefont {Dutt}},
  \bibinfo {author} {\bibfnamefont {Ki~Youl}\ \bibnamefont {Yang}}, \bibinfo
  {author} {\bibfnamefont {Casey~C.}\ \bibnamefont {Wojcik}}, \bibinfo {author}
  {\bibfnamefont {Jelena}\ \bibnamefont {Vu{\v c}kovi{\'c}}}, \ and\ \bibinfo
  {author} {\bibfnamefont {Shanhui}\ \bibnamefont {Fan}},\ }\bibfield  {title}
  {\enquote {\bibinfo {title} {{Generating arbitrary topological windings of a
  non-Hermitian band}},}\ }\href {\doibase 10.1126/science.abf6568} {\bibfield
  {journal} {\bibinfo  {journal} {Science}\ }\textbf {\bibinfo {volume}
  {371}},\ \bibinfo {pages} {1240--1245} (\bibinfo {year} {2021})}\BibitemShut
  {NoStop}%
\bibitem [{\citenamefont {Lee}\ \emph {et~al.}(2019{\natexlab{b}})\citenamefont
  {Lee}, \citenamefont {Li},\ and\ \citenamefont {Gong}}]{ChingHua2019_Hybrid}%
  \BibitemOpen
  \bibfield  {author} {\bibinfo {author} {\bibfnamefont {Ching~Hua}\
  \bibnamefont {Lee}}, \bibinfo {author} {\bibfnamefont {Linhu}\ \bibnamefont
  {Li}}, \ and\ \bibinfo {author} {\bibfnamefont {Jiangbin}\ \bibnamefont
  {Gong}},\ }\bibfield  {title} {\enquote {\bibinfo {title} {{Hybrid
  Higher-Order Skin-Topological Modes in Nonreciprocal Systems}},}\ }\href
  {\doibase 10.1103/PhysRevLett.123.016805} {\bibfield  {journal} {\bibinfo
  {journal} {Phys. Rev. Lett.}\ }\textbf {\bibinfo {volume} {123}},\ \bibinfo
  {pages} {016805} (\bibinfo {year} {2019}{\natexlab{b}})}\BibitemShut
  {NoStop}%
\bibitem [{\citenamefont {Ezawa}(2019{\natexlab{a}})}]{EzawaInterface2019}%
  \BibitemOpen
  \bibfield  {author} {\bibinfo {author} {\bibfnamefont {Motohiko}\
  \bibnamefont {Ezawa}},\ }\bibfield  {title} {\enquote {\bibinfo {title}
  {{Non-Hermitian boundary and interface states in nonreciprocal higher-order
  topological metals and electrical circuits}},}\ }\href {\doibase
  10.1103/PhysRevB.99.121411} {\bibfield  {journal} {\bibinfo  {journal} {Phys.
  Rev. B}\ }\textbf {\bibinfo {volume} {99}},\ \bibinfo {pages} {121411}
  (\bibinfo {year} {2019}{\natexlab{a}})}\BibitemShut {NoStop}%
\bibitem [{\citenamefont {Ezawa}(2019{\natexlab{b}})}]{EzawaPRB2019}%
  \BibitemOpen
  \bibfield  {author} {\bibinfo {author} {\bibfnamefont {Motohiko}\
  \bibnamefont {Ezawa}},\ }\bibfield  {title} {\enquote {\bibinfo {title}
  {{Non-Hermitian higher-order topological states in nonreciprocal and
  reciprocal systems with their electric-circuit realization}},}\ }\href
  {\doibase 10.1103/PhysRevB.99.201411} {\bibfield  {journal} {\bibinfo
  {journal} {Phys. Rev. B}\ }\textbf {\bibinfo {volume} {99}},\ \bibinfo
  {pages} {201411} (\bibinfo {year} {2019}{\natexlab{b}})}\BibitemShut
  {NoStop}%
\bibitem [{\citenamefont {Hofmann}\ \emph {et~al.}(2020)\citenamefont
  {Hofmann}, \citenamefont {Helbig}, \citenamefont {Schindler}, \citenamefont
  {Salgo}, \citenamefont {Brzezi\ifmmode~\acute{n}\else \'{n}\fi{}ska},
  \citenamefont {Greiter}, \citenamefont {Kiessling}, \citenamefont {Wolf},
  \citenamefont {Vollhardt}, \citenamefont {Kaba\ifmmode~\check{s}\else
  \v{s}\fi{}i}, \citenamefont {Lee}, \citenamefont {Bilu\ifmmode \check{s}\else
  \v{s}\fi{}i\ifmmode~\acute{c}\else \'{c}\fi{}}, \citenamefont {Thomale},\
  and\ \citenamefont {Neupert}}]{Titus2020}%
  \BibitemOpen
  \bibfield  {author} {\bibinfo {author} {\bibfnamefont {Tobias}\ \bibnamefont
  {Hofmann}}, \bibinfo {author} {\bibfnamefont {Tobias}\ \bibnamefont
  {Helbig}}, \bibinfo {author} {\bibfnamefont {Frank}\ \bibnamefont
  {Schindler}}, \bibinfo {author} {\bibfnamefont {Nora}\ \bibnamefont {Salgo}},
  \bibinfo {author} {\bibfnamefont {Marta}\ \bibnamefont
  {Brzezi\ifmmode~\acute{n}\else \'{n}\fi{}ska}}, \bibinfo {author}
  {\bibfnamefont {Martin}\ \bibnamefont {Greiter}}, \bibinfo {author}
  {\bibfnamefont {Tobias}\ \bibnamefont {Kiessling}}, \bibinfo {author}
  {\bibfnamefont {David}\ \bibnamefont {Wolf}}, \bibinfo {author}
  {\bibfnamefont {Achim}\ \bibnamefont {Vollhardt}}, \bibinfo {author}
  {\bibfnamefont {Anton}\ \bibnamefont {Kaba\ifmmode~\check{s}\else
  \v{s}\fi{}i}}, \bibinfo {author} {\bibfnamefont {Ching~Hua}\ \bibnamefont
  {Lee}}, \bibinfo {author} {\bibfnamefont {Ante}\ \bibnamefont {Bilu\ifmmode
  \check{s}\else \v{s}\fi{}i\ifmmode~\acute{c}\else \'{c}\fi{}}}, \bibinfo
  {author} {\bibfnamefont {Ronny}\ \bibnamefont {Thomale}}, \ and\ \bibinfo
  {author} {\bibfnamefont {Titus}\ \bibnamefont {Neupert}},\ }\bibfield
  {title} {\enquote {\bibinfo {title} {{Reciprocal skin effect and its
  realization in a topolectrical circuit}},}\ }\href {\doibase
  10.1103/PhysRevResearch.2.023265} {\bibfield  {journal} {\bibinfo  {journal}
  {Phys. Rev. Research}\ }\textbf {\bibinfo {volume} {2}},\ \bibinfo {pages}
  {023265} (\bibinfo {year} {2020})}\BibitemShut {NoStop}%
\bibitem [{\citenamefont {Liu}\ \emph {et~al.}(2019)\citenamefont {Liu},
  \citenamefont {Zhang}, \citenamefont {Ai}, \citenamefont {Gong},
  \citenamefont {Kawabata}, \citenamefont {Ueda},\ and\ \citenamefont
  {Nori}}]{Nori2019}%
  \BibitemOpen
  \bibfield  {author} {\bibinfo {author} {\bibfnamefont {Tao}\ \bibnamefont
  {Liu}}, \bibinfo {author} {\bibfnamefont {Yu-Ran}\ \bibnamefont {Zhang}},
  \bibinfo {author} {\bibfnamefont {Qing}\ \bibnamefont {Ai}}, \bibinfo
  {author} {\bibfnamefont {Zongping}\ \bibnamefont {Gong}}, \bibinfo {author}
  {\bibfnamefont {Kohei}\ \bibnamefont {Kawabata}}, \bibinfo {author}
  {\bibfnamefont {Masahito}\ \bibnamefont {Ueda}}, \ and\ \bibinfo {author}
  {\bibfnamefont {Franco}\ \bibnamefont {Nori}},\ }\bibfield  {title} {\enquote
  {\bibinfo {title} {{Second-Order Topological Phases in Non-Hermitian
  Systems}},}\ }\href {\doibase 10.1103/PhysRevLett.122.076801} {\bibfield
  {journal} {\bibinfo  {journal} {Phys. Rev. Lett.}\ }\textbf {\bibinfo
  {volume} {122}},\ \bibinfo {pages} {076801} (\bibinfo {year}
  {2019})}\BibitemShut {NoStop}%
\bibitem [{\citenamefont {Yoshida}\ \emph {et~al.}(2020)\citenamefont
  {Yoshida}, \citenamefont {Mizoguchi},\ and\ \citenamefont
  {Hatsugai}}]{Yoshida2020}%
  \BibitemOpen
  \bibfield  {author} {\bibinfo {author} {\bibfnamefont {Tsuneya}\ \bibnamefont
  {Yoshida}}, \bibinfo {author} {\bibfnamefont {Tomonari}\ \bibnamefont
  {Mizoguchi}}, \ and\ \bibinfo {author} {\bibfnamefont {Yasuhiro}\
  \bibnamefont {Hatsugai}},\ }\bibfield  {title} {\enquote {\bibinfo {title}
  {{Mirror skin effect and its electric circuit simulation}},}\ }\href
  {\doibase 10.1103/PhysRevResearch.2.022062} {\bibfield  {journal} {\bibinfo
  {journal} {Phys. Rev. Research}\ }\textbf {\bibinfo {volume} {2}},\ \bibinfo
  {pages} {022062(R)} (\bibinfo {year} {2020})}\BibitemShut {NoStop}%
\bibitem [{\citenamefont {Kawabata}\ \emph {et~al.}(2020)\citenamefont
  {Kawabata}, \citenamefont {Sato},\ and\ \citenamefont
  {Shiozaki}}]{Kawabata2020}%
  \BibitemOpen
  \bibfield  {author} {\bibinfo {author} {\bibfnamefont {Kohei}\ \bibnamefont
  {Kawabata}}, \bibinfo {author} {\bibfnamefont {Masatoshi}\ \bibnamefont
  {Sato}}, \ and\ \bibinfo {author} {\bibfnamefont {Ken}\ \bibnamefont
  {Shiozaki}},\ }\bibfield  {title} {\enquote {\bibinfo {title} {{Higher-order
  non-Hermitian skin effect}},}\ }\href {\doibase 10.1103/PhysRevB.102.205118}
  {\bibfield  {journal} {\bibinfo  {journal} {Phys. Rev. B}\ }\textbf {\bibinfo
  {volume} {102}},\ \bibinfo {pages} {205118} (\bibinfo {year}
  {2020})}\BibitemShut {NoStop}%
\bibitem [{\citenamefont {Scheibner}\ \emph {et~al.}(2020)\citenamefont
  {Scheibner}, \citenamefont {Irvine},\ and\ \citenamefont
  {Vitelli}}]{Vincenzo2020}%
  \BibitemOpen
  \bibfield  {author} {\bibinfo {author} {\bibfnamefont {Colin}\ \bibnamefont
  {Scheibner}}, \bibinfo {author} {\bibfnamefont {William T.~M.}\ \bibnamefont
  {Irvine}}, \ and\ \bibinfo {author} {\bibfnamefont {Vincenzo}\ \bibnamefont
  {Vitelli}},\ }\bibfield  {title} {\enquote {\bibinfo {title} {{Non-Hermitian
  Band Topology and Skin Modes in Active Elastic Media}},}\ }\href {\doibase
  10.1103/PhysRevLett.125.118001} {\bibfield  {journal} {\bibinfo  {journal}
  {Phys. Rev. Lett.}\ }\textbf {\bibinfo {volume} {125}},\ \bibinfo {pages}
  {118001} (\bibinfo {year} {2020})}\BibitemShut {NoStop}%
\bibitem [{\citenamefont {Zhang}\ \emph
  {et~al.}(2019{\natexlab{a}})\citenamefont {Zhang}, \citenamefont
  {Rosendo~L\'opez}, \citenamefont {Cheng}, \citenamefont {Liu},\ and\
  \citenamefont {Christensen}}]{SonicSOTI_2019}%
  \BibitemOpen
  \bibfield  {author} {\bibinfo {author} {\bibfnamefont {Zhiwang}\ \bibnamefont
  {Zhang}}, \bibinfo {author} {\bibfnamefont {Mar\'{\i}a}\ \bibnamefont
  {Rosendo~L\'opez}}, \bibinfo {author} {\bibfnamefont {Ying}\ \bibnamefont
  {Cheng}}, \bibinfo {author} {\bibfnamefont {Xiaojun}\ \bibnamefont {Liu}}, \
  and\ \bibinfo {author} {\bibfnamefont {Johan}\ \bibnamefont {Christensen}},\
  }\bibfield  {title} {\enquote {\bibinfo {title} {{Non-Hermitian Sonic
  Second-Order Topological Insulator}},}\ }\href {\doibase
  10.1103/PhysRevLett.122.195501} {\bibfield  {journal} {\bibinfo  {journal}
  {Phys. Rev. Lett.}\ }\textbf {\bibinfo {volume} {122}},\ \bibinfo {pages}
  {195501} (\bibinfo {year} {2019}{\natexlab{a}})}\BibitemShut {NoStop}%
\bibitem [{\citenamefont {Luo}\ and\ \citenamefont
  {Zhang}(2019)}]{ChuanweiPRL2019}%
  \BibitemOpen
  \bibfield  {author} {\bibinfo {author} {\bibfnamefont {Xi-Wang}\ \bibnamefont
  {Luo}}\ and\ \bibinfo {author} {\bibfnamefont {Chuanwei}\ \bibnamefont
  {Zhang}},\ }\bibfield  {title} {\enquote {\bibinfo {title} {{Higher-Order
  Topological Corner States Induced by Gain and Loss}},}\ }\href {\doibase
  10.1103/PhysRevLett.123.073601} {\bibfield  {journal} {\bibinfo  {journal}
  {Phys. Rev. Lett.}\ }\textbf {\bibinfo {volume} {123}},\ \bibinfo {pages}
  {073601} (\bibinfo {year} {2019})}\BibitemShut {NoStop}%
\bibitem [{\citenamefont {Edvardsson}\ \emph {et~al.}(2019)\citenamefont
  {Edvardsson}, \citenamefont {Kunst},\ and\ \citenamefont
  {Bergholtz}}]{KunstPRB2019}%
  \BibitemOpen
  \bibfield  {author} {\bibinfo {author} {\bibfnamefont {Elisabet}\
  \bibnamefont {Edvardsson}}, \bibinfo {author} {\bibfnamefont {Flore~K.}\
  \bibnamefont {Kunst}}, \ and\ \bibinfo {author} {\bibfnamefont {Emil~J.}\
  \bibnamefont {Bergholtz}},\ }\bibfield  {title} {\enquote {\bibinfo {title}
  {{Non-Hermitian extensions of higher-order topological phases and their
  biorthogonal bulk-boundary correspondence}},}\ }\href {\doibase
  10.1103/PhysRevB.99.081302} {\bibfield  {journal} {\bibinfo  {journal} {Phys.
  Rev. B}\ }\textbf {\bibinfo {volume} {99}},\ \bibinfo {pages} {081302}
  (\bibinfo {year} {2019})}\BibitemShut {NoStop}%
\bibitem [{\citenamefont {Ma}\ and\ \citenamefont
  {Hughes}(2020)}]{HughesQSHE2020}%
  \BibitemOpen
  \bibfield  {author} {\bibinfo {author} {\bibfnamefont {Yuhao}\ \bibnamefont
  {Ma}}\ and\ \bibinfo {author} {\bibfnamefont {Taylor~L.}\ \bibnamefont
  {Hughes}},\ }\bibfield  {title} {\enquote {\bibinfo {title} {{The Quantum
  Skin Hall Effect}},}\ }\href@noop {} {\  (\bibinfo {year} {2020})},\ \Eprint
  {http://arxiv.org/abs/2008.02284} {arXiv:2008.02284} \BibitemShut {NoStop}%
\bibitem [{\citenamefont {Okugawa}\ \emph {et~al.}(2020)\citenamefont
  {Okugawa}, \citenamefont {Takahashi},\ and\ \citenamefont
  {Yokomizo}}]{YokomizoPRB2020}%
  \BibitemOpen
  \bibfield  {author} {\bibinfo {author} {\bibfnamefont {Ryo}\ \bibnamefont
  {Okugawa}}, \bibinfo {author} {\bibfnamefont {Ryo}\ \bibnamefont
  {Takahashi}}, \ and\ \bibinfo {author} {\bibfnamefont {Kazuki}\ \bibnamefont
  {Yokomizo}},\ }\bibfield  {title} {\enquote {\bibinfo {title} {{Second-order
  topological non-Hermitian skin effects}},}\ }\href {\doibase
  10.1103/PhysRevB.102.241202} {\bibfield  {journal} {\bibinfo  {journal}
  {Phys. Rev. B}\ }\textbf {\bibinfo {volume} {102}},\ \bibinfo {pages}
  {241202} (\bibinfo {year} {2020})}\BibitemShut {NoStop}%
\bibitem [{\citenamefont {Palacios}\ \emph {et~al.}(2020)\citenamefont
  {Palacios}, \citenamefont {Tchoumakov}, \citenamefont {Guix}, \citenamefont
  {Pagonabarraga}, \citenamefont {Sánchez},\ and\ \citenamefont
  {Grushin}}]{Lucas2020}%
  \BibitemOpen
  \bibfield  {author} {\bibinfo {author} {\bibfnamefont {Lucas~S.}\
  \bibnamefont {Palacios}}, \bibinfo {author} {\bibfnamefont {Serguei}\
  \bibnamefont {Tchoumakov}}, \bibinfo {author} {\bibfnamefont {Maria}\
  \bibnamefont {Guix}}, \bibinfo {author} {\bibfnamefont {Ignasio}\
  \bibnamefont {Pagonabarraga}}, \bibinfo {author} {\bibfnamefont {Samuel}\
  \bibnamefont {Sánchez}}, \ and\ \bibinfo {author} {\bibfnamefont
  {Adolfo~G.}\ \bibnamefont {Grushin}},\ }\bibfield  {title} {\enquote
  {\bibinfo {title} {{Guided accumulation of active particles by topological
  design of a second-order skin effect}},}\ }\href@noop {} {\  (\bibinfo {year}
  {2020})},\ \Eprint {http://arxiv.org/abs/arXiv:2012.14496} {arXiv:2012.14496}
  \BibitemShut {NoStop}%
\bibitem [{\citenamefont {Song}\ \emph {et~al.}(2020)\citenamefont {Song},
  \citenamefont {Liu}, \citenamefont {Zheng}, \citenamefont {Zhang},
  \citenamefont {Wang},\ and\ \citenamefont {Lu}}]{SyntheticPhcPRA_2020}%
  \BibitemOpen
  \bibfield  {author} {\bibinfo {author} {\bibfnamefont {Yiling}\ \bibnamefont
  {Song}}, \bibinfo {author} {\bibfnamefont {Weiwei}\ \bibnamefont {Liu}},
  \bibinfo {author} {\bibfnamefont {Lingzhi}\ \bibnamefont {Zheng}}, \bibinfo
  {author} {\bibfnamefont {Yicong}\ \bibnamefont {Zhang}}, \bibinfo {author}
  {\bibfnamefont {Bing}\ \bibnamefont {Wang}}, \ and\ \bibinfo {author}
  {\bibfnamefont {Peixiang}\ \bibnamefont {Lu}},\ }\bibfield  {title} {\enquote
  {\bibinfo {title} {{Two-dimensional non-Hermitian Skin Effect in a Synthetic
  Photonic Lattice}},}\ }\href {\doibase 10.1103/PhysRevApplied.14.064076}
  {\bibfield  {journal} {\bibinfo  {journal} {Phys. Rev. Applied}\ }\textbf
  {\bibinfo {volume} {14}},\ \bibinfo {pages} {064076} (\bibinfo {year}
  {2020})}\BibitemShut {NoStop}%
\bibitem [{\citenamefont {Fu}\ \emph {et~al.}(2021)\citenamefont {Fu},
  \citenamefont {Hu},\ and\ \citenamefont {Wan}}]{YongxuPRB2021}%
  \BibitemOpen
  \bibfield  {author} {\bibinfo {author} {\bibfnamefont {Yongxu}\ \bibnamefont
  {Fu}}, \bibinfo {author} {\bibfnamefont {Jihan}\ \bibnamefont {Hu}}, \ and\
  \bibinfo {author} {\bibfnamefont {Shaolong}\ \bibnamefont {Wan}},\ }\bibfield
   {title} {\enquote {\bibinfo {title} {Non-hermitian second-order skin and
  topological modes},}\ }\href {\doibase 10.1103/PhysRevB.103.045420}
  {\bibfield  {journal} {\bibinfo  {journal} {Phys. Rev. B}\ }\textbf {\bibinfo
  {volume} {103}},\ \bibinfo {pages} {045420} (\bibinfo {year}
  {2021})}\BibitemShut {NoStop}%
\bibitem [{\citenamefont {Yu}\ \emph {et~al.}(2021)\citenamefont {Yu},
  \citenamefont {Jung},\ and\ \citenamefont {Shvets}}]{GennadyPRB2021}%
  \BibitemOpen
  \bibfield  {author} {\bibinfo {author} {\bibfnamefont {Yang}\ \bibnamefont
  {Yu}}, \bibinfo {author} {\bibfnamefont {Minwoo}\ \bibnamefont {Jung}}, \
  and\ \bibinfo {author} {\bibfnamefont {Gennady}\ \bibnamefont {Shvets}},\
  }\bibfield  {title} {\enquote {\bibinfo {title} {{Zero-energy corner states
  in a non-Hermitian quadrupole insulator}},}\ }\href {\doibase
  10.1103/PhysRevB.103.L041102} {\bibfield  {journal} {\bibinfo  {journal}
  {Phys. Rev. B}\ }\textbf {\bibinfo {volume} {103}},\ \bibinfo {pages}
  {L041102} (\bibinfo {year} {2021})}\BibitemShut {NoStop}%
\bibitem [{\citenamefont {Song}\ \emph {et~al.}(2021)\citenamefont {Song},
  \citenamefont {Wang},\ and\ \citenamefont {Wang}}]{SongFei2021}%
  \BibitemOpen
  \bibfield  {author} {\bibinfo {author} {\bibfnamefont {Fei}\ \bibnamefont
  {Song}}, \bibinfo {author} {\bibfnamefont {Hong-Yi}\ \bibnamefont {Wang}}, \
  and\ \bibinfo {author} {\bibfnamefont {Zhong}\ \bibnamefont {Wang}},\
  }\bibfield  {title} {\enquote {\bibinfo {title} {{Non-Bloch PT symmetry
  breaking: Universal threshold and dimensional surprise}},}\ }\href@noop {}
  {\bibfield  {journal} {\bibinfo  {journal} {arXiv:2102.02230}\ } (\bibinfo
  {year} {2021})}\BibitemShut {NoStop}%
\bibitem [{\citenamefont {Kato}(2013)}]{Kato2013}%
  \BibitemOpen
  \bibfield  {author} {\bibinfo {author} {\bibfnamefont {Tosio}\ \bibnamefont
  {Kato}},\ }\href@noop {} {\emph {\bibinfo {title} {{Perturbation theory for
  linear operators}}}},\ Vol.\ \bibinfo {volume} {132}\ (\bibinfo  {publisher}
  {Springer Science \& Business Media},\ \bibinfo {year} {2013})\BibitemShut
  {NoStop}%
\bibitem [{\citenamefont {Miri}\ and\ \citenamefont
  {Al{\`u}}(2019)}]{Miri2019}%
  \BibitemOpen
  \bibfield  {author} {\bibinfo {author} {\bibfnamefont {Mohammad-Ali}\
  \bibnamefont {Miri}}\ and\ \bibinfo {author} {\bibfnamefont {Andrea}\
  \bibnamefont {Al{\`u}}},\ }\bibfield  {title} {\enquote {\bibinfo {title}
  {{Exceptional points in optics and photonics}},}\ }\href {\doibase
  10.1126/science.aar7709} {\bibfield  {journal} {\bibinfo  {journal}
  {Science}\ }\textbf {\bibinfo {volume} {363}},\ \bibinfo {pages} {eaar7709}
  (\bibinfo {year} {2019})}\BibitemShut {NoStop}%
\bibitem [{\citenamefont {{\"O}zdemir}\ \emph {et~al.}(2019)\citenamefont
  {{\"O}zdemir}, \citenamefont {Rotter}, \citenamefont {Nori},\ and\
  \citenamefont {Yang}}]{YangLan2019}%
  \BibitemOpen
  \bibfield  {author} {\bibinfo {author} {\bibfnamefont {{\c S}.~K.}\
  \bibnamefont {{\"O}zdemir}}, \bibinfo {author} {\bibfnamefont
  {S.}~\bibnamefont {Rotter}}, \bibinfo {author} {\bibfnamefont
  {F.}~\bibnamefont {Nori}}, \ and\ \bibinfo {author} {\bibfnamefont
  {L.}~\bibnamefont {Yang}},\ }\bibfield  {title} {\enquote {\bibinfo {title}
  {{Parity--time symmetry and exceptional points in photonics}},}\ }\href
  {\doibase 10.1038/s41563-019-0304-9} {\bibfield  {journal} {\bibinfo
  {journal} {Nature Materials}\ }\textbf {\bibinfo {volume} {18}},\ \bibinfo
  {pages} {783--798} (\bibinfo {year} {2019})}\BibitemShut {NoStop}%
\bibitem [{\citenamefont {Lin}\ \emph {et~al.}(2011)\citenamefont {Lin},
  \citenamefont {Ramezani}, \citenamefont {Eichelkraut}, \citenamefont
  {Kottos}, \citenamefont {Cao},\ and\ \citenamefont
  {Christodoulides}}]{Lin2011_Invisibility}%
  \BibitemOpen
  \bibfield  {author} {\bibinfo {author} {\bibfnamefont {Zin}\ \bibnamefont
  {Lin}}, \bibinfo {author} {\bibfnamefont {Hamidreza}\ \bibnamefont
  {Ramezani}}, \bibinfo {author} {\bibfnamefont {Toni}\ \bibnamefont
  {Eichelkraut}}, \bibinfo {author} {\bibfnamefont {Tsampikos}\ \bibnamefont
  {Kottos}}, \bibinfo {author} {\bibfnamefont {Hui}\ \bibnamefont {Cao}}, \
  and\ \bibinfo {author} {\bibfnamefont {Demetrios~N.}\ \bibnamefont
  {Christodoulides}},\ }\bibfield  {title} {\enquote {\bibinfo {title}
  {{Unidirectional Invisibility Induced by $\mathcal{P}\mathcal{T}$-Symmetric
  Periodic Structures}},}\ }\href {\doibase 10.1103/PhysRevLett.106.213901}
  {\bibfield  {journal} {\bibinfo  {journal} {Phys. Rev. Lett.}\ }\textbf
  {\bibinfo {volume} {106}},\ \bibinfo {pages} {213901} (\bibinfo {year}
  {2011})}\BibitemShut {NoStop}%
\bibitem [{\citenamefont {Lee}(2016)}]{LeePRL2016}%
  \BibitemOpen
  \bibfield  {author} {\bibinfo {author} {\bibfnamefont {Tony~E.}\ \bibnamefont
  {Lee}},\ }\bibfield  {title} {\enquote {\bibinfo {title} {{Anomalous Edge
  State in a Non-Hermitian Lattice}},}\ }\href {\doibase
  10.1103/PhysRevLett.116.133903} {\bibfield  {journal} {\bibinfo  {journal}
  {Phys. Rev. Lett.}\ }\textbf {\bibinfo {volume} {116}},\ \bibinfo {pages}
  {133903} (\bibinfo {year} {2016})}\BibitemShut {NoStop}%
\bibitem [{\citenamefont {Molina}\ and\ \citenamefont
  {Gonz\'alez}(2018)}]{MolinaPRL2018}%
  \BibitemOpen
  \bibfield  {author} {\bibinfo {author} {\bibfnamefont {Rafael~A.}\
  \bibnamefont {Molina}}\ and\ \bibinfo {author} {\bibfnamefont {Jos\'e}\
  \bibnamefont {Gonz\'alez}},\ }\bibfield  {title} {\enquote {\bibinfo {title}
  {{Surface and 3D Quantum Hall Effects from Engineering of Exceptional Points
  in Nodal-Line Semimetals}},}\ }\href {\doibase
  10.1103/PhysRevLett.120.146601} {\bibfield  {journal} {\bibinfo  {journal}
  {Phys. Rev. Lett.}\ }\textbf {\bibinfo {volume} {120}},\ \bibinfo {pages}
  {146601} (\bibinfo {year} {2018})}\BibitemShut {NoStop}%
\bibitem [{\citenamefont {Carlstr\"om}\ and\ \citenamefont
  {Bergholtz}(2018)}]{BergholtzPRA2018}%
  \BibitemOpen
  \bibfield  {author} {\bibinfo {author} {\bibfnamefont {Johan}\ \bibnamefont
  {Carlstr\"om}}\ and\ \bibinfo {author} {\bibfnamefont {Emil~J.}\ \bibnamefont
  {Bergholtz}},\ }\bibfield  {title} {\enquote {\bibinfo {title} {Exceptional
  links and twisted fermi ribbons in non-hermitian systems},}\ }\href {\doibase
  10.1103/PhysRevA.98.042114} {\bibfield  {journal} {\bibinfo  {journal} {Phys.
  Rev. A}\ }\textbf {\bibinfo {volume} {98}},\ \bibinfo {pages} {042114}
  (\bibinfo {year} {2018})}\BibitemShut {NoStop}%
\bibitem [{\citenamefont {Cerjan}\ \emph {et~al.}(2018)\citenamefont {Cerjan},
  \citenamefont {Xiao}, \citenamefont {Yuan},\ and\ \citenamefont
  {Fan}}]{Cerjan2018}%
  \BibitemOpen
  \bibfield  {author} {\bibinfo {author} {\bibfnamefont {Alexander}\
  \bibnamefont {Cerjan}}, \bibinfo {author} {\bibfnamefont {Meng}\ \bibnamefont
  {Xiao}}, \bibinfo {author} {\bibfnamefont {Luqi}\ \bibnamefont {Yuan}}, \
  and\ \bibinfo {author} {\bibfnamefont {Shanhui}\ \bibnamefont {Fan}},\
  }\bibfield  {title} {\enquote {\bibinfo {title} {{Effects of non-Hermitian
  perturbations on Weyl Hamiltonians with arbitrary topological charges}},}\
  }\href {\doibase 10.1103/PhysRevB.97.075128} {\bibfield  {journal} {\bibinfo
  {journal} {Phys. Rev. B}\ }\textbf {\bibinfo {volume} {97}},\ \bibinfo
  {pages} {075128} (\bibinfo {year} {2018})}\BibitemShut {NoStop}%
\bibitem [{\citenamefont {Kawabata}\ \emph
  {et~al.}(2019{\natexlab{b}})\citenamefont {Kawabata}, \citenamefont
  {Bessho},\ and\ \citenamefont {Sato}}]{Sato2019_EP}%
  \BibitemOpen
  \bibfield  {author} {\bibinfo {author} {\bibfnamefont {Kohei}\ \bibnamefont
  {Kawabata}}, \bibinfo {author} {\bibfnamefont {Takumi}\ \bibnamefont
  {Bessho}}, \ and\ \bibinfo {author} {\bibfnamefont {Masatoshi}\ \bibnamefont
  {Sato}},\ }\bibfield  {title} {\enquote {\bibinfo {title} {{Classification of
  Exceptional Points and Non-Hermitian Topological Semimetals}},}\ }\href
  {\doibase 10.1103/PhysRevLett.123.066405} {\bibfield  {journal} {\bibinfo
  {journal} {Phys. Rev. Lett.}\ }\textbf {\bibinfo {volume} {123}},\ \bibinfo
  {pages} {066405} (\bibinfo {year} {2019}{\natexlab{b}})}\BibitemShut
  {NoStop}%
\bibitem [{\citenamefont {Moors}\ \emph {et~al.}(2019)\citenamefont {Moors},
  \citenamefont {Zyuzin}, \citenamefont {Zyuzin}, \citenamefont {Tiwari},\ and\
  \citenamefont {Schmidt}}]{ThomasPRB2019}%
  \BibitemOpen
  \bibfield  {author} {\bibinfo {author} {\bibfnamefont {Kristof}\ \bibnamefont
  {Moors}}, \bibinfo {author} {\bibfnamefont {Alexander~A.}\ \bibnamefont
  {Zyuzin}}, \bibinfo {author} {\bibfnamefont {Alexander~Yu.}\ \bibnamefont
  {Zyuzin}}, \bibinfo {author} {\bibfnamefont {Rakesh~P.}\ \bibnamefont
  {Tiwari}}, \ and\ \bibinfo {author} {\bibfnamefont {Thomas~L.}\ \bibnamefont
  {Schmidt}},\ }\bibfield  {title} {\enquote {\bibinfo {title}
  {{Disorder-driven exceptional lines and Fermi ribbons in tilted nodal-line
  semimetals}},}\ }\href {\doibase 10.1103/PhysRevB.99.041116} {\bibfield
  {journal} {\bibinfo  {journal} {Phys. Rev. B}\ }\textbf {\bibinfo {volume}
  {99}},\ \bibinfo {pages} {041116} (\bibinfo {year} {2019})}\BibitemShut
  {NoStop}%
\bibitem [{\citenamefont {Yang}\ \emph {et~al.}(2021)\citenamefont {Yang},
  \citenamefont {Schnyder}, \citenamefont {Hu},\ and\ \citenamefont
  {Chiu}}]{Zhesen2019_EP}%
  \BibitemOpen
  \bibfield  {author} {\bibinfo {author} {\bibfnamefont {Zhesen}\ \bibnamefont
  {Yang}}, \bibinfo {author} {\bibfnamefont {A.~P.}\ \bibnamefont {Schnyder}},
  \bibinfo {author} {\bibfnamefont {Jiangping}\ \bibnamefont {Hu}}, \ and\
  \bibinfo {author} {\bibfnamefont {Ching-Kai}\ \bibnamefont {Chiu}},\
  }\bibfield  {title} {\enquote {\bibinfo {title} {{Fermion Doubling Theorems
  in Two-Dimensional Non-Hermitian Systems for Fermi Points and Exceptional
  Points}},}\ }\href {\doibase 10.1103/PhysRevLett.126.086401} {\bibfield
  {journal} {\bibinfo  {journal} {Phys. Rev. Lett.}\ }\textbf {\bibinfo
  {volume} {126}},\ \bibinfo {pages} {086401} (\bibinfo {year}
  {2021})}\BibitemShut {NoStop}%
\bibitem [{\citenamefont {Xue}\ \emph {et~al.}(2020)\citenamefont {Xue},
  \citenamefont {Wang}, \citenamefont {Zhang},\ and\ \citenamefont
  {Chong}}]{Chong2020}%
  \BibitemOpen
  \bibfield  {author} {\bibinfo {author} {\bibfnamefont {Haoran}\ \bibnamefont
  {Xue}}, \bibinfo {author} {\bibfnamefont {Qiang}\ \bibnamefont {Wang}},
  \bibinfo {author} {\bibfnamefont {Baile}\ \bibnamefont {Zhang}}, \ and\
  \bibinfo {author} {\bibfnamefont {Y.~D.}\ \bibnamefont {Chong}},\ }\bibfield
  {title} {\enquote {\bibinfo {title} {{Non-Hermitian Dirac Cones}},}\ }\href
  {\doibase 10.1103/PhysRevLett.124.236403} {\bibfield  {journal} {\bibinfo
  {journal} {Phys. Rev. Lett.}\ }\textbf {\bibinfo {volume} {124}},\ \bibinfo
  {pages} {236403} (\bibinfo {year} {2020})}\BibitemShut {NoStop}%
\bibitem [{\citenamefont {Zhen}\ \emph {et~al.}(2015)\citenamefont {Zhen},
  \citenamefont {Hsu}, \citenamefont {Igarashi}, \citenamefont {Lu},
  \citenamefont {Kaminer}, \citenamefont {Pick}, \citenamefont {Chua},
  \citenamefont {Joannopoulos},\ and\ \citenamefont {Solja{\v
  c}i{\'c}}}]{ZhenBo2015}%
  \BibitemOpen
  \bibfield  {author} {\bibinfo {author} {\bibfnamefont {Bo}~\bibnamefont
  {Zhen}}, \bibinfo {author} {\bibfnamefont {Chia~Wei}\ \bibnamefont {Hsu}},
  \bibinfo {author} {\bibfnamefont {Yuichi}\ \bibnamefont {Igarashi}}, \bibinfo
  {author} {\bibfnamefont {Ling}\ \bibnamefont {Lu}}, \bibinfo {author}
  {\bibfnamefont {Ido}\ \bibnamefont {Kaminer}}, \bibinfo {author}
  {\bibfnamefont {Adi}\ \bibnamefont {Pick}}, \bibinfo {author} {\bibfnamefont
  {Song-Liang}\ \bibnamefont {Chua}}, \bibinfo {author} {\bibfnamefont
  {John~D.}\ \bibnamefont {Joannopoulos}}, \ and\ \bibinfo {author}
  {\bibfnamefont {Marin}\ \bibnamefont {Solja{\v c}i{\'c}}},\ }\bibfield
  {title} {\enquote {\bibinfo {title} {{Spawning rings of exceptional points
  out of Dirac cones}},}\ }\href {\doibase 10.1038/nature14889} {\bibfield
  {journal} {\bibinfo  {journal} {Nature}\ }\textbf {\bibinfo {volume} {525}},\
  \bibinfo {pages} {354--358} (\bibinfo {year} {2015})}\BibitemShut {NoStop}%
\bibitem [{\citenamefont {Zhou}\ \emph {et~al.}(2018)\citenamefont {Zhou},
  \citenamefont {Peng}, \citenamefont {Yoon}, \citenamefont {Hsu},
  \citenamefont {Nelson}, \citenamefont {Fu}, \citenamefont {Joannopoulos},
  \citenamefont {Solja{\v c}i{\'c}},\ and\ \citenamefont {Zhen}}]{Zhou2018}%
  \BibitemOpen
  \bibfield  {author} {\bibinfo {author} {\bibfnamefont {Hengyun}\ \bibnamefont
  {Zhou}}, \bibinfo {author} {\bibfnamefont {Chao}\ \bibnamefont {Peng}},
  \bibinfo {author} {\bibfnamefont {Yoseob}\ \bibnamefont {Yoon}}, \bibinfo
  {author} {\bibfnamefont {Chia~Wei}\ \bibnamefont {Hsu}}, \bibinfo {author}
  {\bibfnamefont {Keith~A.}\ \bibnamefont {Nelson}}, \bibinfo {author}
  {\bibfnamefont {Liang}\ \bibnamefont {Fu}}, \bibinfo {author} {\bibfnamefont
  {John~D.}\ \bibnamefont {Joannopoulos}}, \bibinfo {author} {\bibfnamefont
  {Marin}\ \bibnamefont {Solja{\v c}i{\'c}}}, \ and\ \bibinfo {author}
  {\bibfnamefont {Bo}~\bibnamefont {Zhen}},\ }\bibfield  {title} {\enquote
  {\bibinfo {title} {{Observation of bulk Fermi arc and polarization half
  charge from paired exceptional points}},}\ }\href {\doibase
  10.1126/science.aap9859} {\bibfield  {journal} {\bibinfo  {journal}
  {Science}\ }\textbf {\bibinfo {volume} {359}},\ \bibinfo {pages} {1009--1012}
  (\bibinfo {year} {2018})}\BibitemShut {NoStop}%
\bibitem [{\citenamefont {Zhang}\ \emph
  {et~al.}(2019{\natexlab{b}})\citenamefont {Zhang}, \citenamefont {Ding},
  \citenamefont {Zhou}, \citenamefont {Xu},\ and\ \citenamefont
  {Jin}}]{ZhangXufeng2019}%
  \BibitemOpen
  \bibfield  {author} {\bibinfo {author} {\bibfnamefont {Xufeng}\ \bibnamefont
  {Zhang}}, \bibinfo {author} {\bibfnamefont {Kun}\ \bibnamefont {Ding}},
  \bibinfo {author} {\bibfnamefont {Xianjing}\ \bibnamefont {Zhou}}, \bibinfo
  {author} {\bibfnamefont {Jing}\ \bibnamefont {Xu}}, \ and\ \bibinfo {author}
  {\bibfnamefont {Dafei}\ \bibnamefont {Jin}},\ }\bibfield  {title} {\enquote
  {\bibinfo {title} {{Experimental Observation of an Exceptional Surface in
  Synthetic Dimensions with Magnon Polaritons}},}\ }\href {\doibase
  10.1103/PhysRevLett.123.237202} {\bibfield  {journal} {\bibinfo  {journal}
  {Phys. Rev. Lett.}\ }\textbf {\bibinfo {volume} {123}},\ \bibinfo {pages}
  {237202} (\bibinfo {year} {2019}{\natexlab{b}})}\BibitemShut {NoStop}%
\bibitem [{\citenamefont {Cerjan}\ \emph {et~al.}(2019)\citenamefont {Cerjan},
  \citenamefont {Huang}, \citenamefont {Wang}, \citenamefont {Chen},
  \citenamefont {Chong},\ and\ \citenamefont {Rechtsman}}]{Cerjan2019}%
  \BibitemOpen
  \bibfield  {author} {\bibinfo {author} {\bibfnamefont {Alexander}\
  \bibnamefont {Cerjan}}, \bibinfo {author} {\bibfnamefont {Sheng}\
  \bibnamefont {Huang}}, \bibinfo {author} {\bibfnamefont {Mohan}\ \bibnamefont
  {Wang}}, \bibinfo {author} {\bibfnamefont {Kevin~P.}\ \bibnamefont {Chen}},
  \bibinfo {author} {\bibfnamefont {Yidong}\ \bibnamefont {Chong}}, \ and\
  \bibinfo {author} {\bibfnamefont {Mikael~C.}\ \bibnamefont {Rechtsman}},\
  }\bibfield  {title} {\enquote {\bibinfo {title} {{Experimental realization of
  a Weyl exceptional ring}},}\ }\href {\doibase 10.1038/s41566-019-0453-z}
  {\bibfield  {journal} {\bibinfo  {journal} {Nature Photonics}\ }\textbf
  {\bibinfo {volume} {13}},\ \bibinfo {pages} {623--628} (\bibinfo {year}
  {2019})}\BibitemShut {NoStop}%
\bibitem [{\citenamefont {Yang}\ and\ \citenamefont
  {Hu}(2019)}]{HopflinkPRB2019}%
  \BibitemOpen
  \bibfield  {author} {\bibinfo {author} {\bibfnamefont {Zhesen}\ \bibnamefont
  {Yang}}\ and\ \bibinfo {author} {\bibfnamefont {Jiangping}\ \bibnamefont
  {Hu}},\ }\bibfield  {title} {\enquote {\bibinfo {title} {{Non-Hermitian
  Hopf-link exceptional line semimetals}},}\ }\href {\doibase
  10.1103/PhysRevB.99.081102} {\bibfield  {journal} {\bibinfo  {journal} {Phys.
  Rev. B}\ }\textbf {\bibinfo {volume} {99}},\ \bibinfo {pages} {081102}
  (\bibinfo {year} {2019})}\BibitemShut {NoStop}%
\bibitem [{\citenamefont {Tang}\ \emph {et~al.}(2020)\citenamefont {Tang},
  \citenamefont {Jiang}, \citenamefont {Ding}, \citenamefont {Xiao},
  \citenamefont {Zhang}, \citenamefont {Chan},\ and\ \citenamefont
  {Ma}}]{Tang2020}%
  \BibitemOpen
  \bibfield  {author} {\bibinfo {author} {\bibfnamefont {Weiyuan}\ \bibnamefont
  {Tang}}, \bibinfo {author} {\bibfnamefont {Xue}\ \bibnamefont {Jiang}},
  \bibinfo {author} {\bibfnamefont {Kun}\ \bibnamefont {Ding}}, \bibinfo
  {author} {\bibfnamefont {Yi-Xin}\ \bibnamefont {Xiao}}, \bibinfo {author}
  {\bibfnamefont {Zhao-Qing}\ \bibnamefont {Zhang}}, \bibinfo {author}
  {\bibfnamefont {C.~T.}\ \bibnamefont {Chan}}, \ and\ \bibinfo {author}
  {\bibfnamefont {Guancong}\ \bibnamefont {Ma}},\ }\bibfield  {title} {\enquote
  {\bibinfo {title} {{Exceptional nexus with a hybrid topological
  invariant}},}\ }\href {\doibase 10.1126/science.abd8872} {\bibfield
  {journal} {\bibinfo  {journal} {Science}\ }\textbf {\bibinfo {volume}
  {370}},\ \bibinfo {pages} {1077--1080} (\bibinfo {year} {2020})}\BibitemShut
  {NoStop}%
\bibitem [{\citenamefont {Denner}\ \emph {et~al.}(2020)\citenamefont {Denner},
  \citenamefont {Skurativska}, \citenamefont {Schindler}, \citenamefont
  {Fischer}, \citenamefont {Thomale}, \citenamefont {Bzdušek},\ and\
  \citenamefont {Neupert}}]{ETI_2020}%
  \BibitemOpen
  \bibfield  {author} {\bibinfo {author} {\bibfnamefont {M.~Michael}\
  \bibnamefont {Denner}}, \bibinfo {author} {\bibfnamefont {Anastasiia}\
  \bibnamefont {Skurativska}}, \bibinfo {author} {\bibfnamefont {Frank}\
  \bibnamefont {Schindler}}, \bibinfo {author} {\bibfnamefont {Mark~H.}\
  \bibnamefont {Fischer}}, \bibinfo {author} {\bibfnamefont {Ronny}\
  \bibnamefont {Thomale}}, \bibinfo {author} {\bibfnamefont {Tomáš}\
  \bibnamefont {Bzdušek}}, \ and\ \bibinfo {author} {\bibfnamefont {Titus}\
  \bibnamefont {Neupert}},\ }\bibfield  {title} {\enquote {\bibinfo {title}
  {{Exceptional Topological Insulators}},}\ }\href@noop {} {\  (\bibinfo {year}
  {2020})},\ \Eprint {http://arxiv.org/abs/arXiv:2008.01090} {arXiv:2008.01090}
  \BibitemShut {NoStop}%
\bibitem [{\citenamefont {Yang}\ \emph
  {et~al.}(2020{\natexlab{b}})\citenamefont {Yang}, \citenamefont {Chiu},
  \citenamefont {Fang},\ and\ \citenamefont {Hu}}]{JonesPRL2020}%
  \BibitemOpen
  \bibfield  {author} {\bibinfo {author} {\bibfnamefont {Zhesen}\ \bibnamefont
  {Yang}}, \bibinfo {author} {\bibfnamefont {Ching-Kai}\ \bibnamefont {Chiu}},
  \bibinfo {author} {\bibfnamefont {Chen}\ \bibnamefont {Fang}}, \ and\
  \bibinfo {author} {\bibfnamefont {Jiangping}\ \bibnamefont {Hu}},\ }\bibfield
   {title} {\enquote {\bibinfo {title} {{Jones Polynomial and Knot Transitions
  in Hermitian and non-Hermitian Topological Semimetals}},}\ }\href {\doibase
  10.1103/PhysRevLett.124.186402} {\bibfield  {journal} {\bibinfo  {journal}
  {Phys. Rev. Lett.}\ }\textbf {\bibinfo {volume} {124}},\ \bibinfo {pages}
  {186402} (\bibinfo {year} {2020}{\natexlab{b}})}\BibitemShut {NoStop}%
\bibitem [{\citenamefont {Dembowski}\ \emph {et~al.}(2001)\citenamefont
  {Dembowski}, \citenamefont {Gr\"af}, \citenamefont {Harney}, \citenamefont
  {Heine}, \citenamefont {Heiss}, \citenamefont {Rehfeld},\ and\ \citenamefont
  {Richter}}]{Dembowski2001}%
  \BibitemOpen
  \bibfield  {author} {\bibinfo {author} {\bibfnamefont {C.}~\bibnamefont
  {Dembowski}}, \bibinfo {author} {\bibfnamefont {H.-D.}\ \bibnamefont
  {Gr\"af}}, \bibinfo {author} {\bibfnamefont {H.~L.}\ \bibnamefont {Harney}},
  \bibinfo {author} {\bibfnamefont {A.}~\bibnamefont {Heine}}, \bibinfo
  {author} {\bibfnamefont {W.~D.}\ \bibnamefont {Heiss}}, \bibinfo {author}
  {\bibfnamefont {H.}~\bibnamefont {Rehfeld}}, \ and\ \bibinfo {author}
  {\bibfnamefont {A.}~\bibnamefont {Richter}},\ }\bibfield  {title} {\enquote
  {\bibinfo {title} {{Experimental Observation of the Topological Structure of
  Exceptional Points}},}\ }\href {\doibase 10.1103/PhysRevLett.86.787}
  {\bibfield  {journal} {\bibinfo  {journal} {Phys. Rev. Lett.}\ }\textbf
  {\bibinfo {volume} {86}},\ \bibinfo {pages} {787--790} (\bibinfo {year}
  {2001})}\BibitemShut {NoStop}%
\bibitem [{\citenamefont {Dembowski}\ \emph {et~al.}(2003)\citenamefont
  {Dembowski}, \citenamefont {Dietz}, \citenamefont {Gr\"af}, \citenamefont
  {Harney}, \citenamefont {Heine}, \citenamefont {Heiss},\ and\ \citenamefont
  {Richter}}]{RichterPRL2003}%
  \BibitemOpen
  \bibfield  {author} {\bibinfo {author} {\bibfnamefont {C.}~\bibnamefont
  {Dembowski}}, \bibinfo {author} {\bibfnamefont {B.}~\bibnamefont {Dietz}},
  \bibinfo {author} {\bibfnamefont {H.-D.}\ \bibnamefont {Gr\"af}}, \bibinfo
  {author} {\bibfnamefont {H.~L.}\ \bibnamefont {Harney}}, \bibinfo {author}
  {\bibfnamefont {A.}~\bibnamefont {Heine}}, \bibinfo {author} {\bibfnamefont
  {W.~D.}\ \bibnamefont {Heiss}}, \ and\ \bibinfo {author} {\bibfnamefont
  {A.}~\bibnamefont {Richter}},\ }\bibfield  {title} {\enquote {\bibinfo
  {title} {{Observation of a Chiral State in a Microwave Cavity}},}\ }\href
  {\doibase 10.1103/PhysRevLett.90.034101} {\bibfield  {journal} {\bibinfo
  {journal} {Phys. Rev. Lett.}\ }\textbf {\bibinfo {volume} {90}},\ \bibinfo
  {pages} {034101} (\bibinfo {year} {2003})}\BibitemShut {NoStop}%
\bibitem [{\citenamefont {Regensburger}\ \emph {et~al.}(2012)\citenamefont
  {Regensburger}, \citenamefont {Bersch}, \citenamefont {Miri}, \citenamefont
  {Onishchukov}, \citenamefont {Christodoulides},\ and\ \citenamefont
  {Peschel}}]{Regensburger2012}%
  \BibitemOpen
  \bibfield  {author} {\bibinfo {author} {\bibfnamefont {Alois}\ \bibnamefont
  {Regensburger}}, \bibinfo {author} {\bibfnamefont {Christoph}\ \bibnamefont
  {Bersch}}, \bibinfo {author} {\bibfnamefont {Mohammad-Ali}\ \bibnamefont
  {Miri}}, \bibinfo {author} {\bibfnamefont {Georgy}\ \bibnamefont
  {Onishchukov}}, \bibinfo {author} {\bibfnamefont {Demetrios~N.}\ \bibnamefont
  {Christodoulides}}, \ and\ \bibinfo {author} {\bibfnamefont {Ulf}\
  \bibnamefont {Peschel}},\ }\bibfield  {title} {\enquote {\bibinfo {title}
  {{Parity--time synthetic photonic lattices}},}\ }\href {\doibase
  10.1038/nature11298} {\bibfield  {journal} {\bibinfo  {journal} {Nature}\
  }\textbf {\bibinfo {volume} {488}},\ \bibinfo {pages} {167--171} (\bibinfo
  {year} {2012})}\BibitemShut {NoStop}%
\bibitem [{\citenamefont {Feng}\ \emph {et~al.}(2013)\citenamefont {Feng},
  \citenamefont {Xu}, \citenamefont {Fegadolli}, \citenamefont {Lu},
  \citenamefont {Oliveira}, \citenamefont {Almeida}, \citenamefont {Chen},\
  and\ \citenamefont {Scherer}}]{FengNM2013}%
  \BibitemOpen
  \bibfield  {author} {\bibinfo {author} {\bibfnamefont {Liang}\ \bibnamefont
  {Feng}}, \bibinfo {author} {\bibfnamefont {Ye-Long}\ \bibnamefont {Xu}},
  \bibinfo {author} {\bibfnamefont {William~S.}\ \bibnamefont {Fegadolli}},
  \bibinfo {author} {\bibfnamefont {Ming-Hui}\ \bibnamefont {Lu}}, \bibinfo
  {author} {\bibfnamefont {Jos{\'e}E.~B.}\ \bibnamefont {Oliveira}}, \bibinfo
  {author} {\bibfnamefont {Vilson~R.}\ \bibnamefont {Almeida}}, \bibinfo
  {author} {\bibfnamefont {Yan-Feng}\ \bibnamefont {Chen}}, \ and\ \bibinfo
  {author} {\bibfnamefont {Axel}\ \bibnamefont {Scherer}},\ }\bibfield  {title}
  {\enquote {\bibinfo {title} {{Experimental demonstration of a unidirectional
  reflectionless parity-time metamaterial at optical frequencies}},}\ }\href
  {\doibase 10.1038/nmat3495} {\bibfield  {journal} {\bibinfo  {journal}
  {Nature Materials}\ }\textbf {\bibinfo {volume} {12}},\ \bibinfo {pages}
  {108--113} (\bibinfo {year} {2013})}\BibitemShut {NoStop}%
\bibitem [{\citenamefont {Wiersig}(2014)}]{WiersigPRL2014}%
  \BibitemOpen
  \bibfield  {author} {\bibinfo {author} {\bibfnamefont {Jan}\ \bibnamefont
  {Wiersig}},\ }\bibfield  {title} {\enquote {\bibinfo {title} {{Enhancing the
  Sensitivity of Frequency and Energy Splitting Detection by Using Exceptional
  Points: Application to Microcavity Sensors for Single-Particle Detection}},}\
  }\href {\doibase 10.1103/PhysRevLett.112.203901} {\bibfield  {journal}
  {\bibinfo  {journal} {Phys. Rev. Lett.}\ }\textbf {\bibinfo {volume} {112}},\
  \bibinfo {pages} {203901} (\bibinfo {year} {2014})}\BibitemShut {NoStop}%
\bibitem [{\citenamefont {Gao}\ \emph {et~al.}(2015)\citenamefont {Gao},
  \citenamefont {Estrecho}, \citenamefont {Bliokh}, \citenamefont {Liew},
  \citenamefont {Fraser}, \citenamefont {Brodbeck}, \citenamefont {Kamp},
  \citenamefont {Schneider}, \citenamefont {H{\"o}fling}, \citenamefont
  {Yamamoto}, \citenamefont {Nori}, \citenamefont {Kivshar}, \citenamefont
  {Truscott}, \citenamefont {Dall},\ and\ \citenamefont
  {Ostrovskaya}}]{Gao2015}%
  \BibitemOpen
  \bibfield  {author} {\bibinfo {author} {\bibfnamefont {T.}~\bibnamefont
  {Gao}}, \bibinfo {author} {\bibfnamefont {E.}~\bibnamefont {Estrecho}},
  \bibinfo {author} {\bibfnamefont {K.~Y.}\ \bibnamefont {Bliokh}}, \bibinfo
  {author} {\bibfnamefont {T.~C.~H.}\ \bibnamefont {Liew}}, \bibinfo {author}
  {\bibfnamefont {M.~D.}\ \bibnamefont {Fraser}}, \bibinfo {author}
  {\bibfnamefont {S.}~\bibnamefont {Brodbeck}}, \bibinfo {author}
  {\bibfnamefont {M.}~\bibnamefont {Kamp}}, \bibinfo {author} {\bibfnamefont
  {C.}~\bibnamefont {Schneider}}, \bibinfo {author} {\bibfnamefont
  {S.}~\bibnamefont {H{\"o}fling}}, \bibinfo {author} {\bibfnamefont
  {Y.}~\bibnamefont {Yamamoto}}, \bibinfo {author} {\bibfnamefont
  {F.}~\bibnamefont {Nori}}, \bibinfo {author} {\bibfnamefont {Y.~S.}\
  \bibnamefont {Kivshar}}, \bibinfo {author} {\bibfnamefont {A.~G.}\
  \bibnamefont {Truscott}}, \bibinfo {author} {\bibfnamefont {R.~G.}\
  \bibnamefont {Dall}}, \ and\ \bibinfo {author} {\bibfnamefont {E.~A.}\
  \bibnamefont {Ostrovskaya}},\ }\bibfield  {title} {\enquote {\bibinfo {title}
  {{Observation of non-Hermitian degeneracies in a chaotic exciton-polariton
  billiard}},}\ }\href {\doibase 10.1038/nature15522} {\bibfield  {journal}
  {\bibinfo  {journal} {Nature}\ }\textbf {\bibinfo {volume} {526}},\ \bibinfo
  {pages} {554--558} (\bibinfo {year} {2015})}\BibitemShut {NoStop}%
\bibitem [{\citenamefont {Xu}\ \emph {et~al.}(2016)\citenamefont {Xu},
  \citenamefont {Mason}, \citenamefont {Jiang},\ and\ \citenamefont
  {Harris}}]{XuNature2016}%
  \BibitemOpen
  \bibfield  {author} {\bibinfo {author} {\bibfnamefont {H.}~\bibnamefont
  {Xu}}, \bibinfo {author} {\bibfnamefont {D.}~\bibnamefont {Mason}}, \bibinfo
  {author} {\bibfnamefont {Luyao}\ \bibnamefont {Jiang}}, \ and\ \bibinfo
  {author} {\bibfnamefont {J.~G.~E.}\ \bibnamefont {Harris}},\ }\bibfield
  {title} {\enquote {\bibinfo {title} {{Topological energy transfer in an
  optomechanical system with exceptional points}},}\ }\href {\doibase
  10.1038/nature18604} {\bibfield  {journal} {\bibinfo  {journal} {Nature}\
  }\textbf {\bibinfo {volume} {537}},\ \bibinfo {pages} {80--83} (\bibinfo
  {year} {2016})}\BibitemShut {NoStop}%
\bibitem [{\citenamefont {Doppler}\ \emph {et~al.}(2016)\citenamefont
  {Doppler}, \citenamefont {Mailybaev}, \citenamefont {B{\"o}hm}, \citenamefont
  {Kuhl}, \citenamefont {Girschik}, \citenamefont {Libisch}, \citenamefont
  {Milburn}, \citenamefont {Rabl}, \citenamefont {Moiseyev},\ and\
  \citenamefont {Rotter}}]{Doppler2016}%
  \BibitemOpen
  \bibfield  {author} {\bibinfo {author} {\bibfnamefont {J{\"o}rg}\
  \bibnamefont {Doppler}}, \bibinfo {author} {\bibfnamefont {Alexei~A.}\
  \bibnamefont {Mailybaev}}, \bibinfo {author} {\bibfnamefont {Julian}\
  \bibnamefont {B{\"o}hm}}, \bibinfo {author} {\bibfnamefont {Ulrich}\
  \bibnamefont {Kuhl}}, \bibinfo {author} {\bibfnamefont {Adrian}\ \bibnamefont
  {Girschik}}, \bibinfo {author} {\bibfnamefont {Florian}\ \bibnamefont
  {Libisch}}, \bibinfo {author} {\bibfnamefont {Thomas~J.}\ \bibnamefont
  {Milburn}}, \bibinfo {author} {\bibfnamefont {Peter}\ \bibnamefont {Rabl}},
  \bibinfo {author} {\bibfnamefont {Nimrod}\ \bibnamefont {Moiseyev}}, \ and\
  \bibinfo {author} {\bibfnamefont {Stefan}\ \bibnamefont {Rotter}},\
  }\bibfield  {title} {\enquote {\bibinfo {title} {{Dynamically encircling an
  exceptional point for asymmetric mode switching}},}\ }\href {\doibase
  10.1038/nature18605} {\bibfield  {journal} {\bibinfo  {journal} {Nature}\
  }\textbf {\bibinfo {volume} {537}},\ \bibinfo {pages} {76--79} (\bibinfo
  {year} {2016})}\BibitemShut {NoStop}%
\bibitem [{\citenamefont {Liu}\ \emph {et~al.}(2016)\citenamefont {Liu},
  \citenamefont {Zhang}, \citenamefont {\"Ozdemir}, \citenamefont {Peng},
  \citenamefont {Jing}, \citenamefont {L\"u}, \citenamefont {Li}, \citenamefont
  {Yang}, \citenamefont {Nori},\ and\ \citenamefont {Liu}}]{FrancoPRL2016}%
  \BibitemOpen
  \bibfield  {author} {\bibinfo {author} {\bibfnamefont {Zhong-Peng}\
  \bibnamefont {Liu}}, \bibinfo {author} {\bibfnamefont {Jing}\ \bibnamefont
  {Zhang}}, \bibinfo {author} {\bibfnamefont {\ifmmode \mbox{\c{S}}\else
  \c{S}\fi{}ahin~Kaya}\ \bibnamefont {\"Ozdemir}}, \bibinfo {author}
  {\bibfnamefont {Bo}~\bibnamefont {Peng}}, \bibinfo {author} {\bibfnamefont
  {Hui}\ \bibnamefont {Jing}}, \bibinfo {author} {\bibfnamefont {Xin-You}\
  \bibnamefont {L\"u}}, \bibinfo {author} {\bibfnamefont {Chun-Wen}\
  \bibnamefont {Li}}, \bibinfo {author} {\bibfnamefont {Lan}\ \bibnamefont
  {Yang}}, \bibinfo {author} {\bibfnamefont {Franco}\ \bibnamefont {Nori}}, \
  and\ \bibinfo {author} {\bibfnamefont {Yu-xi}\ \bibnamefont {Liu}},\
  }\bibfield  {title} {\enquote {\bibinfo {title} {{Metrology with
  $\mathcal{PT}$-Symmetric Cavities: Enhanced Sensitivity near the
  $\mathcal{PT}$-Phase Transition}},}\ }\href {\doibase
  10.1103/PhysRevLett.117.110802} {\bibfield  {journal} {\bibinfo  {journal}
  {Phys. Rev. Lett.}\ }\textbf {\bibinfo {volume} {117}},\ \bibinfo {pages}
  {110802} (\bibinfo {year} {2016})}\BibitemShut {NoStop}%
\bibitem [{\citenamefont {Hodaei}\ \emph {et~al.}(2017)\citenamefont {Hodaei},
  \citenamefont {Hassan}, \citenamefont {Wittek}, \citenamefont
  {Garcia-Gracia}, \citenamefont {El-Ganainy}, \citenamefont
  {Christodoulides},\ and\ \citenamefont {Khajavikhan}}]{Hodaei2017}%
  \BibitemOpen
  \bibfield  {author} {\bibinfo {author} {\bibfnamefont {Hossein}\ \bibnamefont
  {Hodaei}}, \bibinfo {author} {\bibfnamefont {Absar~U.}\ \bibnamefont
  {Hassan}}, \bibinfo {author} {\bibfnamefont {Steffen}\ \bibnamefont
  {Wittek}}, \bibinfo {author} {\bibfnamefont {Hipolito}\ \bibnamefont
  {Garcia-Gracia}}, \bibinfo {author} {\bibfnamefont {Ramy}\ \bibnamefont
  {El-Ganainy}}, \bibinfo {author} {\bibfnamefont {Demetrios~N.}\ \bibnamefont
  {Christodoulides}}, \ and\ \bibinfo {author} {\bibfnamefont {Mercedeh}\
  \bibnamefont {Khajavikhan}},\ }\bibfield  {title} {\enquote {\bibinfo {title}
  {{Enhanced sensitivity at higher-order exceptional points}},}\ }\href
  {\doibase 10.1038/nature23280} {\bibfield  {journal} {\bibinfo  {journal}
  {Nature}\ }\textbf {\bibinfo {volume} {548}},\ \bibinfo {pages} {187--191}
  (\bibinfo {year} {2017})}\BibitemShut {NoStop}%
\bibitem [{\citenamefont {Chen}\ \emph {et~al.}(2017)\citenamefont {Chen},
  \citenamefont {Kaya~{\"O}zdemir}, \citenamefont {Zhao}, \citenamefont
  {Wiersig},\ and\ \citenamefont {Yang}}]{ChenNature2017}%
  \BibitemOpen
  \bibfield  {author} {\bibinfo {author} {\bibfnamefont {Weijian}\ \bibnamefont
  {Chen}}, \bibinfo {author} {\bibfnamefont {{\c S}ahin}\ \bibnamefont
  {Kaya~{\"O}zdemir}}, \bibinfo {author} {\bibfnamefont {Guangming}\
  \bibnamefont {Zhao}}, \bibinfo {author} {\bibfnamefont {Jan}\ \bibnamefont
  {Wiersig}}, \ and\ \bibinfo {author} {\bibfnamefont {Lan}\ \bibnamefont
  {Yang}},\ }\bibfield  {title} {\enquote {\bibinfo {title} {{Exceptional
  points enhance sensing in an optical microcavity}},}\ }\href {\doibase
  10.1038/nature23281} {\bibfield  {journal} {\bibinfo  {journal} {Nature}\
  }\textbf {\bibinfo {volume} {548}},\ \bibinfo {pages} {192--196} (\bibinfo
  {year} {2017})}\BibitemShut {NoStop}%
\bibitem [{\citenamefont {Yoon}\ \emph {et~al.}(2018)\citenamefont {Yoon},
  \citenamefont {Choi}, \citenamefont {Hahn}, \citenamefont {Kim},
  \citenamefont {Song}, \citenamefont {Yang}, \citenamefont {Lee},
  \citenamefont {Kim}, \citenamefont {Lee}, \citenamefont {Shin}, \citenamefont
  {Lee},\ and\ \citenamefont {Berini}}]{YoonNature2018}%
  \BibitemOpen
  \bibfield  {author} {\bibinfo {author} {\bibfnamefont {Jae~Woong}\
  \bibnamefont {Yoon}}, \bibinfo {author} {\bibfnamefont {Youngsun}\
  \bibnamefont {Choi}}, \bibinfo {author} {\bibfnamefont {Choloong}\
  \bibnamefont {Hahn}}, \bibinfo {author} {\bibfnamefont {Gunpyo}\ \bibnamefont
  {Kim}}, \bibinfo {author} {\bibfnamefont {Seok~Ho}\ \bibnamefont {Song}},
  \bibinfo {author} {\bibfnamefont {Ki-Yeon}\ \bibnamefont {Yang}}, \bibinfo
  {author} {\bibfnamefont {Jeong~Yub}\ \bibnamefont {Lee}}, \bibinfo {author}
  {\bibfnamefont {Yongsung}\ \bibnamefont {Kim}}, \bibinfo {author}
  {\bibfnamefont {Chang~Seung}\ \bibnamefont {Lee}}, \bibinfo {author}
  {\bibfnamefont {Jai~Kwang}\ \bibnamefont {Shin}}, \bibinfo {author}
  {\bibfnamefont {Hong-Seok}\ \bibnamefont {Lee}}, \ and\ \bibinfo {author}
  {\bibfnamefont {Pierre}\ \bibnamefont {Berini}},\ }\bibfield  {title}
  {\enquote {\bibinfo {title} {{Time-asymmetric loop around an exceptional
  point over the full optical communications band}},}\ }\href {\doibase
  10.1038/s41586-018-0523-2} {\bibfield  {journal} {\bibinfo  {journal}
  {Nature}\ }\textbf {\bibinfo {volume} {562}},\ \bibinfo {pages} {86--90}
  (\bibinfo {year} {2018})}\BibitemShut {NoStop}%
\bibitem [{\citenamefont {Hasan}\ and\ \citenamefont {Kane}(2010)}]{Kane2010}%
  \BibitemOpen
  \bibfield  {author} {\bibinfo {author} {\bibfnamefont {M.~Z.}\ \bibnamefont
  {Hasan}}\ and\ \bibinfo {author} {\bibfnamefont {C.~L.}\ \bibnamefont
  {Kane}},\ }\bibfield  {title} {\enquote {\bibinfo {title} {{Colloquium:
  Topological insulators}},}\ }\href {\doibase 10.1103/RevModPhys.82.3045}
  {\bibfield  {journal} {\bibinfo  {journal} {Rev. Mod. Phys.}\ }\textbf
  {\bibinfo {volume} {82}},\ \bibinfo {pages} {3045--3067} (\bibinfo {year}
  {2010})}\BibitemShut {NoStop}%
\bibitem [{\citenamefont {Armitage}\ \emph {et~al.}(2018)\citenamefont
  {Armitage}, \citenamefont {Mele},\ and\ \citenamefont
  {Vishwanath}}]{Ashvin2018}%
  \BibitemOpen
  \bibfield  {author} {\bibinfo {author} {\bibfnamefont {N.~P.}\ \bibnamefont
  {Armitage}}, \bibinfo {author} {\bibfnamefont {E.~J.}\ \bibnamefont {Mele}},
  \ and\ \bibinfo {author} {\bibfnamefont {Ashvin}\ \bibnamefont
  {Vishwanath}},\ }\bibfield  {title} {\enquote {\bibinfo {title} {{Weyl and
  Dirac semimetals in three-dimensional solids}},}\ }\href {\doibase
  10.1103/RevModPhys.90.015001} {\bibfield  {journal} {\bibinfo  {journal}
  {Rev. Mod. Phys.}\ }\textbf {\bibinfo {volume} {90}},\ \bibinfo {pages}
  {015001} (\bibinfo {year} {2018})}\BibitemShut {NoStop}%
\end{thebibliography}%


\begin{thebibliography}{7}%
\makeatletter
\providecommand \@ifxundefined [1]{%
 \@ifx{#1\undefined}
}%
\providecommand \@ifnum [1]{%
 \ifnum #1\expandafter \@firstoftwo
 \else \expandafter \@secondoftwo
 \fi
}%
\providecommand \@ifx [1]{%
 \ifx #1\expandafter \@firstoftwo
 \else \expandafter \@secondoftwo
 \fi
}%
\providecommand \natexlab [1]{#1}%
\providecommand \enquote  [1]{``#1''}%
\providecommand \bibnamefont  [1]{#1}%
\providecommand \bibfnamefont [1]{#1}%
\providecommand \citenamefont [1]{#1}%
\providecommand \href@noop [0]{\@secondoftwo}%
\providecommand \href [0]{\begingroup \@sanitize@url \@href}%
\providecommand \@href[1]{\@@startlink{#1}\@@href}%
\providecommand \@@href[1]{\endgroup#1\@@endlink}%
\providecommand \@sanitize@url [0]{\catcode `\\12\catcode `\$12\catcode
  `\&12\catcode `\#12\catcode `\^12\catcode `\_12\catcode `\%12\relax}%
\providecommand \@@startlink[1]{}%
\providecommand \@@endlink[0]{}%
\providecommand \url  [0]{\begingroup\@sanitize@url \@url }%
\providecommand \@url [1]{\endgroup\@href {#1}{\urlprefix }}%
\providecommand \urlprefix  [0]{URL }%
\providecommand \Eprint [0]{\href }%
\providecommand \doibase [0]{http://dx.doi.org/}%
\providecommand \selectlanguage [0]{\@gobble}%
\providecommand \bibinfo  [0]{\@secondoftwo}%
\providecommand \bibfield  [0]{\@secondoftwo}%
\providecommand \translation [1]{[#1]}%
\providecommand \BibitemOpen [0]{}%
\providecommand \bibitemStop [0]{}%
\providecommand \bibitemNoStop [0]{.\EOS\space}%
\providecommand \EOS [0]{\spacefactor3000\relax}%
\providecommand \BibitemShut  [1]{\csname bibitem#1\endcsname}%
\let\auto@bib@innerbib\@empty
\bibitem [{\citenamefont {Loomis}\ and\ \citenamefont
  {Sternberg}(1968)}]{SM_loomis1968advanced}%
  \BibitemOpen
  \bibfield  {author} {\bibinfo {author} {\bibfnamefont {Lynn~Harold}\
  \bibnamefont {Loomis}}\ and\ \bibinfo {author} {\bibfnamefont {Shlomo}\
  \bibnamefont {Sternberg}},\ }\href@noop {} {\emph {\bibinfo {title}
  {{Advanced calculus}}}}\ (\bibinfo  {publisher} {World Scientific},\ \bibinfo
  {year} {1968})\BibitemShut {NoStop}%
\bibitem [{\citenamefont {Zhang}\ \emph {et~al.}(2020)\citenamefont {Zhang},
  \citenamefont {Yang},\ and\ \citenamefont {Fang}}]{SM_CorrespondenceKai}%
  \BibitemOpen
  \bibfield  {author} {\bibinfo {author} {\bibfnamefont {Kai}\ \bibnamefont
  {Zhang}}, \bibinfo {author} {\bibfnamefont {Zhesen}\ \bibnamefont {Yang}}, \
  and\ \bibinfo {author} {\bibfnamefont {Chen}\ \bibnamefont {Fang}},\
  }\bibfield  {title} {\enquote {\bibinfo {title} {{Correspondence between
  Winding Numbers and Skin Modes in Non-Hermitian Systems}},}\ }\href {\doibase
  10.1103/PhysRevLett.125.126402} {\bibfield  {journal} {\bibinfo  {journal}
  {Phys. Rev. Lett.}\ }\textbf {\bibinfo {volume} {125}},\ \bibinfo {pages}
  {126402} (\bibinfo {year} {2020})}\BibitemShut {NoStop}%
\bibitem [{\citenamefont {Okuma}\ \emph {et~al.}(2020)\citenamefont {Okuma},
  \citenamefont {Kawabata}, \citenamefont {Shiozaki},\ and\ \citenamefont
  {Sato}}]{SM_Sato}%
  \BibitemOpen
  \bibfield  {author} {\bibinfo {author} {\bibfnamefont {Nobuyuki}\
  \bibnamefont {Okuma}}, \bibinfo {author} {\bibfnamefont {Kohei}\ \bibnamefont
  {Kawabata}}, \bibinfo {author} {\bibfnamefont {Ken}\ \bibnamefont
  {Shiozaki}}, \ and\ \bibinfo {author} {\bibfnamefont {Masatoshi}\
  \bibnamefont {Sato}},\ }\bibfield  {title} {\enquote {\bibinfo {title}
  {{Topological Origin of Non-Hermitian Skin Effects}},}\ }\href {\doibase
  10.1103/PhysRevLett.124.086801} {\bibfield  {journal} {\bibinfo  {journal}
  {Phys. Rev. Lett.}\ }\textbf {\bibinfo {volume} {124}},\ \bibinfo {pages}
  {086801} (\bibinfo {year} {2020})}\BibitemShut {NoStop}%
\bibitem [{\citenamefont {Yokomizo}\ and\ \citenamefont
  {Murakami}(2019)}]{SM_Murakami}%
  \BibitemOpen
  \bibfield  {author} {\bibinfo {author} {\bibfnamefont {Kazuki}\ \bibnamefont
  {Yokomizo}}\ and\ \bibinfo {author} {\bibfnamefont {Shuichi}\ \bibnamefont
  {Murakami}},\ }\bibfield  {title} {\enquote {\bibinfo {title} {{Non-Bloch
  Band Theory of Non-Hermitian Systems}},}\ }\href {\doibase
  10.1103/PhysRevLett.123.066404} {\bibfield  {journal} {\bibinfo  {journal}
  {Phys. Rev. Lett.}\ }\textbf {\bibinfo {volume} {123}},\ \bibinfo {pages}
  {066404} (\bibinfo {year} {2019})}\BibitemShut {NoStop}%
\bibitem [{\citenamefont {Denner}\ \emph {et~al.}(2020)\citenamefont {Denner},
  \citenamefont {Skurativska}, \citenamefont {Schindler}, \citenamefont
  {Fischer}, \citenamefont {Thomale}, \citenamefont {Bzdušek},\ and\
  \citenamefont {Neupert}}]{SM_ETI2020}%
  \BibitemOpen
  \bibfield  {author} {\bibinfo {author} {\bibfnamefont {M.~Michael}\
  \bibnamefont {Denner}}, \bibinfo {author} {\bibfnamefont {Anastasiia}\
  \bibnamefont {Skurativska}}, \bibinfo {author} {\bibfnamefont {Frank}\
  \bibnamefont {Schindler}}, \bibinfo {author} {\bibfnamefont {Mark~H.}\
  \bibnamefont {Fischer}}, \bibinfo {author} {\bibfnamefont {Ronny}\
  \bibnamefont {Thomale}}, \bibinfo {author} {\bibfnamefont {Tomáš}\
  \bibnamefont {Bzdušek}}, \ and\ \bibinfo {author} {\bibfnamefont {Titus}\
  \bibnamefont {Neupert}},\ }\bibfield  {title} {\enquote {\bibinfo {title}
  {{Exceptional Topological Insulators}},}\ }\href@noop {} {\  (\bibinfo {year}
  {2020})},\ \Eprint {http://arxiv.org/abs/arXiv:2008.01090} {arXiv:2008.01090}
  \BibitemShut {NoStop}%
\bibitem [{\citenamefont {Yang}\ \emph {et~al.}(2020)\citenamefont {Yang},
  \citenamefont {Chiu}, \citenamefont {Fang},\ and\ \citenamefont
  {Hu}}]{SM_yang2020jones}%
  \BibitemOpen
  \bibfield  {author} {\bibinfo {author} {\bibfnamefont {Zhesen}\ \bibnamefont
  {Yang}}, \bibinfo {author} {\bibfnamefont {Ching-Kai}\ \bibnamefont {Chiu}},
  \bibinfo {author} {\bibfnamefont {Chen}\ \bibnamefont {Fang}}, \ and\
  \bibinfo {author} {\bibfnamefont {Jiangping}\ \bibnamefont {Hu}},\ }\bibfield
   {title} {\enquote {\bibinfo {title} {{Jones Polynomial and Knot Transitions
  in Hermitian and non-Hermitian Topological Semimetals}},}\ }\href {\doibase
  10.1103/PhysRevLett.124.186402} {\bibfield  {journal} {\bibinfo  {journal}
  {Phys. Rev. Lett.}\ }\textbf {\bibinfo {volume} {124}},\ \bibinfo {pages}
  {186402} (\bibinfo {year} {2020})}\BibitemShut {NoStop}%
\bibitem [{\citenamefont {Yang}\ \emph {et~al.}(2021)\citenamefont {Yang},
  \citenamefont {Schnyder}, \citenamefont {Hu},\ and\ \citenamefont
  {Chiu}}]{SM_YangPRL}%
  \BibitemOpen
  \bibfield  {author} {\bibinfo {author} {\bibfnamefont {Zhesen}\ \bibnamefont
  {Yang}}, \bibinfo {author} {\bibfnamefont {A.~P.}\ \bibnamefont {Schnyder}},
  \bibinfo {author} {\bibfnamefont {Jiangping}\ \bibnamefont {Hu}}, \ and\
  \bibinfo {author} {\bibfnamefont {Ching-Kai}\ \bibnamefont {Chiu}},\
  }\bibfield  {title} {\enquote {\bibinfo {title} {{Fermion Doubling Theorems
  in Two-Dimensional Non-Hermitian Systems for Fermi Points and Exceptional
  Points}},}\ }\href {\doibase 10.1103/PhysRevLett.126.086401} {\bibfield
  {journal} {\bibinfo  {journal} {Phys. Rev. Lett.}\ }\textbf {\bibinfo
  {volume} {126}},\ \bibinfo {pages} {086401} (\bibinfo {year}
  {2021})}\BibitemShut {NoStop}%
\end{thebibliography}
\clearpage
\onecolumngrid
\newpage
\renewcommand{\theequation}{S\arabic{equation}}
\renewcommand{\thefigure}{S\arabic{figure}}
\renewcommand{\thetable}{S\arabic{table}}
\setcounter{equation}{0}
\setcounter{figure}{0}
\setcounter{table}{0}

\begin{center}
    {\bf \large Supplemental Material for ``Universal non-Hermitian skin effect in two and higher dimensions" }
\end{center}


\section{The proof of the theorem}\label{secI}

In this section, we will prove the following theorem appeared in the main text: 
\begin{itemize}
	\item [~]{\bf{Theorem}}: In the thermodynamic limit, the skin effect is present in a hamiltonian having open boundary of arbitrary shape, if the periodic-boundary spectral area of the same hamiltonian is nonzero; vice versa, the skin effect is absent for all possible shapes of open boundary, if the spectral area is zero.
\end{itemize} 
Here the spectral area refers to the area of the region covered by the periodic boundary spectrum on the complex plane. In the following contents, we will first show some numerical verifications of the theorem, and then prove the theorem in two dimensional systems, and finally extend the proof to three-dimensional cases. 

\subsection{Some numerical examples of the theorem}

\begin{figure}[b]
	\begin{centering}
		\includegraphics[width=.9\linewidth]{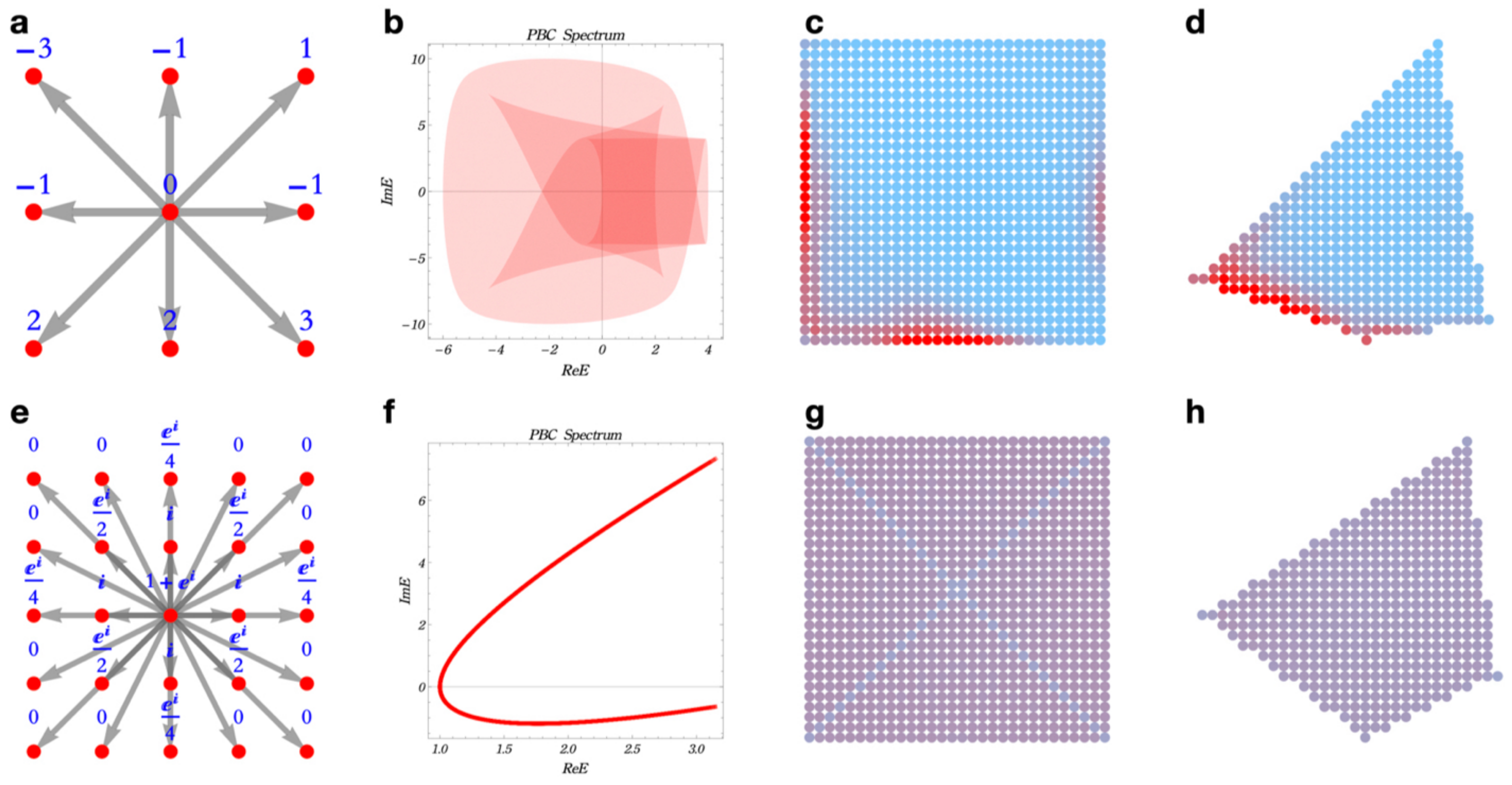}
		\par\end{centering}
	\protect\caption{\label{fig:S1} Some numerical examples of the theorem. (a-d) show an example having skin effect, and (e-h) show an example without skin effect. (a) and (e) show the corresponding hoping parameters of the hamiltonian shown in Eq.~(\ref{Ham2}). (b) and (f) show the periodic boundary spectrum. (c-d) and (g-h) show the distribution of $W(x)$ in Eq.~(\ref{Wx}) under different open boundary geometries.}
\end{figure}

In this subsection, we provide some numerical examples of the theorem. In order to simplify the discussion, we consider the following single-band model
\begin{equation}
	\mathcal{H}(\mathbf{k})=\sum_{i,j}t_{ij}\beta_x^{-i}\beta_y^{-j},\qquad \beta_{x/y}=e^{ik_{x/y}},
	\label{Ham2}
\end{equation}
where $t_{ij}$ represents the hoping parameter that an electron or particle hopes from site $(m,n)$ to site $(m,n)+(i,j)$. 

The two rows in Fig.~\ref{fig:S1} represent two different models. In the first one, the hoping parameters are shown in Fig.~\ref{fig:S1} (a), and the periodic boundary spectrum, i.e. $E(\mathbf{k})=\mathcal{H}(\mathbf{k})$ with $\mathbf{k}$ belonging to the Brillouin zone (BZ), is shown in Fig.~\ref{fig:S1} (b). Here the color strength represents the cover times of $E_0\in E(\mathbf{k})$ when $\mathbf{k}$ sweeps over the whole BZ. One can notice that the spectral area of the first model is nonzero. As a result, the open boundary eigenstates show the localization behaviors, namely, the emergence of skin effect, as shown in Fig.~\ref{fig:S1} (c-d). In order to illustrate the localization properties,
\begin{equation}
W(x)=\frac{1}{N}\sum_n|\psi_n(x)|^2
\label{Wx}
\end{equation}
is plotted in Fig.~\ref{fig:S1} (c-d), , where $\psi_n(x)$ is the $n$-th normalized eigenstate of the open boundary hamiltonian, and $N$ is the number of open boundary eigenstates. For the second example, since the spectral area is zero, as shown in Fig.~\ref{fig:S1} (f), there is no skin effect. Indeed, $W(x)$ shown in Fig.~\ref{fig:S1} (g-h) are extended. 

In the following subsection, we will prove the theorem in two-dimensional systems. Our strategy of the proof is illustrated in Fig.~\ref{fig:S2}(a). The equivalence relation between ``spectral area'' and ``the universal skin effect'' is linked by ``spectral winding" (see the following discussion). 

\subsection{The proof of the theorem in two-dimensions}\label{section_IA}

\subsubsection{Spectral area and spectral winding}

\begin{figure}[t]
	\begin{centering}
		\includegraphics[width=.9\linewidth]{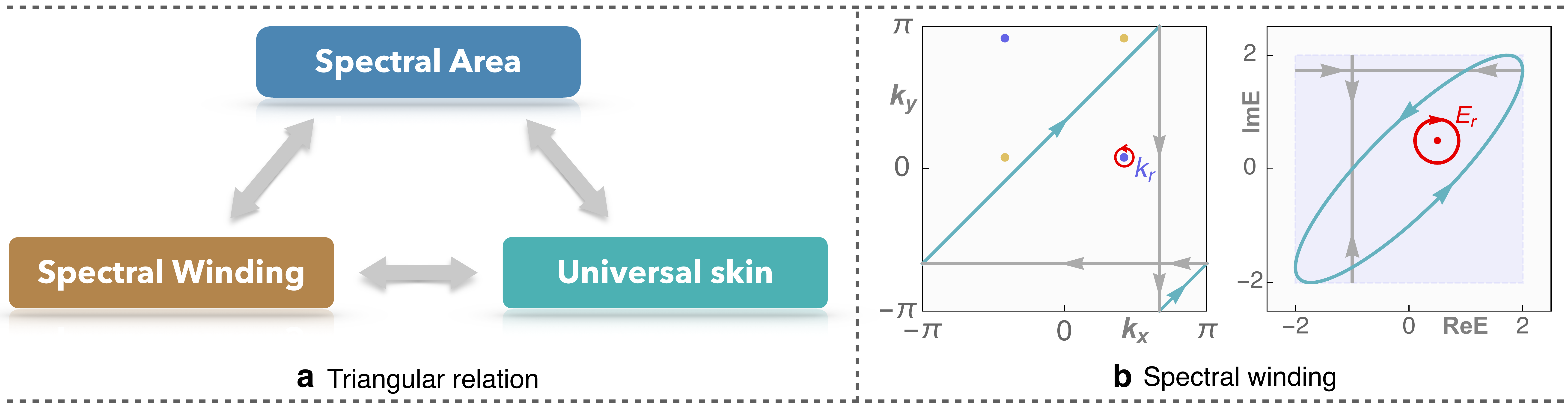}
		\par\end{centering}
	\protect\caption{\label{fig:S2} (a) shows the equivalence relation between spectral area, spectral winding and the universal skin effect. Each equivalence relation is sufficient and necessary. (b) illustrates the spectral winding for hamiltonian Eq.~\ref{SM_2D1BHam}. Here the light blue region represents the periodic boundary spectrum, and the paths on BZ corresponds to the spectral loops (or arcs) on the complex plane with the same color.}
\end{figure}

We begin with a two-dimensional single-band non-Hermitian model with periodic boundary in both $x$ and $y$ directions
\begin{equation}
\mathcal{H}(\mathbf{k})=u(\mathbf{k}) + i v(\mathbf{k}),
\label{Ham1}
\end{equation}
where $u$ and $v$ are real functions about $\mathbf{k}=(k_x,k_y)$. For any $\mathbf{k}_r\in \rm{BZ}$, one can define the following winding number 
\begin{equation}\label{SM_Spec_Winding}
	\nu(\mathbf{k}_r)=\oint_{\Gamma_{\mathbf{k}_r}} \frac{d\mathbf{k}}{2\pi i} \cdot \nabla_{\mathbf{k}} \log\det [\mathcal{H}(\mathbf{k})-E_r], \,\,\,\, \mathbf{k}_r \in \rm{BZ},
\end{equation}
where $\Gamma_{\mathbf{k}_r}$ represents the infinitesimal counterclockwise loop enclosing $\mathbf{k}_r$. Here $E_r=\mathcal{H}(\mathbf{k}_r)$ represents the reference energy, which is shown in Fig.~\ref{fig:S2} (b) with red point. We note that for different $\mathbf{k}_r$, the reference energy is different. This topological invariant describes the spectral winding on the complex plane. As shown in Fig.~\ref{fig:S2} (b), if $\nu(\mathbf{k}_r)$ is nonzero, the image of $\Gamma_{\mathbf{k}_r}$, i.e. $\mathcal{H}(\Gamma_{\mathbf{k}_r})$, forms a closed loop that encloses $E_r$. 

First we prove the following two statements, 
\begin{itemize}
	\item [1.]if there are some $\mathbf{k}_r$ points in the BZ with nonzero topological charge, the spectral area must be nonzero;
	\item [2.] if all the $\mathbf{k}_r$ points on the BZ have zero topological charge, the spectral area must be zero. 
\end{itemize}
The above two statement can be represented by the equivalence relation between ``spectral winding" and ``spectral area" shown in Fig.~\ref{fig:S2} (a). 

Based on the definition of winding number, the statement 1 is obvious. Therefore, we only need to prove the statement 2. In order to show this, we expand the hamiltonian at the point $\mathbf{k}_r\equiv(k_{x}^r,k_y^r)$ as follows
\begin{equation}\label{SM_Taylor}
    \mathcal{H}(\mathbf{k})-\mathcal{H}(\mathbf{k}_r)  \approx \partial_x \mathcal{H}(\mathbf{k}_r) q_x + \partial_y \mathcal{H}(\mathbf{k}_r) q_y, 
\end{equation}
where $q_{x},q_y$ are the displacements $\mathbf{k}-\mathbf{k}_r$ in $x, y$ directions respectively. For a generic $\mathbf{k}_r$ point, the first derivative of $\mathcal{H}$ does not vanish, i.e., the coefficients of $q_x$ and $q_y$ in Eq.~(\ref{SM_Taylor}) cannot be zero at the same time. The reason is that in order to make $\partial_x \mathcal{H}(\mathbf{k}_r)=\partial_y \mathcal{H}(\mathbf{k}_r)=0$, four independent real conditions, $\Re[\partial_x \mathcal{H}(\mathbf{k}_r)]=\Im[\partial_x \mathcal{H}(\mathbf{k}_r)]=\Re[\partial_y \mathcal{H}(\mathbf{k}_r)]=\Im[\partial_y \mathcal{H}(\mathbf{k}_r)]=0$, need to be satisfied. However, in two dimensions, there are only two free parameters $(k_x,k_y)$, which cannot satisfy the above four equations generally.

In order to calculate the topological charge of a generic $\mathbf{k}_r$ point, according to Eq.~(\ref{Ham1}), one can define 
\begin{equation}
	C(\mathbf{k}_r)=\begin{pmatrix}
		\partial_x u(\mathbf{k}_r) & \partial_y u(\mathbf{k}_r) \\ 	\partial_x v(\mathbf{k}_r) & \partial_y v(\mathbf{k}_r)
	\end{pmatrix} ,
\label{ZCC}
\end{equation}
where the notation $\partial_{x/y}$ refers to $\partial/\partial_{k_{x/y}}$. 
When $\det[C(\mathbf{k}_r)]\neq0$, the topological charge of $\mathbf{k}_r$ is the sign of the determinant of $C(\mathbf{k}_r)$, expressed as 
\begin{equation}
	\nu(\mathbf{k}_r)=\mathrm{sgn}[\det[C(\mathbf{k}_r)]].
\end{equation} 
Therefore, a sufficient and necessary condition for the zero charge for each $\mathbf{k}\in \rm{BZ}$ is 
\begin{equation}\label{ZeroChargeCond}
\det[C(\mathbf{k})]=\partial_x u(\mathbf{k})\partial_y v(\mathbf{k})-\partial_y u(\mathbf{k})\partial_x v(\mathbf{k})=0. 
\end{equation}
A theorem (the corollary of theorem 13.2) in Ref.~\cite{SM_loomis1968advanced} tells us that if $C(\mathbf{k})\neq0$ and $\det[C(\mathbf{k})]=0$, for an open set $S$, then, $u(\mathbf{k})$ and $v(\mathbf{k})$ have a functional dependent relation with $\mathbf{k}\in S\subset \rm{BZ}$. Applying the theorem to the entire BZ (except for some isolated points at which the first derivative vanishes), one can reexpress the single-band hamiltonian that satisfies Eq.~(\ref{ZeroChargeCond}) as 
\begin{equation}
	\mathcal{H}(\mathbf{k})= P[h(\mathbf{k})],
\end{equation}
where $h(\mathbf{k})$ is a real and periodic function of $\mathbf{k}$, and $P$ is a complex polynomial of $h$. Since $h(\mathbf{k})$ is a real periodic function, its image must be an arc on the real axis, e.g. $h(\mathbf{k})\in[h_1,h_2]$, where $h_{1/2}$ are real numbers. Therefore, the image of $P[h(\mathbf{k})]$ is also an arc on the complex plane. This completes the proof of the statement 2 for single-band models. 

Generalizing the above discussion to the multi-band case, for each $\mathbf{k}_r\in \rm{BZ}$, the topological charge defined for the $m$-th band is 
\begin{equation}\label{MultiBandCharge}
	\begin{aligned}
	\nu_m(\mathbf{k}_r)&=\oint_{\Gamma_{\mathbf{k}_r}} \frac{d\mathbf{k}}{2\pi i} \cdot \nabla_{\mathbf{k}} \log\det [\mathcal{H}(\mathbf{k})-E_m(\mathbf{k}_r)] \\
	&=\sum_n\oint_{\Gamma_{\mathbf{k}_r}} \frac{d\mathbf{k}}{2\pi i} \cdot \nabla_{\mathbf{k}} \log [E_n(\mathbf{k})-E_m(\mathbf{k}_r)], 
	\end{aligned}
\end{equation}
where $E_m(\mathbf{k}_r)$ is the energy of the $m$-th band with the momentum $\mathbf{k}_r$. For the second equal sign in Eq.~(\ref{MultiBandCharge}), we have used $\det[\mathcal{H}(\mathbf{k})-E_m(\mathbf{k}_r)]=\prod_n[E_n(\mathbf{k})-E_m(\mathbf{k}_r)]$. For Eq.~(\ref{MultiBandCharge}), if $\mathbf{k}_r$ is not the degeneracy point, only $n=m$ term in the summation has contributions to the topological charge. Therefore, the Eq.~(\ref{MultiBandCharge}) further becomes 
\begin{equation}
	\nu_m(\mathbf{k}_r)=\oint_{\Gamma_{\mathbf{k}_r}} \frac{d\mathbf{k}}{2\pi i} \cdot \nabla_{\mathbf{k}} \log [E_m(\mathbf{k})-E_m(\mathbf{k}_r)].
\end{equation}
Using the similar approaches in the single-band case, one can conclude that, the the real and imaginary parts of $E_m(\mathbf{k})$ are locally functional dependent on the neighbourhood of $\mathbf{k}\in \rm{BZ}$. As a result, the spectrum of $E_m(\mathbf{k})$ must be an arc. The above conclusion applies for each band. So far, we have proved that if each $\mathbf{k}$ point on the BZ has zero topological charge, the spectral area must be zero.

\subsubsection{Spectral winding and the universal skin effect}

In the above contents, we have proved the equivalence relation between ``spectral winding" and ``spectral area". Here, we will prove the equivalence condition between ``spectral winding" and ``universal skin effect", as shown in Fig.~\ref{fig:S2} (a). 

We first notice that the BZ in two-dimensional systems can be covered by a set of straight lines of any slope, labeled as $\{L_s\}$. Here, the subscript $s$ indicates the slope of the set $\{L_s\}$, and $L_s$ represents a generic straight line belonging to $\{L_s\}$. For example, if we fix $k_y(k_x)$ and change $k_x(k_y)$ from $0$ to $2\pi$, we get a horizontal (vertical) straight line with the slope being $0$ ($\infty$) on BZ, and the set of all horizontal or vertical straight lines ($\{L_{0}\}$ or $\{L_{\infty}\}$) covers the entire BZ. Particularly, an inclined straight line goes out from one side of BZ and again enters from another side as shown in Fig.~\ref{fig:S2}(b). 
Since the straight lines on the BZ are periodic, one can define the spectral winding number for each straight lines with respect to the prescribed reference energy $E_r$. 
\begin{equation}
\nu(L_s,E_r)=\oint_{L_s} \frac{d\mathbf{k}}{2\pi i} \cdot \nabla_{\mathbf{k}} \log\det [\mathcal{H}(\mathbf{k})-E_r]. 
\end{equation}
Obviously, if all the $\mathbf{k}$ points on the BZ have zero topological charge, $\nu(L_s,E_r)$ must be zero for arbitrary $L_s$ and $E_r$. Otherwise, one can always find some $L_s$, such that $\nu(L_s,E_r)$ is nonzero. Next we briefly prove the latter statement. Assuming that each straight line in $\{L_{0}\}$ and $\{L_{\infty}\}$ has zero spectral winding, and there is a $k_r$ point carrying nonzero topological charge on the $\rm{BZ}$. For a generic inclined straight line, one can always find corresponding horizontal and vertical straight lines, such that together with the inclined straight line to form a closed path enclosing $\mathbf{k}_r$ on BZ. Therefore, the closed path has nonzero winding number with respect to $E_r$. Due to the zero spectral winding of the horizontal and vertical straight lines as we assumed, hence a generic inclined straight line must have nonzero spectral winding number. 

Let's use an example to show this. Consider the following hamiltonian 
\begin{equation}
    \mathcal{H}(k_x,k_y)=2\cos{k_x}+2 i \sin{k_y}.
    \label{SM_2D1BHam}
\end{equation}
As shown in Fig.~\ref{fig:S2}, the spectrum of the hamiltonian along each horizontal or vertical straight line (gray lines) on BZ has zero winding number with respect to any reference energy on the spectral area (lightblue square region). However, once we choose the straight line with darker cyan color, these three straight lines (two gray lines with the darker cyan line) together form a closed path that encloses $\mathbf{k}_r$. Therefore, the closed path has nonzero spectral winding number with respect to $E_r$. Due to the zero spectral winding of two gray lines, the darker cyan straight line must have nonzero winding number for $E_r$ as illustrated in Fig.~\ref{fig:S2}(b). 

If the system has no skin effect under a specific parallelepiped open boundary, the winding number $\nu_m(L_m,E_r)$ along each straight lines that are perpendicular to the boundary cut directions should be zero. (This is a conjecture that we cannot exactly proved currently. However, we believe this statement is true as all the numerical results we obtained obey this conclusion. Furthermore, it is also a natural generalization of the one-dimensional results~\cite{SM_CorrespondenceKai,SM_Sato} and has been mentioned or applied in some recent works~\cite{SM_Murakami,SM_ETI2020}) More generally, if the system has no skin effect under any parallelepiped open boundary, then the spectral windings of all straight lines on BZ are required to be zero, which is satisfied when the spectral area of the system is zero. Therefore, nonzero spectral area means that there must be skin effect under certain open boundaries, namely, the existence of the universal skin effect. 

\subsection{The proof of the theorem in three-dimensions}

In this section, we extend the above proof of two-dimensional systems into three dimensions. We obtain the similar conclusion that nonzero spectral area is equivalent to the existence of the universal skin effect. 

Consider a general three-dimensional single-band tight-binding hamiltonian, which consists of real- and imaginary-part functions
\begin{equation}
	\mathcal{H}(k_x,k_y,k_z)=u(k_x,k_y,k_z)+i v(k_x,k_y,k_z).
\end{equation}
We choose a generic $\mathbf{k}_r$ point and use its energy $\mathcal{H}(\mathbf{k}_r)$ as the reference energy. For a given reference energy $E_r$, we can obtain a one-dimensional curve in the there-dimensional BZ by solving the following two real equations,
\begin{equation}
	\begin{split}
		u(k_x,k_y,k_z) = \Re(E_r); \\
		v(k_x,k_y,k_z) = \Im(E_r).
	\end{split}
\end{equation}
Each equation determines a surface, and the intersection of two surfaces is one-dimensional curve in there-dimensional BZ. The tangent direction of the curve at $\mathbf{k}_r$ is perpendicular to the normal vector of the two surfaces at this $\mathbf{k}_r$ point. The tangent vector at $\mathbf{k}_r$ is expressed as 
\begin{equation}
	\bm{T}_{\mathbf{k}_r} = \bm{\nabla}u(\mathbf{k}_r) \times \bm{\nabla}v(\mathbf{k}_r),
\end{equation}
where $\bm{\nabla}u(\mathbf{k}_r)$ represents the gradient of $u$. We choose the local coordinate system ($R^3$ space) with $\mathbf{k}_r$ as the origin, and the gradient is reexpressed as
\begin{equation}
	\bm{\nabla}u(\mathbf{k}_r)=\partial_x u(\mathbf{k}_r) q_x + \partial_y u(\mathbf{k}_r) q_y + \partial_z u(\mathbf{k}_r) q_z,
\end{equation}
where $q_i\equiv (\frac{\mathbf{k}-\bm{k_0}}{|\mathbf{k}-\bm{k_0}|})_i$, $\mathbf{k}$ and $\bm{k_0}$ represent two vectors in the global coordinate system. 

Next we expand the hamiltonian into Taylor series around the origin of the local coordinate system,
\begin{equation}
	\mathcal{H}(\mathbf{k})-\mathcal{H}(\mathbf{k}_r) = \sum_{i=x,,y,z}\partial_i \mathcal{H}(\mathbf{k}_r) q_i + o(|\bm{q}|),
\end{equation}
where the subscription $i$ represents the partial differential to $x,y,z$. And $q_i$ represents the deviation of $k_i$ from $k_{r,i}$ and the last term is the infinitesimal of higher order of $|\bm{q}|$. Obvious, the zero winding condition requires 
\begin{equation}
	\begin{split}
		\partial_x u(\mathbf{k}_r)\partial_y v(\mathbf{k}_r)-\partial_x v(\mathbf{k}_r)\partial_y u(\mathbf{k}_r) = 0 ;\\ 
		\partial_x v(\mathbf{k}_r)\partial_z u(\mathbf{k}_r)-\partial_x u(\mathbf{k}_r)\partial_z v(\mathbf{k}_r) = 0 ;\\
		\partial_y u(\mathbf{k}_r)\partial_z v(\mathbf{k}_r)-\partial_y v(\mathbf{k}_r)\partial_z u(\mathbf{k}_r) = 0,
	\end{split}
\end{equation}
or equivalently, 
\begin{equation}
\bm{T}_{k_r} = \bm{0}.
\label{NoCharge}
\end{equation}

Next, we prove that if all $\mathbf{k}$ points in three-dimensional BZ satisfy $\bm{T}_{k} = \bm{0}$, then the entire 3D periodic-boundary spectrum must be an arc in the complex plane. We define a two-tuple function $W(\mathbf{k}):=[u(\mathbf{k})\;\;\; v(\mathbf{k})]^t$ with three variables, the exterior derivative of the vector-valued function is expressed as 
\begin{equation}
	dW = 
	\begin{pmatrix}
		\partial_x u(\mathbf{k}) &\partial_y u(\mathbf{k}) &\partial_z u(\mathbf{k})\\ 
		\partial_x v(\mathbf{k}) &\partial_y v(\mathbf{k}) &\partial_z v(\mathbf{k})
	\end{pmatrix}.
\end{equation}
Eq.~(\ref{NoCharge}) implies that the rank of $dW$ less than 2 (the number of components of $W$). To be precise, there are the following cases. (i.) Both the gradients of $u$ and $v$ are not zero vector, and they are linearly dependent on each other. (ii.) One of the gradients of $u$ and $v$ is zero vector. (iii.) Both the gradients of $u$ and $v$ are zero vector. In all these cases, we can obtain the final conclusion that $u$ and $v$ are linearly functional dependent on each other. Therefore, the spectrum must be arcs on the complex plane. 

A 3D BZ can be divided into a series of plane systems, each plane corresponds to a two-dimensional subsystem. If the spectral area of a three-dimensional system is nonzero, then for each reference energy on the spectral area, its preimage (1D ring) has nonzero topological charge. Equivalently, the two-dimensional subsystem, of which the BZ (2D plane) has intersections with the ring, also has nonzero topological charge for the intersecting $k$ points. Hence, the 2D subsystem has nonzero spectral area, and has the universal skin effect. Correspondingly, we come to the same conclusion in 3D systems that nonzero spectral area signifies the existence of the universal skin effect. 

\section{A physical explanation for the theorem}\label{secII}

\begin{figure}[t]
	\begin{centering}
	\includegraphics[width=.95\linewidth]{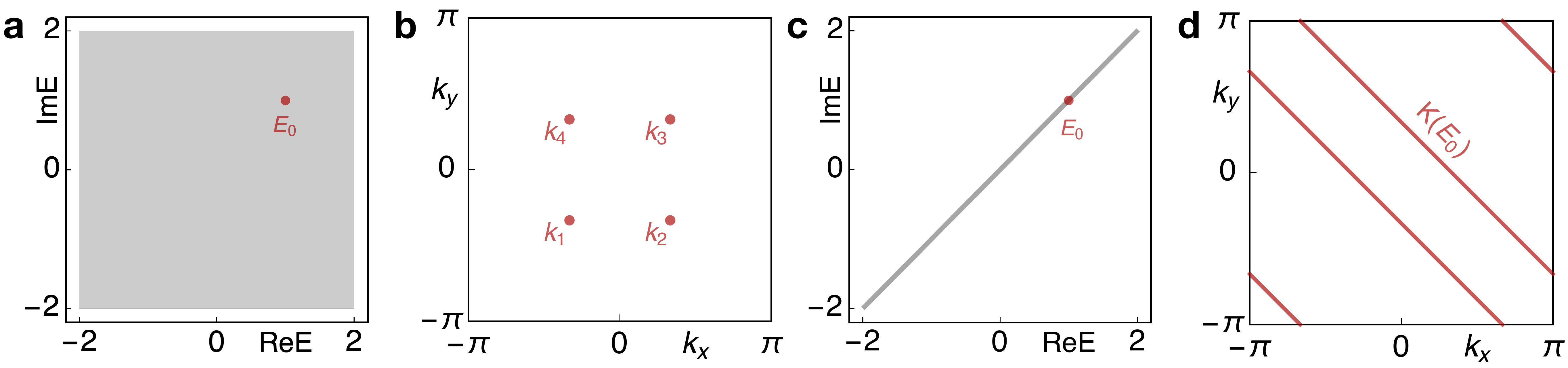}
	\par\end{centering}
	\protect\caption{\label{fig:S3} 
	(a) shows the periodic-boundary spectrum of Eq.~(\ref{intuimodel_1}) with gray color, and the  pre-images of $E_0=1+i$ (red point in (a)) are the four red points in (b). The periodic-boundary spectrum of Eq.~(\ref{intuimodel_2}) is the gray line in (c), and $\mathbf{k}(E_0=1+i)$ is plotted by the red lines in (d). 
	} 
\end{figure}

Here, we use some examples to illustrate the intuition that motivates the theorem.
Consider the following one-dimensional model 
\begin{equation}
	\mathcal{H}_0(k)=2\cos k
\end{equation}
placed on a chain of length $L$.
Under the periodic boundary condition, the two Bloch waves $e^{ik_0x}$ and $e^{-ik_0x}$ have the same energy $E(k_0)=2\cos{k_0}$.
When the system has open boundary condition, the Bloch wave $e^{ik_0x}$ will be reflected to $e^{-ik_0x}$ with a $\pi$-phase shift.
Their linear superposition $e^{ik_0x}-e^{-ik_0x}$ is an eigenstate with energy $2\cos{k_0}$ that satisfies the zero boundary condition at $x=0,L$, thus being an open-boundary eigenstate.
When the system is added a momentum-dependent dissipation, 
\begin{equation}
	\mathcal{H}(k)=2\cos k+i\sin k,
\end{equation}
the spectrum $E(k)$ becomes complex and forms an ellipse in the complex plane. 
In this case, the degeneracy is broken, e.g. $E(k)\neq{}E(-k)$, which implies the open-boundary eigenstates are no longer the linear superposition of the extended Bloch waves.
This implies the emergence of skin effect. 

Extend the above arguments to two dimensions, and we can provide a physical explanation for the theorem proved in the section~\ref{secI}. 

Formally, we consider a single-band model 
\begin{equation}
	\mathcal{H}(\mathbf{k})=\mathcal{H}_0(\mathbf{k})+i\Gamma(\mathbf{k}).
\end{equation}
When the real and imaginary parts of which are functionally independent, the hamiltonian will have a non-zero spectral area. 
For a given eigenvalue $E_0$ of the Bloch hamiltonian, by solving $\mathcal{H}_0(\mathbf{k})=\Re E_0$ and $\Gamma(\mathbf{k})=\Im E_0$, one can obtain a finite set of pre-images of $E_0$, i.e, $\mathbf{K}({E_0})=\{\mathbf{k}_1,...,\mathbf{k}_m\}$, which includes all Bloch waves having energy $E_0$.
Now suppose that one of the Bloch waves $\mathbf{k}_i\in\mathbf{K}({E_0})$ is incident on the boundary, depending on the normal direction of the boundary, the corresponding momentum of the reflected wave can be arbitrary. 
However, the number of elements of $\mathbf{K}({E_0})$ is finite, and as such cannot support so many reflection channels.
This failure of reflection mechanism at a generic boundary means the failure in forming an open boundary eigenstate from Bloch waves, which implies the emergence of skin effect under a generic open-boundary geometry. 
However, the spectrum collapses into an arc (zero spectral area) if the real and imaginary parts of the hamiltonian are functionally dependent, and the number of the corresponding solutions of $\mathcal{H}(\mathbf{k})=E_0$ is infinite. 
It means that there are infinite reflection channels to satisfy the open boundary of any shape, and an open boundary eigenstate can be formed from superimposing all Bloch-wave channels. 

Concretely, we choose two examples to demonstrate the above arguments. The first example is
\begin{equation}\label{intuimodel_1}
	\mathcal{H}(\mathbf{k}) = 2\cos{k_x} + 2i \cos {k_y},
\end{equation}
of which the spectral area is nonzero shown in Fig.~\ref{fig:S3}(a). For a given eigenvalue $E_0=1+i$, by solving $2\cos{k_x}=1$ and $2\cos{k_y}=1$, we can obtain a finite set of pre-images of $E_0$, that is, $\mathbf{K}(E_0)=\{k_1,k_2,k_3,k_4\}$ [red points in Fig.~\ref{fig:S3}(b)]. The finite solutions of $\mathcal{H}(\mathbf{k})=E_0$ cannot support so many reflection channels, that is to say, cannot form an open-boundary eigenstate on a generic geometry by superimposing these Bloch waves specified by $k_{i=1,2,3,4}$. Therefore, the hamiltonian Eq.~(\ref{intuimodel_1}) has skin effect under open-boundary geometry of a generic shape. The second example reads 
\begin{equation}\label{intuimodel_2}
	\mathcal{H}(\mathbf{k}) = 2\cos{(k_x+k_y)} + 2i \cos {(k_x+k_y)},
\end{equation}
the periodic-boundary spectrum of which is an arc [the gray line in Fig.~\ref{fig:S3}(c)]. The set of pre-images of $E_0=1+i$ has infinite elements [the red lines in Fig.~\ref{fig:S3}(d)], which means that there are infinite ways of superimposing these Bloch waves to satisfy the open boundary condition of any shape. Therefore, the hamiltonian Eq.~(\ref{intuimodel_2}) has no skin effect under any open-boundary geometry. 

\section{Corner-skin effect and Geometry-dependent-skin effect}\label{secIII}

In this section, we will provide some examples to demonstrate the characteristics of the two manifestations of the universal skin effect. 
We discuss the role of symmetry on the universal skin effect. 
We define the current functional to explain the appearance of corner-skin effect. 
In addition, we numerically verified that geometry-dependent-skin effect obeys the volume law, which is a significant feature to distinguish skin modes form conventional boundary states.

\subsection{Symmetry and the universal skin effect}

\begin{figure}[t]
    \begin{centering}
    \includegraphics[width=.9\linewidth]{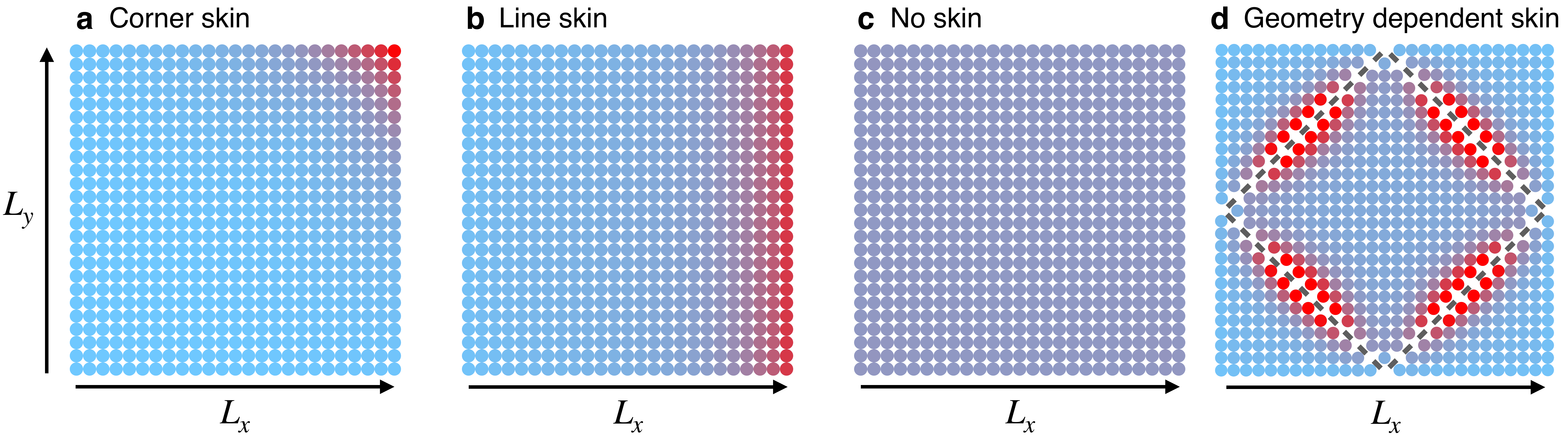}
    \par\end{centering}
    \protect\caption{\label{fig:S4} The distribution of $W(x)$ for Hamiltonian Eq.~(\ref{SM_SymModel}) with different parameters and on different geometries. The system size is $L_x=L_y=25$. The probability density is proportional to the opacity of the red color. (a) shows the corner-skin effect with $t_1=t_2=1,w=0$; (b) shows line skin with $t_1=1,t_2=w=0$; (c) has no skin effect with $t_1=t_2=0,w=1$, and geometry-dependent-skin effect appears in (d) under triangle and diamond geometries.}
\end{figure}

We have proved that if the system has nonzero spectral area, the universal skin effect will occur under a general open boundary condition. According to the symmetry restriction, the universal skin effect has two manifestations, that is, corner-skin effect and geometry-dependent-skin effect. Consider a system with nonzero spectral area, if its Hamiltonian has no any symmetry, the skin modes are localized at one or several vertices on an open geometry of any shape, and the number of modes are proportional to the volume of the system. If the Hamiltonian has certain spatial symmetries, such as mirror symmetry, the corner-skin effect will be forbidden. The reason is that the corner-skin effect has the nature of non-reciprocity, which is incompatible with mirror symmetry. However, if we change the open boundary geometry such that the symmetry is broken on the boundary, then geometry-dependent-skin modes will appear, which is a unique but universal phenomenon in higher-dimensional systems. We take a concrete example to demonstrate the role of symmetry on the skin effect. Consider a tight-binding model, of which the periodic-boundary Hamiltonian reads 
\begin{equation}\label{SM_SymModel}
    \begin{split}
    H(k_x,k_y) &= h_0(k_x,k_y) + i h_1(k_x,k_y)   \\
    &=\sin{k_x} \sigma_x + \sin{k_y} \sigma_y + (2 - \cos{k_x} - \cos{k_y}) \sigma_z  + i [t_1 \sin{k_x}+ t_2 \sin{k_y}  + w (\cos{k_x} - \cos{k_y})] \sigma_z,
    \end{split}
\end{equation}
where $h_0$ and $h_1$ are the Hermitian and non-Hermitian parts, respectively. The Hermitian part is in a gapless phase with a Dirac point at $k_x=k_y=0$, and has inversion symmetry $\sigma_z h_0(k_x,k_y) \sigma_z= h_0(-k_x,-k_y)$. 

If we only add the $w$ non-Hermitian term ($t_1=t_2=0;w=1$), the combined mirror and non-Hermitian time-reversal symmetry $M_x T = A_t$ and $M_y T = \sigma_z A_t$ are preserved ($A_t$ representing transpose operator),
\begin{equation}
    \begin{split}
    (M_x T) H(k_x,k_y) (M_x T)^{-1} = H(k_x,-k_y); \\
    (M_y T) H(k_x,k_y) (M_y T)^{-1} = H(-k_x,k_y). 
    \end{split} 
\end{equation}
There is no skin effect under open boundary with square geometry as shown in Fig.~\ref{fig:S4}(c). However, if we cut the square lattice into triangles and diamond lattices, the skin effect will retrieve on the boundaries that destroy the two symmetries, which is shown in Fig.~\ref{fig:S4}(d). 

If we only turn on $t_1$ non-Hermitian term ($t_1=1;t_2=w=0$), the Hamiltonian preserves $M_xT$ symmetry but destroys $M_yT$ symmetry. In this case, the system has skin modes along $x$ direction as shown in Fig.~\ref{fig:S4}(b), the number of the skin modes are proportional to the volume of the system. If we add $t_1$ and $t_2$ non-Hermitian terms ($t_1=t_2=1; w=0$), the Hamiltonian destroys the two symmetries, and skin modes will be concentrated on the corner as shown in Fig.~\ref{fig:S4}(a). 

\subsection{Current functional}
In this section, we first simply introduce the concept of current functional, then discuss the restrictions of point groups in two and three dimensions on the current functional, a quantity faithfully depicting the appearance of corner-skin effect, which demonstrates that corner-skin effect is only compatible with point groups $C_m$ and $C_{2,3,4,6,2v,3v,4v,6v}$. 

\subsubsection{An introduction to the current functional}

We define the current functional to depict the corner-skin effect in $d$ dimensions. 
\begin{equation}\label{SM_Current}
	J_\alpha[n]=\sum_i J_{i,\alpha}[n]=\sum_i\oint_{\rm{BZ}}{dk^d}n(E_i,E^\ast_i)\partial_{k_\alpha}{E}_i(\mathbf{k}),
\end{equation}
where $i$ is band index, and $k_{\alpha}$ is a vector, expressed as $k_{\alpha} = \sum_{i} k_i \mathbf{e}_i$ in $d$-dimensional momentum space with unit vector basis $\mathbf{e}_i$. Here $n(E,E^{\ast})$ is a distribution function depending on $E$ and $E^{\ast}$, but does not depend on $k$ explicitly, such as the Bose distribution $n(E,E^{\ast})=(e^{\Re E(k)/k_BT}-1)^{-1}$. 
If there exists a $n(E,E^{\ast})$ such that the current functional is nonzero for any $\alpha$, then the system must have the corner-skin effect. If for any possible $n(E,E^{\ast})$ and $\alpha$, the current functional is zero, then the system has no corner-skin effect. 

For example, we take a single-band tight-binding model [Eq.~(1) in the main text] as 
\begin{equation}
	\mathcal{H}(\mathbf{k})=[5(\cos{k_x}+\cos{2k_x})-i(\sin{k_x}+3\sin{2k_x}) 
	+5\cos{k_y}+i\sin{k_y}]/2,
\end{equation}
and $n(E,E^{\ast})$ as $\Im{[\mathcal{H}(\mathbf{k})]}$. In this case, $J_x$ is equal to $25\pi^2/2$ and $J_y$ is equal to $-5\pi^2/2$. Hence, the system has corner-skin effect. Another example of tight-binding model [Eq.~(3) in the main text] reads
\begin{equation}
	\mathcal{H}(\mathbf{k})=2\cos{k_x}+2i\cos{k_y}, 
\end{equation}
of which the current functional $J_x=J_y=0$ regardless of the choose of $n(E,E^{\ast})$. Therefore, the hamiltonian has no corner-skin effect (although hosts the geometry-dependent-skin effect due to the existence of two mirror symmetries). Meanwhile, the example also reminds us that certain symmetries may prohibit the corner-skin effect. Next, we will systematically analyze the interplay of point-group symmetries and the corner-skin effect. 

\subsubsection{Corner-skin effect under point groups}
Here we consider the hamiltonian with only spatial symmetries (without any anti-unitary symmetry such as non-Hermitian time-reversal symmetry), and investigate the restrictions of the point-group symmetries on the current functional, and further conclude that corner-skin effect is only compatible with point groups $C_m$ and $C_{2,3,4,6,2v,3v,4v,6v}$. 


\textbf{Inversion:} Consider a system that only has inversion symmetry $I$, and each band satisfies $E_i(\mathbf{k}) = E_i(I\,\mathbf{k})=E_i(-\mathbf{k})$. The current functional for $i$-th band can be expressed as 
\begin{equation}\label{SM_CurFunComp}
	J_{i,\alpha}[n] = \oint_{\rm{BZ}}{dk^d}n(E_i,E^\ast_i)\partial_{k_\alpha}{E}_i(\mathbf{k}),
\end{equation}
which is invariant by replacing $\mathbf{k}$ with $\mathbf{k'}\equiv-\mathbf{k}$. After the transformation, Eq.~(\ref{SM_CurFunComp}) becomes 
\begin{equation}
	\begin{split}
	J_{i,\alpha}[n] & = \oint_{\rm{BZ}'} (-1)^d {d k'^{d}}n(E_i,E^\ast_i)\partial_{-k'_\alpha}{E}_i(-\mathbf{k}') = (-1)^d \oint_{\rm{BZ}} (-1)^d {d k'^{d}} n(E_i,E^\ast_i) \partial_{-k'_\alpha}{E}_i(-\mathbf{k}') \\
	& = -\oint_{\rm{BZ}} {d k'^{d}}n(E_i,E^\ast_i) \partial_{k'_\alpha}{E}_i(\mathbf{k}') = -\oint_{\rm{BZ}} {d k^{d}}n(E_i,E^\ast_i) \partial_{k_\alpha}{E}_i(\mathbf{k}) =  - J_{i,\alpha}[n] = 0.
	\end{split}
\end{equation} 
It means that if the hamiltonian has only inversion symmetry, the current functional for each band must be zero regardless of the choose of $n(E, E^*)$. Equivalently, the corner-skin effect must vanish in the system with inversion symmetry, and is incompatible with the point groups including inversion symmetry, such as $C_{i,3i,2h,4h,6h}$, $D_{3d,2h,4h,6h}$, $T_h$ and $O_h$. 

\textbf{Rotation:} Consider a system that is invariant under a point group including rotation operator $R$, then $E_i(\mathbf{k})=E_i(R\,\mathbf{k})$. The Eq.~(\ref{SM_CurFunComp}) is also invariant under the transformation from $\mathbf{k}$ to $\mathbf{k}'\equiv R^{-1}\,\mathbf{k}$ = $\sum_{ij} \mathbf{e}_i \rm{c}_{ij}k_j$, where $k_j$ is the $j$-th component (along $\mathbf{e}_j$) of $\mathbf{k}$. After the transformation, the current functional becomes 
\begin{equation}\label{SM_Rotation}
	\begin{split}
	J_{i,\alpha}[n] &= \oint_{\rm{BZ}'} \det{[J_{\mathbf{k},\mathbf{k}'}]} {dk'^d}n(E_i,E^\ast_i)\partial_{Rk'_\alpha}{E}_i(R\,\mathbf{k}') = \oint_{\rm{BZ}}  dk^{'d} n(E,E^{\ast}) \partial_{R\, k^{'}_{\alpha}}E_i(\mathbf{k'}) \\
	&= \oint_{\rm{BZ}} dk^{d} n(E,E^{\ast}) \partial_{R\, k_{\alpha}}E_i(\mathbf{k}) = \oint_{\rm{BZ}} dk^{d} n(E,E^{\ast}) \partial_{k_{\alpha}}E_i(\mathbf{k}), 
	\end{split}
\end{equation}
where $\det{[J_{\mathbf{k},\mathbf{k}'}]}$ in the second term is the determinant of the Jacobian $J_{\mathbf{k},\mathbf{k}'}$ that measures the change of differential volume element under different representations and the sign of $\det{[J_{\mathbf{k},\mathbf{k}'}]}$ is positive because the rotational operator preserves orientation. One can always choose an appropriate basis transformation such that $\det{[J_{\mathbf{k},\mathbf{k}'}]}=1$. In addition, since the Brillouin zone has the same symmetry group as the hamiltonian and rotational operator $R$ does not change the orientation, the integral region $\rm{BZ}$ is invariant under the point group. 
For example, consider a rotation $R$ that rotates $\pi/4$ along $\mathbf{e}_z$ axis, then $k_x=-k'_y$ and $k_y=k'_x$. The Jacobian matrix can be written as 
\begin{equation}
	J_{\mathbf{k},\mathbf{k}'} = \begin{pmatrix}
	\partial_{x'} k_x & \partial_{y'} k_x \\ 
	\partial_{x'} k_y & \partial_{y'} k_y. 
	\end{pmatrix}
\end{equation}
In this case, the determinant of Jacobian $\det{[J_{\mathbf{k},\mathbf{k}'}]}$ is 1. 

The last equation of Eq.~(\ref{SM_Rotation}) requires
\begin{equation}\label{SM_RotRestrict}
	R k_{\alpha} = k_{\alpha},
\end{equation}
which means that the direction of $k_{\alpha}$ is parallel to the rotational axis of $R$, equivalently, the component of $k_{\alpha}$ perpendicular to the rotational axis must be zero. For example, if $R$ is a rotation that rotates $\theta$ along $\mathbf{e}_z$ axis, that is, 
\begin{equation}
	R = \begin{pmatrix}
	\cos{\theta} & -\sin{\theta} & 0 \\
	\sin{\theta} & \cos{\theta} & 0 \\
	0 & 0 & 1
	\end{pmatrix}, 
\end{equation}
then Eq.~(\ref{SM_RotRestrict}) restricts $k_{\alpha}$ as $0 \, \mathbf{e}_x + 0\,  \mathbf{e}_y + k_z \, \mathbf{e}_z$, along $\mathbf{e}_z$ axis. 

As a consequence, if a point group contains two or more rotations with non-parallel rotational axes, the current functional for each band must be zero. If the point group contains only one rotation, a nonzero current functional for each band is allowed, thus corner-skin effect is compatible with the point groups including only one rotation, such as $C_{2,3,4,6}$. 

\textbf{Mirror:} In a similar way, a mirror symmetry requires
\begin{equation}
	M k_{\alpha} = k_{\alpha}
\end{equation}
if nonzero current functional for each band is allowed, which means the corner-skin effect is compatible with the group containing only one mirror symmetry, that is, $C_m$. 

It is notable that an additional rotation symmetry is allowed when the rotational axis lies in the mirror plane, and $k_{\alpha}$ is restricted as 
\begin{equation}
	R M k_{\alpha} = k_{\alpha}.
\end{equation}
Therefore, the corner-skin effect is also compatible with such point groups, i.e., $C_{2v,3v,4v,6v}$. 

In summary, we have explicitly shown that corner-skin effect does not exist under most point groups because the current functional is restricted to zero under such point groups, regardless of the choice of $n(E, E^{\ast})$, but is allowed to appear under these point groups, i.e., 
\begin{equation}
	\{ C_m, C_2,C_3,C_4,C_6,C_{2v},C_{3v},C_{4v},C_{6v} \}. 
\end{equation}

\subsection{Volume Law}

\begin{figure}[t]
    \begin{centering}
    \includegraphics[width=.9\linewidth]{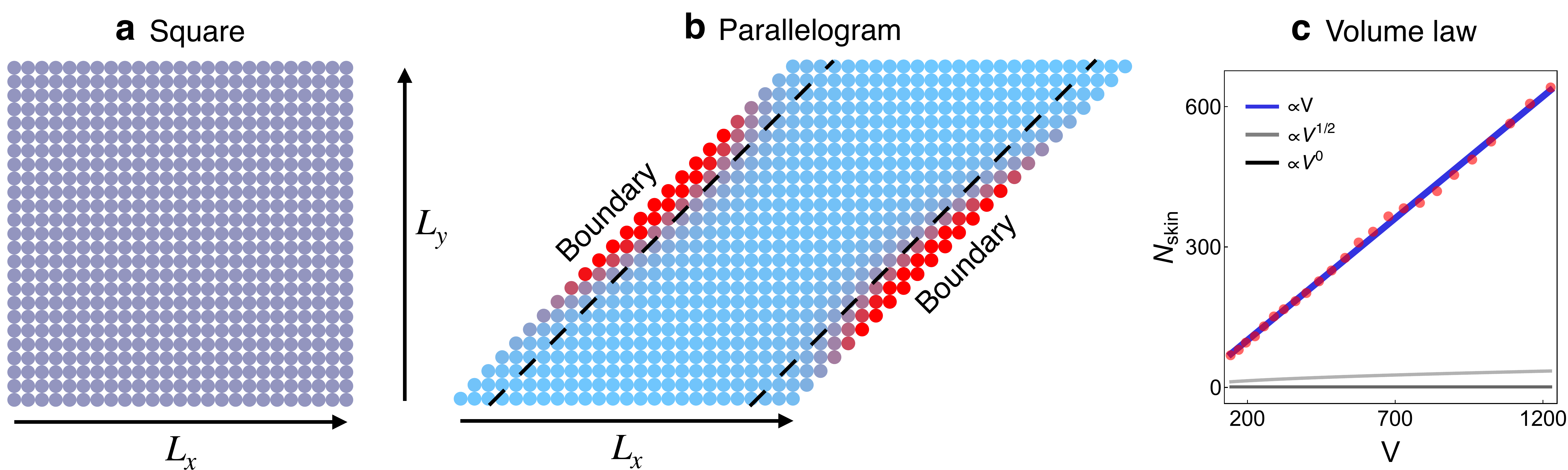}
    \par\end{centering}
    \protect\caption{\label{fig:S5} The norm squared of all wave functions of the Hamiltonian Eq.~(\ref{SM_2D1BHam}) on square lattice (a) and parallelogram lattice (b) is plotted. The system size is chosen as $L_x=L_y=25$. The volume law is shown in (c), in which blue line represents $N_{skin}\propto V$, gray line $N_{skin}\propto\sqrt{V}$ and black line $N_{skin}\propto 1 $. }
\end{figure}

We numerically show that the geometry-dependent-skin effect satisfies the volume law, that is, the increase in the number of skin modes is proportional to the increase in volume of the system,
\begin{equation}
    \delta N_{skin} \propto \delta V.
\end{equation}
For example, If the shape of the open boundary is a parallelogram whose side-lengths are $L_x$ and $L_y$ as shown in Fig.~\ref{fig:S5}, then, $V=L_x L_y$. Our criterion for judging a mode as a skin mode is to check whether ninety percent of the probability density of this mode lies within the boundary we appointed. 

Consider a tight-binding model with periodic-boundary Hamiltonian $\mathcal{H}(k_x,k_y)=2\cos{k_x}+2 i \sin{k_y}$, there is no skin effect under square geometry Fig.~\ref{fig:S5}(a), but the skin effect appears under parallelogram geometry Fig.~\ref{fig:S5}(b) due to the spectral area being nonzero, which is geometry-dependent-skin effect. The distributions of $W(\bm{x})$ under different open boundaries are plotted. For parallelogram geometry, we specify the thickness of the boundary to be the width of three unit cells, and use black dashed lines to distinguish the boundary from the bulk. If ninety percent of the probability density of a mode lies in the boundary, we count it as a skin mode. We count the number of skin modes for different volumes ($\mathcal{L}_x\mathcal{L}_y$), and the fitting curve (blue curve in Fig.~\ref{fig:S5}(c)) shows that the two are in a linear relation, that is, $\delta N_{skin} = 0.52 \delta V$. The volume law of geometry-dependent-skin effect has been verified numerically.

\section{Exceptional semimetals}\label{secIV}

In this section, we will prove the corollary of our theorem, that is all stable exceptional semimetals imply the universal skin effect. We first review the topological charge of non-Hermitian band degeneracies. 

\subsection{Non-Hermitian band degeneracy}

Consider a general $m$-band non-Hermitian Bloch Hamiltonian (with periodic boundary condition),
\begin{equation}
	\mathcal{H}(\mathbf{k})=\sum_{s=1}^{m^2-1}\left[h_s^r(\mathbf{k})+ih_s^i(\mathbf{k})\right ]\Gamma_s,
\end{equation}
where $\Gamma_s$ are the generators of Lei algebra $\mathfrak{s u}(m)$ and $h_s^r(\mathbf{k})$ and $h_s^i(\mathbf{k})$ are real functions of $\mathbf{k}$. When $m=2,3,4$, $\Gamma_s$ are the Pauli, GellMann, and $\gamma$ matrices, respectively. The eigenvalues of $\mathcal{H}(\mathbf{k})$ can be obtained by solving the following characteristic polynomial
\begin{equation}
	f_E (\mathbf{k})=\det[E-\mathcal{H}(\mathbf{k})]=\prod_{i=1}^m [E-E_i(\mathbf{k})], 
\end{equation} 
where $E_i(\mathbf{k})$ is the $i$th eigenvalue of the non-Hermitian Hamiltonian $\mathcal{H} (\mathbf{k} )$. At the degeneracy point $\mathbf{k}_D$, two bands must have the same energy, i.e. 
\begin{equation}
E_i(\mathbf{k}_D)=E_j(\mathbf{k}_D)
\end{equation}
for some $i\neq j$. In Ref.~\cite{SM_yang2020jones,SM_YangPRL}, the authors have shown that the above condition is equivalent to the vanishing of the discriminant of $f_E (\mathbf{k})$, i.e. 
\begin{equation}
\operatorname{Disc}_{E}[\mathcal{H}](\mathbf{k}_D)=0,
\end{equation}
where 
\begin{equation}
	\operatorname{Disc}_{E}[\mathcal{H}](\boldsymbol{k})=\prod_{i<j}\left[E_{i}(\boldsymbol{k})-E_{j}(\boldsymbol{k})\right]^{2}
\end{equation}
is the discriminant of $f_E (\mathbf{k})$. Although the discriminant is defined by the roots of $f_E(\mathbf{k})=0$, it can be computed directly from the determinant of the Sylvester matrix of $f_E(\mathbf{k})$ and $\partial_Ef_E(\mathbf{k})$, which can be expressed by the coefficients of $f_E (\mathbf{k})$. Now we show a concrete example of the discriminant method. 
\begin{itemize}
	\item[~] {\bf Example}: Consider a generic two-band model 
	\begin{equation}
		\mathcal{H}(\mathbf{k})=h_0(\mathbf{k}) \sigma_0 +h_x(\mathbf{k})\sigma_x+h_y(\mathbf{k})\sigma_y+h_z(\mathbf{k})\sigma_z, 
		\label{twoband}
	\end{equation}
	where $h_\mu(\mathbf{k})=h_\mu^r(\mathbf{k})+i h_\mu^i(\mathbf{k})$ are complex functions of $\mathbf{k}$. The characteristic polynomial of the two-band model can be written as 
	\begin{equation}
		f_E(\mathbf{k})=E^2+b(\mathbf{k}) E+c(\mathbf{k}),
		\label{Ex1a}
	\end{equation}
	where $b(\mathbf{k})=-2h_0(\mathbf{k})$ and $c(\mathbf{k})=h_0^2(\mathbf{k})-h_x^2(\mathbf{k})-h_y^2(\mathbf{k})-h_z^2(\mathbf{k})$. Computing the discriminant of polynomial~\eqref{Ex1a} 
	with respect to the energy $E$,  we obtain the following condition for the existence of DPs
	\begin{equation}
		\operatorname{Disc}_{E}[\mathcal{H}](\mathbf{k})=b^2(\mathbf{k})-4c(\mathbf{k})=4[h_x^2(\mathbf{k})+h_y^2(\mathbf{k})+h_z^2(\mathbf{k})]=0.
	\end{equation} 
	This condition can also be obtained from the energy spectrum, that is the two bands $E_\pm = h_0(\mathbf{k}) \pm ([h_x^2(\mathbf{k})+h_y^2(\mathbf{k})+h_z^2(\mathbf{k}))^{1/2}$
	are degenerate, whenever the square root is vanishing. 
\end{itemize}

From the above example, one can notice that the discriminant $\operatorname{Disc}_{E}[\mathcal{H}](\mathbf{k})$ is a complex periodic function of $\mathbf{k}$. Its vanishing is equivalent to the vanishing of the real and imaginary parts, i.e. 
\begin{equation}
	\Re \operatorname{Disc}_{E}[\mathcal{H}](\mathbf{k})=\Im \operatorname{Disc}_{E}[\mathcal{H}](\mathbf{k})=0.
\end{equation}
The solution of the above equation are the non-Hermitian degeneracy points in 2D and lines in 3D. 

\begin{figure}[t]
	\begin{centering}
		\includegraphics[width=.65\linewidth]{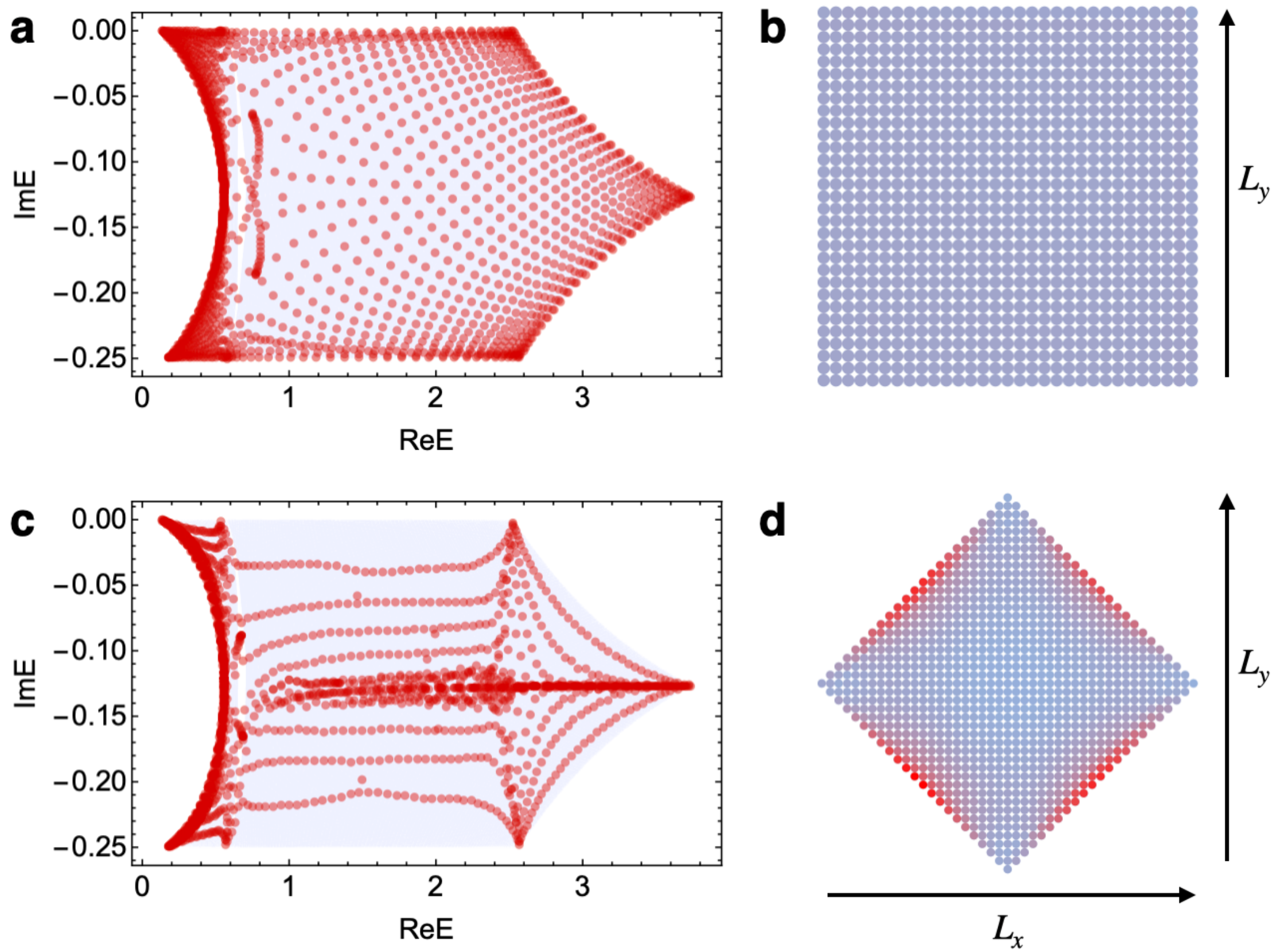}
		\par\end{centering}
	\protect\caption{\label{fig:S6} The periodic-boundary spectrum of the photonic crystal model is shown in (a)(c) with light blue color. Under the square geometry with the systems size $L_x=L_y=31$, the eigenvalues (red points) and the norm squared of all wave functions are shown in (a) and (b), respectively. Under the diamond geometry with the systems size $L_x=L_y=45$, the eigenvalues (red points) and the norm squared of all wave functions are plotted in (c) and (d), respectively. 
	}
\end{figure}

\subsection{Topological charge of non-Hermitian band degeneracies}

In this subsection, we will review the topological charge of the non-Hermitian band degeneracies. Based on the discriminant of the characteristic polynomial, one can define the topological charge of the degeneracy point $\mathbf{k}_D$, i.e. 
\begin{equation}
	\nu\left(\boldsymbol{k}_{D}\right)=\frac{1}{2 \pi i} \oint_{\Gamma\left(\boldsymbol{k}_{D}\right)} d \boldsymbol{k} \cdot \nabla_{\boldsymbol{k}} \ln \operatorname{Disc}_{E}[\mathcal{H}](\boldsymbol{k}).
\end{equation}
where $\Gamma\left(\boldsymbol{k}_{D}\right)$ is a loop encircling the degeneracy point $\mathbf{k}_D$. Since $\operatorname{Disc}_{E}[\mathcal{H}](\boldsymbol{k})$ is single valued, this invariant is quantized, which is called the discriminant number in Ref.~\cite{SM_YangPRL}. Putting 
\begin{equation}
	\operatorname{Disc}_{E}[\mathcal{H}](\boldsymbol{k})=\prod_{i<j}\left[E_{i}(\boldsymbol{k})-E_{j}(\boldsymbol{k})\right]^{2}
\end{equation}
into $\nu(\mathbf{k}_D)$, one can obtain 
\begin{equation}\begin{aligned}
	\nu(\mathbf{k}_D)
	&=
	\frac{1}{2\pi i}\oint_{\Gamma(\mathbf{k}_D)} d\mathbf{k} \cdot \nabla_{\mathbf{k}} \ln \prod_{1 \leq i<j \leq n}\left[E_{i}(\boldsymbol{k})-E_{j}(\boldsymbol{k})\right]^2\\
	&=
	\frac{1}{2\pi i}\sum_{i\neq j}\oint_{\Gamma(\mathbf{k}_D)} d\mathbf{k} \cdot \nabla_{\mathbf{k}} \ln \left[E_{i}(\boldsymbol{k})-E_{j}(\boldsymbol{k})\right]\\
	&=
	\frac{1}{2\pi}\sum_{i \neq j}\oint_{\Gamma(\mathbf{k}_D)} d\mathbf{k} \cdot \nabla_{\mathbf{k}} \arg \left[E_{i}(\boldsymbol{k})-E_{j}(\boldsymbol{k})\right].
\end{aligned}\end{equation}
Therefore, for a two-band system, 
\begin{equation}
\nu(\mathbf{k}_D)=\frac{1}{2\pi}\oint_{\Gamma(\mathbf{k}_D)} d\mathbf{k} \cdot \nabla_{\mathbf{k}} \arg \left[E_{+}(\boldsymbol{k})-E_{-}(\boldsymbol{k})\right]
\end{equation}
which describes the winding of the complex energy between two bands. Now we show a concrete example of the winding number. 
\begin{itemize}
	\item [~] {\bf Example}: Consider the following low energy Hamiltonian around $\mathbf{k}_D$, 
	\begin{equation}
		\mathcal{H}_1(\delta\mathbf{k})=\sigma_++(\delta k_x+i\delta k_y)\sigma_-,
	\end{equation}
    where $\delta\mathbf{k}=\mathbf{k}-\mathbf{k}_D$ and $\sigma_\pm=(\sigma_x\pm i\sigma_y)/2$. The eigenvalues of $\mathcal{H}(\delta\mathbf{k})$ are 
    \begin{equation}
    	E_\pm(\delta\mathbf{k})=\pm \sqrt{\delta k_x+i\delta k_y}.
    \end{equation}
    When $\mathbf{k}=\mathbf{k}_D$, which is equivalent to $\delta k_x=\delta k_y=0$, one can find $E_+(\delta\mathbf{k}=0)=E_-(\delta\mathbf{k}=0)=0$. This means $\mathbf{k}_D$ is a non-Hermitian degeneracy point. Now we choose $\Gamma(\mathbf{k}_D)=\mathbf{k}_D+\delta k_r(\cos\theta,\sin\theta)$, then, 
    \begin{equation}
    	E_\pm(\delta\mathbf{k})=\pm \delta k_r^{1/2}e^{i\theta/2},\qquad \theta\in (-\pi,\pi]. 
    \end{equation}
    One can find that $E_+(\delta\mathbf{k})$ and $E_-(\delta\mathbf{k})$ forms a spectral loop that encloses $E_+(\mathbf{k}_D)=E_-(\mathbf{k}_D)=0$. The winding number $\nu(\mathbf{k}_D)$ describes this topological properties of degeneracy points. 
\end{itemize}

The topological charge $\nu(\mathbf{k}_D)$ can be used to classify the non-Hermitian degeneracies. However, the classification is not complete. As a comparison with $\mathcal{H}_1(\delta\mathbf{k})$, consider the following two low energy Hamiltonians,
\begin{equation}
	\mathcal{H}_2(\delta\mathbf{k})=(\delta k_x+i\delta k_y)^2\sigma_++(\delta k_x-i\delta k_y)\sigma_-.
\end{equation}
Obvious, $\delta\mathbf{k}=0$ is a degeneracy point. One can further prove that its topological charge is $+1$, which is equal to the charge of $\delta\mathbf{k}=0$ in $\mathcal{H}_1(\delta\mathbf{k})$. However, these two degeneracy points have different properties. For example
\begin{equation}
	\mathcal{H}_1(\delta\mathbf{k}=0)=\sigma_+,\qquad 	\mathcal{H}_2(\delta\mathbf{k}=0)=0. 
\end{equation}
One can notice that $\mathcal{H}_1(\delta\mathbf{k}=0)$ is non-diagonal. This type of non-Hermitian degeneracy points are called exceptional points. In Ref.~\cite{SM_YangPRL}, the authors have shown that only the exceptional points with $\nu(\mathbf{k}_D)=\pm1$ are robust in 2D. Any other non-Hermitian band degeneracies are unstable against non-Hermitian perturbations. 

Having clarifying the topological charge of non-Hermitian degeneracies, now we can prove the corollary of our theorem. Since in 2D, the topological charge of the stable exceptional points must be $\pm1$, the corresponding spectrum area must be nonzero. 

\subsection{The photonic crystal model}

In this subsection, we numerically calculate the spectrum and spatial distribution of the wave function, i.e. $W(x)$ in Eq.~\ref{Wx}, for the photonic crystal model in the main text under different geometries. 

It shows that the skin effect disappears under square geometry in Fig.~\ref{fig:S6}(b), and reappears under diamond geometry in Fig.~\ref{fig:S6}(d), which is a characteristic signature of geometry-dependent-skin effect. Here we take the non-Hermitian parameter $\gamma$ as $1/4$. The spectrum under square geometry (red points in Fig.~\ref{fig:S6}(a)) coincides with the spectrum of periodic boundary (light blue region in Fig.~\ref{fig:S6}(a)(c)). We conjecture that the spectrum under diamond geometry (red points in Fig.~\ref{fig:S6}(d)) will also coincide with the periodic-boundary spectrum as the system size increases. However, it clearly shows that the density of states under different geometries is completely different. The dependence of density of states on the choice of boundary geometry is another significant feature of geometry-dependent-skin effect.

\begin{figure}[t]
	\begin{centering}
	\includegraphics[width=.85\linewidth]{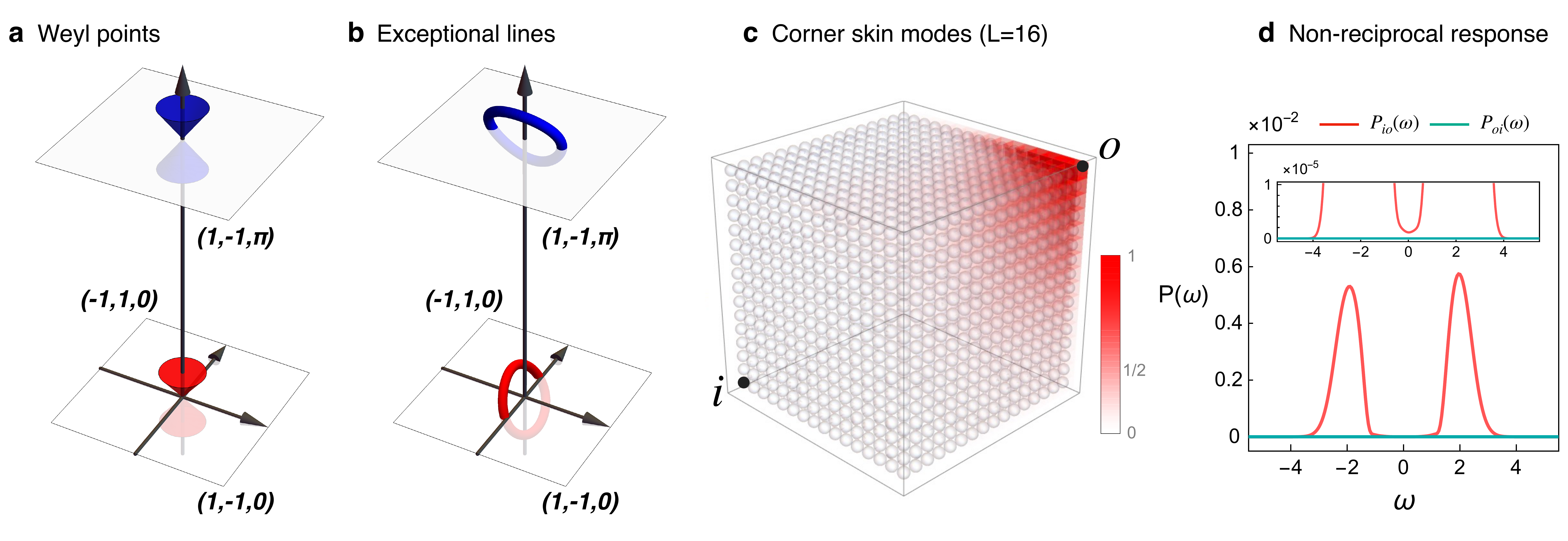}
	\par\end{centering}
	\protect\caption{\label{fig:S7} Two Weyl points (a) of a three-dimensional Weyl semimetal are expanded into two exceptional rings (b) after the addition of non-Hermitian perturbations. The spatial distribution of eigenstate is plotted in (c). 
	The modulus square of the propagator from $i$ to $o$ $P_{io}(\omega)$ and that from $o$ to $i$ $P_{oi}(\omega)$, as functions of $\omega$, are plotted with red color and dark cyan color in (d), respectively.}
\end{figure}

\subsection{The corner-skin effect in a three-dimensional exceptional-line semimetal}

We propose the realization for corner-skin effect in a three-dimensional system with exceptional lines. Consider a Weyl semimetal with non-Hermitian term as a perturbation, of which the periodic-boundary hamiltonian reads 
\begin{equation}\label{WeylModel}
	H(k) = [\bm{d_r}(k) + i \delta \bm{d_i}(k)] \cdot \bm{\sigma},
\end{equation}
where $\bm{d_r}(k)$ and $\bm{d_i}(k)$ are vectors with four components, that is, 
\begin{equation}
	\begin{split}
	&\bm{d_r}(k) = (0,\sin{k_x},\sin{k_y},2-\cos{k_x}-\cos{k_y}+\sin{k_z}), \\
	&\bm{d_i}(k) = (-\sqrt{5},1+\cos{k_z},1-\cos{k_z},\cos{k_z}).
	\end{split}
\end{equation}
The Hermitian part $\bm{d_r}\cdot \bm{\sigma}$ is a Weyl semimetal with two Weyl points. One Weyl point with $+1$ topological charge [red cone in Fig~\ref{fig:S7} (a)] is at $(0,0,0)$, and another with $-1$ topological charge [blue cone in Fig~\ref{fig:S7} (a)] is at $(0,0,\pi)$. 
Upon turning on the non-Hermitian term ($\delta\neq 0$), the Weyl points evolve into two exceptional rings as shown in Fig.~\ref{fig:S7} (b). According to our theorem, the system with exceptional lines must have the universal skin effect. Specially, the system described in Eq.~(\ref{WeylModel}) always has corner-skin effect as shown in Fig.~\ref{fig:S7} (c) with $\delta=1/6$. 

Due to the non-reciprocity of the corner-skin effect, we propose an experimental approach of two-point Green function to detect the corner-skin effect. We give a source at $i=(1,1,1)$ position, and probe it at $o=(16,16,16)$ position in Fig.~\ref{fig:S7} (c). The modulus square of the propagator from $i$ to $o$ is expressed as 
\begin{equation}
	P_{io}(\omega) =\sum_{\alpha,\beta =1,2}|\langle o,\beta| \frac{1}{\omega - \hat{H}} | i,\alpha \rangle |^2,
\end{equation}
where $\alpha,\beta$ label the orbitals of the unit cell. We calculate $P_{io}(\omega)$ and plot it with red color in Fig.~\ref{fig:S7} (d). We do the same process but interchange $i$ and $o$, and $P_{oi}(\omega)$ is plotted with dark cyan color in Fig.~\ref{fig:S7} (d). A significant difference between $P_{io}(\omega)$ and $P_{oi}(\omega)$ demonstrates the non-reciprocity of corner-skin effect. 

%


\end{document}